\pdfoutput=1
\documentclass[11pt]{article}
\usepackage{comment}
\excludecomment{IGNORE}
\usepackage{placeins}
\usepackage{graphicx}
\usepackage{subfigure,amssymb,amsmath}
\usepackage{cite,color,xcolor,url,cancel,ulem,float}
\usepackage{bm}
\usepackage{float}
\usepackage{hyperref}
\hypersetup{
  colorlinks   = true, 
  urlcolor     = blue, 
  linkcolor    = black, 
  citecolor   = blue 
}

\newcommand{\newc}{\newcommand}

\def\Ord{\lower .7ex\hbox{$\;\stackrel{\textstyle <}{\sim}\;$}}
\def\OOrd{\lower .7ex\hbox{$\;\stackrel{\textstyle >}{\sim}\;$}}

\newc{\order}{{\cal O}}

\newc{\be}{\begin{equation}}
\newc{\ee}{\end{equation}}
\newc{\br}{\begin{eqnarray}}
\newc{\er}{\end{eqnarray}}
\newc{\ba}{\begin{array}}
\newc{\ea}{\end{array}}
\newc{\bi}{\begin{itemize}}
\newc{\ei}{\end{itemize}}
\newc{\bn}{\begin{enumerate}}
\newc{\en}{\end{enumerate}}
\newc{\bc}{\begin{center}}
\newc{\ec}{\end{center}}
\newc{\ul}{\underline}
\newc{\ra}{\rightarrow}
\newc{\lra}{\longrightarrow}
\newc{\wt}{\widetilde}
\newc{\til}{\tilde}

\newc{\wh}{\widehat}
\newc{\ti}{\times}
\newc{\Dir}{\kern -6.4pt\Big{/}}
\newc{\Dirin}{\kern -10.4pt\Big{/}\kern 4.4pt}
\newc{\DDir}{\kern -10.6pt\Big{/}}
\newc{\DGir}{\kern -6.0pt\Big{/}}
\newc{\sig}{\sigma}
\newc{\sigmalstop}{\sig_{\lstoppair}}
\newc{\Sig}{\Sigma}  
\newc{\del}{\delta}
\newc{\Del}{\Delta}
\newc{\lam}{\lambda}
\newc{\Lam}{\Lambda}
\newc{\gam}{\gamma}
\newc{\Gam}{\Gamma}
\newc{\eps}{\epsilon}
\newc{\Eps}{\Epsilon}
\newc{\kap}{\kappa}
\newc{\Kap}{\Kappa}
\newc{\modulus}[1]{\left| #1 \right|}
\newc{\eq}[1]{(\ref{eq:#1})}
\newc{\eqs}[2]{(\ref{eq:#1},\ref{eq:#2})}
\newc{\etal}{{\it et al.}\ }
\newc{\ibid}{{\it ibid}.}
\newc{\ibidem}{{\it ibidem}.}
\newc{\eg}{{\it e.g.}\ }
\newc{\ie}{{\it i.e.}\ }

\newc{\nonum}{\nonumber}
\newc{\lab}[1]{\label{eq:#1}}
\newc{\dpr}[2]{({#1}\cdot{#2})}
\newc{\lt}{\stackrel{<}}
\newc{\gt}{\stackrel{>}}
\newc{\lsimeq}{\stackrel{<}{\sim}}
\newc{\gsimeq}{\stackrel{>}{\sim}}
\def\lsim{\buildrel{\scriptscriptstyle <}\over{\scriptscriptstyle\sim}}
\def\gsim{\buildrel{\scriptscriptstyle >}\over{\scriptscriptstyle\sim}}
\def\lapp{\mathrel{\rlap{\raise.5ex\hbox{$<$}}
                    {\lower.5ex\hbox{$\sim$}}}}
\def\gapp{\mathrel{\rlap{\raise.5ex\hbox{$>$}}
                    {\lower.5ex\hbox{$\sim$}}}}
\newc{\half}{\frac{1}{2}}

\newc{\bQ}{\ol{Q}}
\newc{\dota}{\dot{\alpha }}
\newc{\dotb}{\dot{\beta }}
\newc{\dotd}{\dot{\delta }}
\newc{\nindnt}{\noindent}

\newc{\matth}{\mathsurround=0pt}
\def\ML{\ifmmode{{\mathaccent"7E M}_L}
             \else{${\mathaccent"7E M}_L$}\fi}
\def\MR{\ifmmode{{\mathaccent"7E M}_R}
             \else{${\mathaccent"7E M}_R$}\fi}

\newc{\mr}{\mathrm}

\newc{\siminf}{\mbox{$_{\sim}$ {\small {\hspace{-1.em}{$<$}}}    }}
\newc{\simsup}{\mbox{$_{\sim}$ {\small {\hspace{-1.em}{$>$}}}    }}

\newc {\Zboson}{{\mathrm Z}^{0}}
\newc{\thetaw}{\theta_W}
\newc{\mbot}{{m_b}}
\newc{\mtop}{{m_t}}
\newc{\sm}{${\cal {SM}}$}
\newc{\as}{\alpha_s}
\newc{\aem}{\alpha_{em}}

\newc{\ppbar}{\mbox{$p\ol{p}$}}
\newc{\bbbar}{\mbox{$b\ol{b}$}}
\newc{\ccbar}{\mbox{$c\ol{c}$}}
\newc{\ttbar}{\mbox{$t\ol{t}$}}
\newc{\eebar}{\mbox{$e\ol{e}$}}
\newc{\zzero}{\mbox{$Z^0$}}

\newc{\wplus}{\mbox{$W^+$}}
\newc{\wminus}{\mbox{$W^-$}}
\newc{\ellp}{\ell^+}
\newc{\ellm}{\ell^-}
\newc{\elp}{\mbox{$e^+$}}
\newc{\elm}{\mbox{$e^-$}}
\newc{\elpm}{\mbox{$e^{\pm}$}}
\newc{\qbar}     {\mbox{$\ol{q}$}}

\newc{\Ebar}{{\bar E}}
\newc{\Dbar}{{\bar D}}
\newc{\Ubar}{{\bar U}}
\newc{\susy}{{{SUSY}}}
\newc{\msusy}{{{M_{SUSY}}}}

\def\photino{\ifmmode{\mathaccent"7E \gam}\else{$\mathaccent"7E \gam$}\fi}
\def\taugluino{\ifmmode{\tau_{\mathaccent"7E g}}
             \else{$\tau_{\mathaccent"7E g}$}\fi}
\def\mphotino{\ifmmode{m_{\mathaccent"7E \gam}}
             \else{$m_{\mathaccent"7E \gam}$}\fi}
\newc{\gl}   {\mbox{$\wt{g}$}}
\newc{\mgl}  {\mbox{$m_{\gl}$}}
\def \charginopm{{\wt\chi}^{\pm}}

\def \chonep {{\wt\chi_1^+}}

\def \ch2p {{\wt\chi_2^+}}
\def \chonem {{\wt\chi_1^-}}
\def \ch2m {{\wt\chi_2^-}}

\def \chonepm{{\wt\chi_1}^{\pm}}

\def \mchonepm{m_{\chonepm}}

\newc{\dmchi}{\Delta m_{\wt\chi}}

\def \lspone{\wt\chi_1^0}
\def \mlspone{m_{\lspone}}
\def \lsptwo{\wt\chi_2^0}
\def \mlsptwo{m_{\lsptwo}}

\newc{\sele}{\wt{\mathrm e}}
\newc{\sell}{\wt{\ell}}

\newc{\snue}     {\mbox{$ \wt{\nu_e}$}}
\newc{\smu}{\wt{\mu}}
\newc{\stau}{\wt{\tau}}
\newc {\nuL} {\wt{\nu}_L}
\newc {\nuR} {\wt{\nu}_R}
\newc {\snub} {\bar{\wt{\nu}}}
\newc {\eL} {\wt{e}_L}
\newc {\eR} {\wt{e}_R}

\def \stau{\wt\tau}

\def \sq{\wt{q}}

\newc{\msqot}  {\mbox{$m_(\sq_{1,2} )$}}
\newc{\sqbar}    {\mbox{$\bar{\wt{q}}$}}
\newc{\ssb}      {\mbox{$\squark\ol{\squark}$}}
\newc {\qL} {\wt{q}_L}
\newc {\qR} {\wt{q}_R}
\newc {\uL} {\wt{u}_L}
\newc {\uR} {\wt{u}_R}
\def \ul{\wt{u}_L}

\newc {\dL} {\wt{d}_L}
\newc {\dR} {\wt{d}_R}
\newc {\cL} {\wt{c}_L}
\newc {\cR} {\wt{c}_R}
\newc {\sL} {\wt{s}_L}
\newc {\sR} {\wt{s}_R}
\newc {\tL} {\wt{t}_L}
\newc {\tR} {\wt{t}_R}
\newc {\stb} {\ol{\wt{t}}_1}
\newc {\sbot} {\wt{b}_1}
\newc {\msbot} {m_{\sbot}}
\newc {\sbotb} {\ol{\wt{b}}_1}
\newc {\bL} {\wt{b}_L}
\newc {\bR} {\wt{b}_R}

\newc{\csquark}  {\mbox{$\wt{c}$}}
\newc{\csquarkl} {\mbox{$\wt{c}_L$}}
\newc{\mcsl}     {\mbox{$m(\csquarkl)$}}

\newc {\stopl}         {\wt{\mathrm{t}}_{\mathrm L}}
\newc {\stopr}         {\wt{\mathrm{t}}_{\mathrm R}}
\newc {\stoppair}      {\wt{\mathrm{t}}_{1}
\bar{\wt{\mathrm{t}}}_{1}}
\def \lstop{\wt{t}_{1}}

\def \lstoppair{\lstop\lstop^*}

\newc{\tsquark}  {\mbox{$\wt{t}$}}
\newc{\ttb}      {\mbox{$\tsquark\ol{\tsquark}$}}
\newc{\ttbone}   {\mbox{$\tsquark_1\ol{\tsquark}_1$}}

\newc{\mix}{\theta_{\wt t}}
\newc{\cost}{\cos{\theta_{\wt t}}}
\newc{\sint}{\sin{\theta_{\wt t}}}
\newc{\costloop}{\cos{\theta_{\wt t_{loop}}}}

\newc{\mixsbot}{\theta_{\wt b}}

\newc{\tb}{\tan\beta}
\newc{\cb}{\cot\beta}
\newc{\vev}[1]{{\left\langle #1\right\rangle}}

\newc{\mhalf}{m_{1/2}}
\newc{\mzero} {\mbox{$m_0$}}
\newc{\azero} {\mbox{$A_0$}}

\newc{\lb}{\lam}
\newc{\lbp}{\lam^{\prime}}
\newc{\lbpp}{\lam^{\prime\prime}}
\newc{\rpv}{{\not \!\! R_p}}
\newc{\rpvm}{{\not  R_p}}
\newc{\rp}{R_{p}}
\newc{\rpmssm}{{RPC MSSM}}
\newc{\rpvmssm}{{RPV MSSM}}

\newc{\sbyb}{S/$\sqrt B$}
\newc{\pelp}{\mbox{$e^+$}}
\newc{\pelm}{\mbox{$e^-$}}
\newc{\pelpm}{\mbox{$e^{\pm}$}}
\newc{\epem}{\mbox{$e^+e^-$}}
\newc{\lplm}{\mbox{$\ell^+\ell^-$}}

\def\Ecm{\ifmmode{E_{\mathrm{cm}}}\else{$E_{\mathrm{cm}}$}\fi}
\newc{\rts}{\sqrt{s}}
\newc{\rtshat}{\sqrt{\hat s}}
\newc{\gev}{\,GeV}
\newc{\mev}{~{\rm MeV}}
\newc{\tev}  {\mbox{$\;{\rm TeV}$}}
\newc{\gevc} {\mbox{$\;{\rm GeV}/c$}}
\newc{\gevcc}{\mbox{$\;{\rm GeV}/c^2$}}
\newc{\intlum}{\mbox{${ \int {\cal L} \; dt}$}}
\newc{\call}{{\cal L}}

\newc{\ptmiss}{/ \hskip-7pt p_T}

\newc{\PT}{\mbox{$p_T$}}
\newc{\ET}{\mbox{$E_T$}}
\newc{\dedx}{\mbox{${\rm d}E/{\rm d}x$}}
\newc{\ifb}{\mbox{${\rm fb}^{-1}$}}
\newc{\ipb}{\mbox{${\rm pb}^{-1}$}}
\newc{\pb}{~{\rm pb}}
\newc{\fb}{~{\rm fb}}
\newc{\ycut}{y_{\mathrm{cut}}}
\newc{\chis}{\mbox{$\chi^{2}$}}

\def \jet(s){\emph{jet(s) }}

\newc{\mpl}{M_{\rm Pl}}
\newc{\mgut}{M_{GUT}}
\newc{\mw}{M_{W}}
\newc{\mweak}{M_{weak}}
\newc{\mz}{M_{Z}}

\newc{\OPALColl}   {OPAL Collaboration}
\newc{\ALEPHColl}  {ALEPH Collaboration}
\newc{\DELPHIColl} {DELPHI Collaboration}
\newc{\XLColl}     {L3 Collaboration}
\newc{\JADEColl}   {JADE Collaboration}
\newc{\CDFColl}    {CDF Collaboration}
\newc{\DXColl}     {D0 Collaboration}
\newc{\HXColl}     {H1 Collaboration}
\newc{\ZEUSColl}   {ZEUS Collaboration}
\newc{\LEPColl}    {LEP Collaboration}
\newc{\ATLASColl}  {ATLAS Collaboration}
\newc{\CMSColl}    {CMS Collaboration}
\newc{\UAColl}    {UA Collaboration}
\newc{\KAMLANDColl}{KamLAND Collaboration}
\newc{\IMBColl}    {IMB Collaboration}
\newc{\KAMIOColl}  {Kamiokande Collaboration}
\newc{\SKAMIOColl} {Super-Kamiokande Collaboration}
\newc{\SUDANTColl} {Soudan-2 Collaboration}
\newc{\MACROColl}  {MACRO Collaboration}
\newc{\GALLEXColl} {GALLEX Collaboration}
\newc{\GNOColl}    {GNO Collaboration}
\newc{\SAGEColl}  {SAGE Collaboration}
\newc{\SNOColl}  {SNO Collaboration}
\newc{\CHOOZColl}  {CHOOZ Collaboration}
\newc{\PDGColl}  {Particle Data Group Collaboration}

\def\issue(#1,#2,#3){{\bf #1}, #2 (#3)}
\def\iss(#1,#2,#3){{\bf #1} (#3) #2}
\def\ASTR(#1,#2,#3){Astropart.\ Phys. \issue(#1,#2,#3)}
\def\AJ(#1,#2,#3){Astrophysical.\ Jour. \issue(#1,#2,#3)}
\def\AJS(#1,#2,#3){Astrophys.\ J.\ Suppl. \issue(#1,#2,#3)}
\def\APP(#1,#2,#3){Acta.\ Phys.\ Pol. \issue(#1,#2,#3)}
\def\JCAP(#1,#2,#3){Journal\ XX. \issue(#1,#2,#3)} 
\def\SC(#1,#2,#3){Science \issue(#1,#2,#3)}
\def\PRD(#1,#2,#3){Phys.\ Rev.\ D \issue(#1,#2,#3)}
\def\PR(#1,#2,#3){Phys.\ Rev.\ \issue(#1,#2,#3)} 
\def\PRC(#1,#2,#3){Phys.\ Rev.\ C \issue(#1,#2,#3)}
\def\NPB(#1,#2,#3){Nucl.\ Phys.\ B \issue(#1,#2,#3)}
\def\NPPS(#1,#2,#3){Nucl.\ Phys.\ Proc. \ Suppl \issue(#1,#2,#3)}
\def\NJP(#1,#2,#3){New.\ J.\ Phys. \issue(#1,#2,#3)}
\def\JP(#1,#2,#3){J.\ Phys.\issue(#1,#2,#3)}
\def\PL(#1,#2,#3){Phys.\ Lett. \issue(#1,#2,#3)}
\def\ZP(#1,#2,#3){Z.\ Phys. \issue(#1,#2,#3)}
\def\ZPC(#1,#2,#3){Z.\ Phys.\ C  \issue(#1,#2,#3)}
\def\PREP(#1,#2,#3){Phys.\ Rep. \issue(#1,#2,#3)}
\def\PRL(#1,#2,#3){Phys.\ Rev.\ Lett. \issue(#1,#2,#3)}
\def\MPL(#1,#2,#3){Mod.\ Phys.\ Lett. \issue(#1,#2,#3)}
\def\RMP(#1,#2,#3){Rev.\ Mod.\ Phys. \issue(#1,#2,#3)}
\def\SJNP(#1,#2,#3){Sov.\ J.\ Nucl.\ Phys. \issue(#1,#2,#3)}
\def\CPC(#1,#2,#3){Comp.\ Phys.\ Comm. \issue(#1,#2,#3)}
\def\IJMPA(#1,#2,#3){Int.\ J.\ Mod. \ Phys.\ A \issue(#1,#2,#3)}
\def\MPLA(#1,#2,#3){Mod.\ Phys.\ Lett.\ A \issue(#1,#2,#3)}
\def\PTP(#1,#2,#3){Prog.\ Theor.\ Phys. \issue(#1,#2,#3)}
\def\RMP(#1,#2,#3){Rev.\ Mod.\ Phys. \issue(#1,#2,#3)}
\def\NIMA(#1,#2,#3){Nucl.\ Instrum.\ Methods \ A \issue(#1,#2,#3)}
\def\EPJC(#1,#2,#3){Eur.\ Phys.\ J.\ C \issue(#1,#2,#3)}
\def\RPP (#1,#2,#3){Rept.\ Prog.\ Phys. \issue(#1,#2,#3)}
\def\PPNP(#1,#2,#3){ Prog.\ Part.\ Nucl.\ Phys. \issue(#1,#2,#3)}
\newc{\PRDR}[3]{{Phys. Rev. D} {\bf #1}, Rapid  Communications, #2 (#3)}

\def\PLB(#1,#2,#3){Phys.\ Lett.\ B  \iss(#1,#2,#3)}
\def\JHEP(#1,#2,#3){JHEP \iss(#1,#2,#3)}

\def\amu{a_\mu}

\def\tanbeta{{\rm tan}\beta}

\def\amususy{a_\mu^{\rm SUSY}}

\def\gmin2mu{(g-2)_\mu}
\def\gmin2e{(g-2)_e}
\def\gmin2{(g-2)_i}

\catcode`\@=11 

\def\vev#1{\left\langle #1\right\rangle}
\def\lsim{\mathrel{\mathpalette\@versim<}}
\def\gsim{\mathrel{\mathpalette\@versim>}}
\def\@versim#1#2{\vcenter{\offinterlineskip
    \ialign{$\m@th#1\hfil##\hfil$\crcr#2\crcr\sim\crcr } }}
\def\etal{{\em et. al.}}

\def\r2{\sqrt 2}
\def\beq{\begin{equation}}
\def\eeq{\end{equation}}
\def\beqn{\begin{eqnarray}}
\def\eeqn{\end{eqnarray}}
\def\sinW2{\sin^2\theta_W}

\def\mz2{M_{z}^2}
\def\c2b{\cos 2\beta}

\def\m#1{{\tilde m}_#1}

\def\mw#1{{\tilde m}_{\omega #1}}

\def\mz{M_Z}
\def\m0{m_0}
\def\mhalf{m_{\frac{1}{2}}}

\def\cb{\cos\beta}

\def\sec2w{sec^2\theta_W}

\def\amu{a_\mu}

\def\tanbeta{{\rm tan}\beta}
\def\amususy{a_\mu^{\rm SUSY}}
\def\gmin2{(g-2)_\mu}

\catcode`\@=11 

\def\vev#1{\left\langle #1\right\rangle}
\def\lsim{\mathrel{\mathpalette\@versim<}}
\def\gsim{\mathrel{\mathpalette\@versim>}}
\def\@versim#1#2{\vcenter{\offinterlineskip
    \ialign{$\m@th#1\hfil##\hfil$\crcr#2\crcr\sim\crcr } }}
\def\etal{{\em et. al.}}

\def\tb{\tilde b}

\def\tL{\tilde L}

\def \charginopm{{\wt\chi}^{\pm}}

\def \chonep{{\wt\chi_1}^{+}}
\def \chonem{{\wt\chi_1^-}}
\def \chonep2{{\wt\chi_2^+}}
\def \chonem2{{\wt\chi_2^-}}

\def \chonepm{{\wt\chi_1}^{\pm}}

\def \mchonepm{m_{\chonepm}}

\def \lstop{\wt{t}_{1}}

\def \lspone{\wt\chi_1^0}
\def \mlspone{m_{\lspone}}
\def \lsptwo{\wt\chi_2^0}
\def \mlsptwo{m_{\lsptwo}}

\def\PL{Phys. Lett.}
\def\PRL{Phys. Rev. Lett.}

\def\PR{Phys. Rev.}

\def \lsptwo{\wt\chi_2^0}
\def \lspone{\wt\chi_1^0}
\def \chonem {{\wt\chi_1^\pm}}
\def \chargino1 {{\wt\chi_1^\pm}}
\def \chargino2 {{\wt\chi_2^\pm}}
\def \lstop{\wt{t}_{1}}
\def \ch2m {{\wt\chi_2^-}}

\def \chonep {{\wt\chi_1^+}}

\def\lsim{\ ^<\llap{$_\sim$}\ }
\def\gsim{\ ^>\llap{$_\sim$}\ }

\usepackage[numbers,sort&compress]{natbib}
\begin{document}
\thispagestyle{empty}

\vspace{0.5cm}

\begin{center}
{\large
  {\bf Muon and Electron $(g-2)$ Anomalies with Non-Holomorphic Interactions in
    MSSM}}

Md. Isha Ali $^{1}$%
, Manimala Chakraborti$^{2}$%
, Utpal Chattopadhyay$^{1}$%
~and Samadrita Mukherjee$^{3}$%
~\footnote{emails: isha.ali080@gmail.com, mani.chakraborti@gmail.com, tpuc@iacs.res.in,\\samadrita.mukherjee@theory.tifr.res.in}

\vspace{0.5cm}
$^{1}$School of Physical Sciences, Indian Association
for the Cultivation of Science,
2A \& 2B Raja S.C. Mullick Road, Jadavpur,
Kolkata 700 032, India

\vspace{0.1cm}

$^{2}$Astrocent, Nicolaus Copernicus Astronomical Center of the Polish Academy of Sciences, ul.Rektorska 4, 00-614 Warsaw, Poland

\vspace{0.1cm}

$^{3}$Department of Theoretical Physics, Tata Institute of Fundamental
Research,
	1, Homi Bhabha Road, Colaba, Mumbai 400005, India
\end{center}
        
\vspace*{0.1cm}

\begin{abstract}
\noindent
  The recent Fermilab muon $g-2$ result and the same for electron due to fine-structure constant
measurement through ${}^{133}{\rm Cs}$ matter-wave
interferometry are probed in relation to MSSM with non-holomorphic (NH) trilinear soft SUSY breaking terms, referred to as NHSSM.
Supersymmetric contributions to charged lepton $(g-2)_l$ can be enhanced via the new trilinear terms
involving a wrong Higgs coupling with left and
right-handed scalars.  Bino-slepton loop is used to enhance the SUSY contribution to $g-2$ where wino mass stays
at 1.5 TeV and the left and right slepton mass parameters for the first two generations are considered to be
the same.  Unlike many MSSM-based analyses completed before,
the model does not require a light electroweakino, or
light sleptons, or unequal left and right slepton masses, 
or a very large higgsino mass parameter. In absence of popular UV complete
models, we treat the NH terms at par with MSSM soft terms, in a model independent framework of Minimal Effective
Supersymmetry. The first part of the analysis involves the study of
$(g-2)_\mu$ constraint along with the limits from Higgs mass, B-physics, collider data,
direct detection of dark matter (DM), while focusing on a higgsino DM which is underabundant in nature.
We then impose the constraint from electron $g-2$ where a large Yukawa threshold correction (an outcome of NHSSM) and opposite signs of 
trilinear NH coefficients associated with $\mu$ and $e$ fields
are used to satisfy the
dual limits of $\Delta {a_\mu}$ and $\Delta {a_e}$ (where the latter comes with negative sign). 
Varying Yukawa threshold corrections further provide the necessary 
flavor-dependent enhancement of $\Delta {a_e}/m_e^2$ compared to that of $\Delta {a_\mu}/m_\mu^2$.
A larger Yukawa threshold correction through $A^\prime_e$ for $y_e$ also takes away the direct
proportionality of $a_e$ with respect to $\tan\beta$. With a finite intercept, $a_e$ becomes only an
increasing function of $\tan\beta$.
We identified the available parameter space in the two cases while also satisfying the ATLAS data from slepton
pair production searches in the plane of slepton mass parameter and the mass of the lightest neutralino. 
\end{abstract}


\section{Introduction}
\label{intro}
The discovery of the 
Higgs boson\cite{ATLAS:2012yve,CMS:2012qbp} at the Large Hadron Collider (LHC)
almost a decade ago gave the Standard Model (SM) of particle physics\cite{Burgess:2006hbd} a strong foundation.
However, SM has its limitations both in the
theoretical as well as in the observational sides. The gauge hierarchy
problem, matter-antimatter asymmetry, no candidate for dark matter\cite{Bertone:2004pz, Jungman:1995df} are to name a few in this regard.
This demands the existence of a Beyond the Standard Model (BSM) physics. 
Low energy Supersymmetry (SUSY)\cite{Nilles:1983ge,Lykken:1996xt,Wess:1992cp,Drees:2004jm,Baer:2006rs,Binetruy:2012uov,Haber:1984rc,Martin:1997ns,Chung:2003fi}
is especially attractive in this context since it can
address the gauge hierarchy
problem associated with the Standard Model (SM) and also
it is able to provide with particle dark matter candidates. 
Additionally, we must not forget that the Higgs boson is found to have a  
mass of 125 GeV, which is well below the predicted upper limit for
an SM-like Higgs particle of the Minimal Supersymmetric Standard Model
(MSSM)\cite{Drees:2004jm,Baer:2006rs,Binetruy:2012uov,Haber:1984rc,Martin:1997ns,Chung:2003fi}.
Thus, over the past decades, SUSY, with its strong theoretical appeal and its ability to 
influence a variety of observables of phenomenological interest,       
continues to remain as the most attractive candidate for a 
BSM physics. 

Undoubtedly, a BSM physics demands nothing less than
direct observations of new particles at the LHC.
This will then lead us to the
possible new symmetries and interactions present in nature. However, even after a decade of
running of the LHC, we are yet to see a cherished new particle. 
We may as well need 
to accept the fact that BSM particles could perhaps be staying quite far
from our experimental reach. Keeping hope for a
collider discovery, we must at the same time continue to look
for possible indirect signatures of SUSY. This may come from flavor physics,  
electroweak physics precision tests and dark matter.
Concerning the above, we remember that the anomalous
magnetic moment of muon, $a_\mu = \frac{1}{2} (g-2)_\mu$,
stands out prominently over the past two decades showing some 
degree of disagreement (over 2 to 3$\sigma$) 
of the experimental result as obtained in Brookhaven\cite{Muong-2:2006rrc}
with that of SM evaluations performed at different times.
The hadronic vacuum polarization part of the SM result has a large
uncertainty, particularly the lowest order part of the same that requires analysis
in the non-perturbative regime. The non-perturbative aspect may require
input from effective field theory like chiral perturbation theory,
hadronic models, dispersion relations together with experimental data
like $e^+e^- \rightarrow {\rm hadrons} $, and Lattice Quantum
Chromodynamics (LQCD). A comprehensive analysis explaining 
the break-up of different
contributions to the SM result of $a_\mu$ may be seen in
Ref.\cite{Aoyama:2020ynm}\footnote{A chosen set of
references among various important past contributions to $(g-2)_\mu$ 
evaluation may be found in Ref.\cite{G2Theory}}. While
the recent Fermilab $a_\mu$ data\cite{Muong-2:2021vma,Muong-2:2021ojo}
is consistent with the same from Brookhaven,
the difference
$\Delta a_\mu$ has grown larger. The combined data from Fermilab and
Brookhaven show a $4.2 \sigma$ level of discrepancy\cite{Muong-2:2021ojo}
as given below.
\begin{equation}
\Delta a_\mu=a_\mu^{\rm exp}- a_\mu^{\rm SM}=(251 \pm 59)\times 10^{-11}.
\label{delta_amu_value}
\end{equation}
Interestingly, $\Delta a_\mu$ can be ascribed to $a_\mu^{\rm SUSY}$, the SUSY contributions to
$a_\mu$ which in turn will help us to constrain the SUSY model parameter space. 

With further results to come from the Fermilab in the near future and
the data from upcoming experiment JPARC\cite{Otani:2015jra}, muon $g-2$ can
shed light on various BSM physics models.  
In this context, we must point out the   
recent Lattice result\cite{LatticeBMW}
for the hadronic vacuum polarization. This has 
effectively shifted $a^{\rm SM}_\mu$ to move toward $a^{\rm exp}_\mu$ rather
closely causing tension between the dispersive and LQCD modes of
evaluations of the hadronic uncertainty amount within $a^{\rm SM}_\mu$.
We would also like to 
point out that an agreement between $a_\mu^{\rm exp}$ and $a_\mu^{\rm SM}$
may invite issues with global electroweak fits to electroweak
precision observables. 
This is because the existing deviation of the above two $a_\mu$ values is
related to precision electroweak predictions via the common dependence on
hadronic vacuum polarization
effects\cite{Crivellin:2020zul,Keshavarzi:2020bfy,Colangelo:2020lcg}. 
In any case, such important issues will be transparent  
in future, but at this point we will
use Eq.\ref{delta_amu_value} for 
$a_\mu^{\rm SUSY}$. 
 
On the top of $(g-2)_\mu$, we would also include the existing deviation for 
$(g-2)_e$, the anomalous magnetic moment for electron. A smaller but not insignificant discrepancy
exists for the electron $g-2$ anomaly 
arising out of the measurement of the fine-structure constant that used
${}^{133}{\rm Cs}$ matter-wave 
interferometry.  An approximately $2.5\sigma$ level of
discrepancy is given below\cite{Parker:2018vye}, 
\beq
  \Delta a_e=a_e^{\rm exp}- a_e^{\rm SM}=(-8.8 \pm 3.6)\times 10^{-13}.
\label{delta_ae}
\eeq
Unlike the above cases of muon and electron anomalies of
Eqs.\ref{delta_amu_value} and \ref{delta_ae} where they come with
opposite signs, a newer
measurement of fine-structure constant based on ${}^{87}{\rm Rb}$\cite{Morel:2020dww} shows a 1.6$\sigma$ deviation in the positive side. 
\beq
  \Delta {a_e}^{Rb}=a_e^{\rm exp}- a_e^{\rm SM}=(4.8 \pm 3.0)\times 10^{-13}.
\label{delta_aerb}
\eeq

In many new physics models with flavor universality, one finds 
$\frac{m_\mu^2}{m_e^2}\frac{\Delta a_{e}}{\Delta\amu} \simeq 1$.
This is also true in SUSY\footnote{except in cases involving an appreciably
large Yukawa threshold corrections as we will see.}. In contrast to
the above, the measurement values referred to in
Eqs.\ref{delta_amu_value} and \ref{delta_ae} lead
to an appreciably larger negative value for the above quantity,
namely:
\begin{equation}
R_{e,\mu}=\frac{m_\mu^2}{m_e^2}\frac{\Delta a_{e}}{\Delta\amu} \simeq -15.
\label{ratio_eqn}
\end{equation}
Thus, using the two constraints simultaneously leads to a rather difficult situation.
We note that the right hand side of Eq.\ref{ratio_eqn} is only the central value.
Appropriate error estimates of the two magnetic moments may be used for
obtaining the combined uncertainty values. On the other hand, use of
Eqs.\ref{delta_amu_value} and \ref{delta_aerb} lead to $R_{e,\mu} \simeq 8$.
Clearly, the later case of having simultaneous positive values for
the two deviations with also a smaller $R_{e,\mu}$ is easier to accommodate in
 SUSY analyses. In the absence of a resolution of the $\Delta a_e$ puzzle,
 we choose 
to consider the rather difficult ${}^{133}{\rm Cs}$-based value of Eq.\ref{delta_ae}.

This will analyze ${(g-2)}_{\mu,e}$ in the framework of    
Non-standard soft SUSY breaking terms\cite{Girardello:1981wz,Hall:1990ac} 
contrast it with other SUSY-based analyses
that also used the ${}^{133}{\rm Cs}$-based result of Eq.\ref{delta_ae}. 
We will also
show the result of using the ${}^{87}{\rm Rb}$ data briefly just
for the sake of completeness. 

Besides ${(g-2)}_{\mu,e}$ we will also include dark matter
constraints in our analysis.  The plan of our work is given below.
In Section~\ref{nhssmlabel} we will describe
  the Non-holomorphic MSSM (NHSSM) model and its signature on the SUSY
  spectra. We will also discuss the constraints arising from avoiding 
  charge and color breaking (CCB) minima.  Apart from the above, we will
  also mention the difference of status between the non-holomorphic soft parameters with the
  ones of regular MSSM soft terms in the context of ultraviolet (UV) completion.  
In Section~\ref{leptonicg-2} we will discuss the SUSY
  contributions to the magnetic moment of charged leptons. The above will
  also emphasize the role of Yukawa threshold corrections in MSSM that
  may be important for satisfying Eq.\ref{delta_ae}.  
We will then discuss the effect
  of non-holomorphic trilinear interactions on leptonic magnetic
  moments ${(g-2)}_l$ and how the trilinear NH terms may
  provide the necessary threshold effects appropriate for
  Eq.\ref{delta_ae}. 
  We will particularly 
  outline the parameter zone that would be consistent with a higgsino
  dark matter as a multi-component dark matter element with relic density
  obeying only the upper limit from the PLANCK data. Then we will discuss the
  combined case of obeying ${(g-2)}_\mu$ and 
  ${(g-2)}_e$ in MSSM and see how NHSSM effects can generate appropriate
  threshold corrections to $y_e$ and also to some lesser extent
to $y_\mu$. We will discuss the essential points of past MSSM-based analyses
  in contrast to the features of
  NHSSM.  We will see that there is no need to assume a flavor unfriendly
  choice for slepton 
  masses in NHSSM, neither we do we need to consider any superheavy higgsino
  state to generate
  Yukawa threshold corrections. In Section~\ref{resultsection} we will present the results in two separate parts namely ${(g-2)}_\mu$ with 
DM and inclusion of ${(g-2)}_e$ for the combined analysis.
We will also use LHC constraints from
slepton pair-production. Finally, we will conclude in
Section~\ref{conclusionsection}.

\section{MSSM With Non-Holomorphic Soft Terms}
\label{nhssmlabel}
\subsection{MSSM: Superpotential and Soft terms}
 The MSSM Superpotential is given by\cite{Drees:2004jm}, 
  \begin{eqnarray}
    \mathcal{W} = \mu H_D. H_U -Y_{ij}^e H_D.L_i {\bar E}_j -Y_{ij}^d H_D.Q_i {\bar D}_j  -Y_{ij}^u Q_i.H_U
            {\bar U}_j. 
\label{superpotential}
\end{eqnarray}
Here, for two doublet chiral superfields $A$ and $B$, one has
$A.B=\epsilon_{\alpha \beta}A^\alpha B^\beta$, where
$\epsilon_{\alpha \beta}$ is an antisymmetric (Levi-Civita)
tensor in 2-dimension. $Y_{ij}^e$, $Y_{ij}^u$ and $Y_{ij}^d$ are lepton, up and down type of Yukawa matrices respectively.  
 $H_D$ and $H_U$ with hypercharges -1 and 1 respectively 
refer to down and up type of
doublets of Higgs chiral superfields that contain both the Higgs scalars and
and their fermionic partners higgsinos.
$L_i,Q_i$ and $E_i,U_i$ are left handed doublet and right handed singlet
chiral superfields of applicable fermions and their scalar
superpartners.

The MSSM soft terms including the non-holomorphic scalar mass terms and
the holomorphic trilinear coupling terms are as given
below\cite{Drees:2004jm}.
 \begin{eqnarray}
   -{\mathcal{L}}_{soft}
   &=&[{\tilde q}_{iL}.h_u{(Y^u A_u)}_{ij} {\tilde u}^*_{jR}+
   h_d.{\tilde q}_{iL}{(Y^d A_d)}_{ij} {\tilde d}^*_{jR}+
   h_d.{\tilde l}_{iL}{(Y^e A_e)}_{ij} {\tilde e}^*_{jR}+ h.c.] \nonumber \\
   &+&\left(B\mu h_d.h_u + h.c.  \right) + m_d^2 {|h_d|}^2 +m_u^2 {|h_u|}^2
    \nonumber \\
   &+& {\tilde q}_{iL}^*{({M}_{\tilde q}^2)}_{ij} {\tilde q}_{jL}+ {\tilde u}_{iR}^*
   {({M}_{\tilde u}^2)}_{ij}
   {\tilde u}_{jR} +
   {\tilde d}_{iR}^*
   {({M}_{\tilde d}^2)}_{ij}
   {\tilde d}_{jR}
   +{\tilde l}_{iL}^*
   {({M}_{\tilde l}^2)}_{ij}
   {\tilde l}_{jL}+
   +{\tilde e}_{iR}^*
   {({M}_{\tilde e}^2)}_{ij}
   {\tilde e}_{jR} \nonumber \\
   &+& {\rm gaugino~mass~terms.}
\label{mssm_soft_terms}
\end{eqnarray}
Here, $h_d$ and $h_u$ are doublets of Higgs scalar fields. The other 
terms contain mass terms and trilinear terms involving the scalar parts
of the associated matter superfields. Finally, there are Majorana mass terms
involving the gauginos. 
With $v_u,v_d$ as the vacuum expectation values (vevs)
of the neutral components of Higgs scalar fields $h_u$ and $h_d$, one has  
$\tan\beta=v_u/v_d$ and $M_Z^2=\frac{1}{4}(g_Y^2+g_2^2)(v_u^2+v_d^2)$, leading
to $\sqrt{(v_u^2+v_d^2)}\simeq 246$~GeV. The Yukawa couplings and masses are
related via the vevs as
$y_e=\frac{m_e}{(v_d/\sqrt 2)}$, $y_u=\frac{m_u}{(v_u/\sqrt 2)}$ etc. 
The above $y_e$ relates to $Y_{ij}^e$ the leptonic Yukawa matrix as 
$Y_{11}^e=y_e,Y_{22}^e=y_\mu, Y_{33}^e=y_\tau$\footnote{The Yukawa matrices are 
assumed to be diagonal and the trilinear soft terms considered for
leptons given at the low scale 
are also diagonal corresponding to the Yukawa matrices.}.
Similar notation holds good for the quark Yukawa matrices.

\subsection{Non-holomorphic soft terms}
Ref.\cite {Girardello:1981wz} enumerated the MSSM
soft terms shown as $-{\mathcal{L}}_{soft}$ in Eq.\ref{mssm_soft_terms}. 
It also listed a few additional SUSY breaking interactions in a general 
sense that would be regarded as hard SUSY breaking terms in presence of a
gauge singlet scalar field\cite{Bagger:1993ji}. 
On the other hand, in absence of any such singlets as in MSSM,
such terms grouped within $-{\mathcal{L}'}_{soft}$ shown as below,  are
no longer of hard SUSY breaking type\cite{Bagger:1993ji,Hall:1990ac,Martin:1999hc} and these are 
labelled as non-holomorphic {\it soft} SUSY breaking terms or the so-called 
``C-terms'' in SUSY texts\cite{Baer:2006rs}. In MSSM these are regarded
as soft SUSY breaking terms. 
  \begin{eqnarray}
    -{\mathcal{L}'}_{soft} = h_d^c.{\tilde q}_{iL}{(Y^u A_u^\prime)}_{ij}
    {\tilde u}^*_{jR}+
  {\tilde q}_{iL}.h_u^c{(Y^d A_d^\prime)}_{ij} {\tilde d}^*_{jR}+
  {\tilde l}_{iL}.h_u^c {({Y^e A}_e^\prime)}_{ij} {\tilde e}^*_{jR}+
  \mu' {\tilde h}_u.{\tilde h}_d+ h.c.
  \label{nonstandardsoftterms}
  \end{eqnarray}  
  Here, instead of the Higgs scalar doublets one has their conjugates
  $h_d^c$ and $h_u^c$. 
  With appropriate hypercharges, 
  $h_d^c$ couples with the up-type of 
  squarks and $h_u^c$ goes with the down type of squarks and sleptons.
This is why the above is often referred as a scenario with
wrong Higgs coupling.

Ref.\cite{Hall:1990ac} analyzed such terms (along with the MSSM soft terms)
in a model independent way.  Instead of a model based analysis, in an
agnostic point of view the authors named the framework as 
``{\it Minimal Effective Supersymmetry}'' where the new soft 
parameters considered were to be treated at par with the ones
of Eq.\ref{mssm_soft_terms}  and these were left to be 
determined from low energy data only. 

In the quest of a model, Ref.\cite{Martin:1999hc} considered 
a hidden sector based F-type supesymmetry breaking scenario
including two chiral superfields.
The author obtained terms of $-{\mathcal{L}'}_{soft}$ terms to be of the
order $\frac{{|F|}^2}{M^3}\sim \frac{M_W^2}{M}$ indicating 
suppression by the scale of mediation $M$ of SUSY breaking. 
Arising from the same analysis, there is no such supression for the
MSSM soft terms of Eq.\ref{mssm_soft_terms}.
Clearly, according to the model of Ref.\cite{Martin:1999hc}
these are highly suppressed terms
if the scale of mediation is close to the grand unification theory (GUT)
scale or the Planck scale. However, Ref.\cite{Martin:1999hc} also pointed out
that the above does not make the terms of $-{\mathcal{L}'}_{soft}$ irrelevant
for issues starting from a possible incorrectness in using a simplistic form of
spontaneous SUSY breaking to involvement of multiple high scales etc. 

It is known that building models for nonstandard
soft SUSY breaking terms or the ``C-terms'' is
difficult\cite{Baer:2006rs}. 
We cite a few available analyses here. 
Refs.\cite{Chakraborty:2019wav,Chakraborty:2018izc,Nelson:2015cea} may be
seen for analyses with generalized supersoft SUSY 
breaking that have relations to non-holomorphic soft terms. Generating 
non-holomorphic terms based on gauge mediated SUSY breaking may be seen in
Ref.\cite{Haber:2007dj}. Here, the authors discussed the
possibility of finding gauge invariant
supersymmetric direct Yukawa couplings between the Higgs and the messenger
fields. One-loop corrections involving messenger fields in the loop may
lead to the wrong-Higgs gaugino operators\cite{Haber:2007dj}
that may become important
in the low energy theory below the scale of SUSY breaking\footnote{A gauge mediated susy breaking (GMSB) inspired analysis was studied
in Ref.\cite{Chattopadhyay:2017qvh}, however here the bilinear higgsino mass
soft term was assumed to have an unknown origin.}.
Ref.\cite{Buican:2008qe} relates to 
non-holomorphic terms arising out of D-brane instantons stretched
between SUSY breaking and visible sectors. 

It is clear 
that the NH terms of $-{\mathcal{L}'}_{soft}$  are not friendly to
supergravity\cite{Hall:1983iz_sugra,Martin:1999hc} types of scenarios or they may be difficult to be found 
from a popular UV complete theory.      
Similar to Ref.\cite{Baer:2006rs,Martin:1999hc} we are
also of the opinion that the NH terms are hardly irrelevant in spite of
their difficulty in model building and we may consider 
the approach of ``Minimal Effective Supersymmetry'' of
Ref.\cite{Hall:1990ac}. 
Based on the inputs given at a high or a low scale we classify below the past
phenomenological analyses with 
non-standard soft terms that considered no explicit model 
or in other words were consistent with 
the ``Effective'' approach of Ref.\cite{Hall:1990ac}.  
Analyses of Refs.\cite{Jack:1999ud,Jack:nh1,Jack:2004dv,Hetherington:2001bk,Sabanci:2008qp,Solmaz2009,Un:2014afa,Ross:2016pml,Ross:2017kjc,Hicyilmaz:2021onw} used renormalization group evolutions within a Constrained MSSM 
(CMSSM\cite{Drees:2004jm}) like setup. 
Here the NH parameters were considered to be of unknown origin and were at par
with other CMSSM mass and trilinear parameters. Similarly, 
there are works with phenomenological MSSM (pMSSM)\cite{MSSMWorkingGroup:1998fiq} like
inputs\cite{Chattopadhyay:2016ivr,Beuria:2017gtf,Chattopadhyay:2018tqv,Chattopadhyay:2019ycs} where 
the NH parameters given at a low scale were treated at par with the MSSM 
soft SUSY breaking parameters. 
Thus as with previous analyses, we consider the new parameters to be of
  unknown origin given at a low scale. 
  The tree-level Higgs potential is unaffected, 
  but this is not so for the charge 
  and color 
  breaking terms of the scalar potential\cite{Beuria:2017gtf}.
  The presence of the
  higgsino mass soft terms with coupling $\mu^\prime$
   may cause isolation of a fine-tuning  
   from the Higgsino mass $\mu$ since at tree level higgsino mass would
  have components from the superpotential (Eq.\ref{superpotential})
  as well as from the soft term of (Eq.\ref{nonstandardsoftterms})\cite{Ross:2016pml, Ross:2017kjc,Chattopadhyay:2016ivr}. 
The mass matrices for the scalars get modified in the off-diagonal component
involving L-R mixing. For example, a slepton mass matrix may be written as,  
\begin{eqnarray}
\label{slepton_mass}
M_{\tilde{e}}^2=&\left(\begin{matrix}
M_{\tilde{l_L}}^2+M_Z^2(T_{3L}^{\tilde{e}}-Q_e\sin^2\theta_{W})\cos2\beta +m_e^2 &\hspace{1mm}
-m_e({A}_e-(\mu+{A}_e')\tan\beta) \\
\hspace{-3mm} -m_e({A}_e-(\mu+{A}_e')\tan\beta) & \hspace{-6mm}
M_{\tilde{l_R}}^2+M_Z^2 Q_e\sin^2\theta_{W}\cos2\beta +m_e^2
\end{matrix} \right).\hspace{6mm}
\end{eqnarray}
We note that going from MSSM to NHSSM, $\mu\tan\beta$ gets replaced
by $(\mu + {A}_l^\prime)\tan\beta$ in the off-diagonal entries. 
In the electroweakino sector the higgsino mass entries are altered from
$\mu$ to $\mu+\mu'$ leading to the following neutralino and chargino
mass matrices in NHSSM. 
\begin{eqnarray}
\label{neumat}
 M_{\widetilde{\chi^0}}=&\left(\begin{matrix}
                        M_1 & 0 & -M_Z\cos\beta \sin\theta_W & M_Z\sin\beta \sin\theta_W \\
                        0 & M_2 & M_Z\cos\beta \cos\theta_W & -M_Z\sin\beta \cos\theta_W \\
                       -M_Z\cos\beta \sin\theta_W & M_Z\cos\beta \cos\theta_W & 0 & -(\mu+\mu')\\
                        M_Z\sin\beta \sin\theta_W & -M_Z\sin\beta \cos\theta_W & -(\mu+\mu') & 0
                                           \end{matrix} \right).\hspace{6mm}
\end{eqnarray}
\begin{eqnarray}
\label{charmat}
  M_{\widetilde{\chi^{\pm}}}=&\left(\begin{matrix}
                                 M_2 & \sqrt{2} M_W\sin\beta \\
                                 \sqrt{2} M_W\cos\beta & -(\mu+\mu')\\
                                \end{matrix}\right).
\end{eqnarray}
The present analysis will consider vanishing $\mu^\prime$.

\subsection{Charge and Color Breaking:}
Avoiding a Charge and Color Breaking (CCB) minima in
NHSSM\cite{Beuria:2017gtf} while considering both the
holomorphic and non-holomorphic 
trilinear couplings requires a 4-vev scenario, like the vevs for $H_u$, $H_d$,
${\tilde f}_L$ and ${\tilde f}_R$. Here $\tilde f$ stands for the concerned
sfermion. With $A_f$ and $A_f^\prime$ both present
there is no possibility of considering a 3-vev scenario unlike in MSSM.  
A rather straigtforward computation as shown in Ref.\cite{Beuria:2017gtf}
results into the following inequalities for avoiding a CCB minima.
It is seen that unlike MSSM, there
is no D-flat direction so that terms with $g_1^2+g_2^2$ come into the
picture arising out of the D-term potential.
\begin{eqnarray}
\label{ccbconst}
\left[|A_t|+|\mu| + |A_t'|
\right]^2 &<& 3 \left( m_1^2 + m_2^2 +
m_{\tilde{t}_L}^2 + m_{\tilde{t}_R}^2  -2B_{\mu}\right),  \nonumber \\
\left[|A_b|+|\mu| + |A_b'|
\right]^2 &<& 3 \left\{1+\frac{g_1^2+g_2^2}{24y_b^2}\right\}\left
( m_1^2 + m_2^2 + m_{\tilde{b}_L}^2 + m_{\tilde{b}_R}^2  -2B_{\mu}\right),
\nonumber \\
\left[|A_\tau|+|\mu| + |A_\tau'|
\right]^2 &<& 3 \left\{1+\frac{g_1^2+g_2^2}{24y_\tau^2}\right\}\left
( m_1^2 + m_2^2 + m_{\tilde{\tau}_L}^2 + m_{\tilde{\tau}_R}^2
-2B_{\mu}\right),   
\end{eqnarray}
where $m_{1,2}^2=m_{{H_d},{H_u}}^2+\mu^2$. 
One finds that even with a very large $A_e^\prime$ as in our analysis, the
above constraint is easily satisfied because of the $g_1^2+g_2^2$ 
term that becomes very large due to a small $y_e$ in the denominator. This is
entirely different from the MSSM case that has a D-flat direction coming out
in a 3-vev based scenario\footnote{For MSSM one has the following:
  $A_\tau^2 < 3(m_1^2 + m_{\tilde{\tau}_L}^2 + m_{\tilde{\tau}_R}^2)$\cite{Binetruy:2012uov,Drees:1985ie,Chowdhury:2013dka,Chattopadhyay:2014gfa}.
}.

Apart from a global vacuum stability where a lot of MSSM parameter space
can be excluded for a large value of $A_f$ corresponding to the
first two generations of leptons ($f\equiv e,\mu$), we must point out
that for a long-lived universe, the CCB conditions corresponding to
the light fermion cases are readily evaded.
The above is quite commonly used in SO(10) based analyses (e.g
Ref.\cite{Fukuyama:2016mqb}) to label a parameter point valid even
when the absolute stability is affected. This is 
true as long as the CCB inequalities are satisfied for the large Yukawa
coupling cases i.e.
the inequalities involving the third generation trilinear
couplings like $A_t$.
The rate of tunneling from the Standard Model like false vacuum to a CCB
true vacuum is 
proportional to $e^{-a/y^2}$\cite{Chattopadhyay:2014gfa}
(follows from Eq. 9 and 11 of Ref.\cite{Kusenko:1996jn}),
where $a$ is a constant and $y$ is the associated Yukawa
coupling for the colored/charged fields.
If the dangerous third generation of sfermion
constraints are already avoided, the rate of tunneling corresponding to a 
small Yukawa case may be very small.
As mentioned in Ref.\cite{Kusenko:1996jn}
these are the cases where D-term contributions to the potential
cannot be neglected. NHSSM with large values of $A_e^\prime$ also falls into
this class and it is thus additionally consistent with a long-lived universe
consideration. 
Thus for NHSSM, the inqualities involving
$A_e$, $A_e^\prime$ and $A_\mu$, $A_\mu^\prime$ are either satisfied
for absolute stability  
or they can be ignored as in MSSM for a long-lived universe 
leading to cosmological stability\cite{Fukuyama:2016mqb,Kusenko:1996jn}.

\section{Leptonic $(g-2)$ and Yukawa Threshold Corrections in MSSM}
\label{leptonicg-2}
\begin{figure} 
\begin{center}
\includegraphics[width=\textwidth]{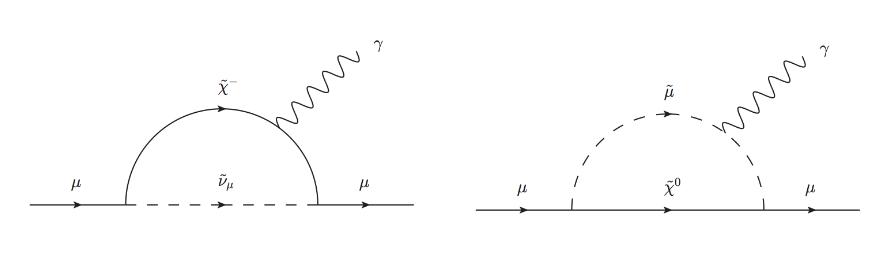}
\end{center}
\caption{One-loop contributions to $\amususy$}
\label{one-loop-diags}
\end{figure}

In MSSM, as shown in Fig.\ref{one-loop-diags}, at the one-loop level
the leading contributions to $\amususy$ come from
${\tilde \chi}^0-\tilde \mu$ and ${\tilde \chi}^\pm -\tilde \nu_\mu$
loops\cite{muong1,Endo_muong}.
The required chirality flip may be found from a SUSY Yukawa coupling of a higgsino to a lepton, an appropriate slepton ($\tilde \mu$) or sneutrino ${\tilde \nu}_\mu$. Otherwise, the chirality flip may be associated at a slepton
$\tilde \mu$ line corresponding to the transition
${\tilde \mu}_L-{\tilde \mu}_R$\cite{Stockinger:2006zn}.
Using Ref.\cite{Martin:2001st} 
the one-loop contributions are given as follows. 

\begin{eqnarray}
a_\mu^{{\tilde \chi}^0} & = & \frac{m_\mu}{16\pi^2}
   \sum_{i,m}\left\{ -\frac{m_\mu}{ 12 m^2_{\tilde\mu_m}}
  (|n^L_{im}|^2+ |n^R_{im}|^2)F^N_1(x_{im}) 
 +\frac{m_{{\tilde \chi}^0_i}}{3 m^2_{\tilde \mu_m}}
    {\rm Re}[n^L_{im}n^R_{im}] F^N_2(x_{im})\right\}\phantom{xxxx}\\
a_{\mu}^{{\tilde \chi}^\pm} & = & \frac{m_\mu}{16\pi^2}\sum_k
  \left\{ \frac{m_\mu}{ 12 m^2_{\tilde\nu_\mu}}
   (|c^L_k|^2+ |c^R_k|^2)F^C_1(x_k)
 +\frac{2m_{{\tilde \chi}^\pm_k}}{3m^2_{\tilde\nu_\mu}}
         {\rm Re}[ c^L_kc^R_k] F^C_2(x_k)\right\}\phantom{xxxx}
\label{electroweakino-contributions}
\end{eqnarray}
where $i$ and $m$ refer to the four neutralino and two chargino states whereas
$k$ indicates the two smuon states. The referred couplings are given by,
\begin{eqnarray}
n^R_{im} & = &  \sqrt{2} g_1 N_{i1} X_{m2} + y_\mu N_{i3} X_{m1},\\
n^L_{im} & = &  {1\over \sqrt{2}} \left (g_2 N_{i2} + g_1 N_{i1}
\right ) X_{m1}^* - y_\mu N_{i3} X^*_{m2},\\
c^R_k & = & y_\mu U_{k2}, \\
c^L_k & = & -g_2V_{k1}.
\end{eqnarray}
The loop functions for the neutralino and chargino loops namely
$F^{N,C}_{1,2}$ may be seen in Ref.\cite{Martin:2001st}.
A simplified result follows when the loops that contribute most are the ones
with chargino-sneutrino and bino-smuon fields\cite{Badziak:2019gaf}.  
\begin{eqnarray}
\label{amu_chargino_loop}
a_{\mu}^{{\tilde \chi}^\pm} &\simeq &
\frac{\alpha_2 \, m^2_\mu \, \mu\,M_{2} \tan\beta}
{4\pi \sin^2\theta_W \, m_{\tilde{\nu}_\mu}^{2}}
\left( \frac{f_{\chi^{\pm}}(M_{2}^2/m_{\tilde{\nu}_{\mu}}^2)
-f_{\chi^{\pm}}(\mu^2/m_{\tilde{\nu}_{\mu}}^2)}{M_2^2-\mu^2} \right) \, ,
\\
\label{amuslep}
a_\mu^{{\tilde \chi}^0}  &\simeq &
\frac{\alpha_1 \, m^2_\mu \, \,M_{1}(\mu \tan\beta- {A}_\mu)}
{4\pi \cos^2\theta_W \, (m_{\tilde{\mu}_R}^2 - m_{\tilde{\mu}_L}^2)}
\left(\frac{f_{\chi^0}(M^2_1/m_{\tilde{\mu}_R}^2)}{m_{\tilde{\mu}_R}^2}
- \frac{f_{\chi^0}(M^2_1/m_{\tilde{\mu}_L}^2)}{m_{\tilde{\mu}_L}^2}\right)\,,
\label{amu_bino_smuon_loop}
\end{eqnarray}
where the loop functions $f$ are as given in Ref.~\cite{Badziak:2019gaf}.

\subsection{Yukawa Threshold Corrections}
\label{thresholddetailsnow}
In MSSM, the Yukawa couplings are modified because of soft interactions.
We would now like to discuss Yukawa threshold corrections for fermions that
affects ${(g-2)}_l$ as a higher order effect.
At one-loop level, the lepton Yukawa coupling in MSSM can be given as,
\begin{equation}
 y_{l} = \frac{m_l}{(v/ \sqrt 2)}\sqrt{1+\tan^2\beta} \frac{1}{1+\Delta_l} \left(1+ O(\cot\beta)\right).
 \label{yukawas}
\end{equation}
Here $\Delta_l$ in MSSM is given by~\cite{Marchetti:2008hw},
\begin{eqnarray}
\Delta_l &=& -\frac{g_2^2\ M_2}{16\pi^2}\tan\beta \Big[
\mu I(m_1,m_2,m_{{{\tilde{\nu}}_l}}) + \mu \frac{1}{2} I(m_1,m_2,m_{\tilde{l}_L})\Big] \nonumber \\
&\quad&
-\frac{g_1^2\ M_1}{16\pi^2}\tan\beta
\Big[\mu I(\mu,M_1,m_{\tilde{l}_R}) -
\mu \frac{1}{2}I(\mu,M_1,m_{\tilde{l}_L})-
\mu I(M_1,m_{\tilde{l}_L},m_{\tilde{l}_R})
\Big].
\label{radcoreqn}
\end{eqnarray}
Here, $m_{1,2}$ are the chargino masses and the loop function $I$ is
as given in Ref.~\cite{Marchetti:2008hw}. As we will discuss
for NHSSM, the last term $\mu I(M_1,m_{\tilde{l}_L},m_{\tilde{l}_R})$ will
be altered via $\mu \rightarrow (\mu+A_l^\prime)$. 
Including these corrections, the one-loop contributions get modified as,
\begin{align}
a_l^{\rm SUSY,1L} \longrightarrow a_l^{\rm SUSY,1L} \frac{1}{1+\Delta_l}.
\label{a_e_scaling}
\end{align}
$\Delta_l$, which is proportional to $\tan\beta$, contains the all order
re-summation of the $\tan\beta$ enhanced contributions\cite{Altmannshofer:2010zt, Carena:1999py}.
We further note that with the assumption that all the SUSY masses
are equal and much larger than the W-boson mass $M_W$,
$\Delta_\mu=-0.0018\tan\beta ~{\rm sign (\mu)}$~\cite{Marchetti:2008hw,Altmannshofer:2010zt}.
$\Delta_\mu$ may be much larger for unequal SUSY particle mass parameters~\cite{Marchetti:2008hw}. 
A larger $\Delta_l$ that is itself proportional to $\tan\beta$,
can even influence $a_l$'s $\tan\beta$-dependence via $y_l$. In
the limit of
radiative mass generations of fermions, $a_l$ can become almost independent of $\tan\beta$\cite{Borzumati:1999sp}.  
Such effects of the radiative generation of mass of fermions
in analyzing $g-2$ were 
considered in Refs.\cite{Borzumati:1999sp,Marchetti:2008hw,Crivellin:2010ty,Crivellin:2011jt, Thalapillil:2014kya,
Bach:2015doa,Baker:2021yli}. The authors pointed out the role of non-holomorphic trilinear interactions 
on lepton Yukawa couplings $y_l$ and $(g-2)_l$.
A similar control of $(g-2)_e$ via  
enhancement of Yukawa threshold corrections in MSSM was used in
Ref.\cite{Endo:2019bcj} where the authors considered very large values of
$\mu$ (up to 500 TeV). We will discuss some details of the
work in Sec.\ref{accommodategualg}.

Following Sec.\ref{nhssmlabel}, an off-diagonal element of a
slepton matrix would look like
$m_l(A_l-(\mu+A_l^\prime)\tan\beta)$, indicating a generic 
alteration of $\mu$ of MSSM going to $\mu+A_l^\prime$ in NHSSM wherever
there is an L-R mixing. The same
will be true in the last term (L-R) of the last line
of Eq. \ref{radcoreqn}. Since for an electron the
aforesaid off-diagonal term is multiplied by $m_e$  which is too small,
one must have a very large 
value of $(\mu+A_e^\prime)$ so as to get a finite L-R mixing effect in relation
to the diagonal terms. With
an electroweak fine-tuning-friendly low $\mu$, we see that $A_l^\prime$
has to be very large so that $\mu$ may be ignored in $\mu+A_l^\prime$.
Thus, with an appropriately large 
${A}^\prime_l$, $\Delta_l$ in NHSSM may
be approximately proportional to ${A}^\prime_l$,
and the effect on $y_l$ can be much larger than what 
may be possible in MSSM due to a potentially large
enhancement factor $1/(1+\Delta_e)$.
We will demonstrate how the radiatively corrected Yukawa 
coupling $y_e$ may become very important  
while discussing the case of $a_e$ in Sec.\ref{resultsection}.
Additionally, the sign correlation of $a_l$ with respect to
$A_l^\prime$ can be ascertained from the relevant bino-slepton loop 
result of Eq.\ref{amu_bino_smuon_loop} when one changes $\mu$ to
$\mu+A_l^\prime$.
Since $\Delta_e$ is proportional to $\tan\beta$
we will see that the threshold corrections to $y_e$ will cause
$a_e$ to be a slowly increasing function of $\tan\beta$ with some intercept,
but unlike MSSM it would no longer be proportional to
$\tan\beta$ (Eq.\ref{a_e_scaling}).

      There have been an appreciable number of SUSY-based
      analyses on ${(g-2)}_\mu$
after the announcement of Fermilab result\cite{Chakraborti:2021kkr,Endo:2021zal,Iwamoto:2021aaf,Gu:2021mjd,
VanBeekveld:2021tgn,Yin:2021mls,Abdughani:2021pdc,Cao:2021tuh,
Chakraborti:2021dli,Ibe:2021cvf,Cox:2021nbo,Han:2021ify,
Heinemeyer:2021opc,Baum:2021qzx,Zhang:2021gun,Ahmed:2021htr,Yang:2021duj,
Athron:2021iuf,Aboubrahim:2021xfi,Baer:2021aax,Altmannshofer:2021hfu,
Aboubrahim:2021phn,Aboubrahim:2021ily,Chakraborti:2021mbr,Dutta:2018fge,Banerjee:2020zvi,
Crivellin:2018qmi,Frank:2021nkq,Cao:2021lmj,Li:2021koa}.
Some of the above works also
involve dark matter constraints. DM relic density could be satisfied via
higgsino or wino type of LSPs provided one considers underabundant
scenarios with the possibility of multiple candidates for DM. In this case,
$\charginopm-{\tilde \nu}_\mu$ loop may contribute dominantly to $a_\mu$. One can similarly consider
bino-wino mixed LSP in this regard. One can also consider bino type of LSP
which obtains correct relic density by
self-annihilation via s-channel H/A boson, with $a_\mu$ constraint
being satisfied via the contribution from the bino-smuon loop with relatively light smuons.
Alternatively, bino-stau coannihilation may be used for
DM relic density generation and $a_\mu$ constraint may be similarly addressed. As we will
see in our work with nonstandard trilinear soft terms,
we consider higgsino to be the LSP (in an underabundant choice for DM).
Regarding $a_\mu$, we use the contribution from the bino-smuon loop
that is enhanced by larger L-R mixing due to the above soft terms. We will
further see that the large threshold corrections (due to the nonstandard soft
terms) to leptonic Yukawas, particularly $y_e$, can be useful to accommodate 
the $a_e$ constraint simultaneously. 

\subsection{Accommodating $a_e$ constraint in addition to $a_\mu$ limits}
\label{accommodategualg}
Eq.\ref{ratio_eqn} summarizes the requirement for a new physics
to accommodate both the constraints. The ratio is required to be not only
large but also negative. We would like to address the 
essential parts of a few past MSSM analyses in this regard. 
Refs.\cite{Badziak:2019gaf,Endo:2019bcj,Dutta:2018fge} analyzed the
$a_\mu$ and $a_e$ constraints in the context of MSSM. Ref.\cite{Dutta:2018fge}
used 1-3 flavor violation in the bino-slepton loop to get the desired
outcome for $a_e$ while satisfying the
$\tau \rightarrow e \gamma$ bound. There was no flavor violation to use for
$a_\mu$. 
The essential focus 
of Ref.\cite{Badziak:2019gaf} was to satisfy $(g-2)_\mu$ and $(g-2)_e$ either
via the lighter chargino-sneutrino loop or via the neutralino-slepton loop
diagrams 
with appropriate signs of the U(1) and SU(2) gaugino masses $M_1$ and $M_2$.
The analysis of Ref.\cite{Badziak:2019gaf} that could only have a light
electroweakino spectra, used 
different mass parameters for the sleptons of the first 
two generations (apart from the sneutrinos). The 
work considered differing right and left slepton mass parameters, 
a highly unfriendly choice for flavor.
Clearly, such large mass splittings in the
sleptons of the first two generations are prone to create 
large flavor-violating off-diagonal entries 
in the slepton mass matrices when lepton matrices become
diagonalized. It can easily give rise to 
lepton flavor violation, which is severely
limited via the $\mu \rightarrow e \gamma$ constraint.
Alignment of slepton and lepton matrices were to be invoked
in order to meet the above constraint. 
Apart from the above, a generally light SUSY spectrum in satisfying the
$(g-2)_l$ magnetic moment data while having strong LHC   
constraints on sparticle masses forces one to have only 
a compressed scenario involving light sleptons and wino-like
chargino along with a bino-like LSP.
Ref.\cite{Li:2021koa} also used differing mass parameters between
the two generations of sleptons, whereas an overabundant bino DM was avoided by
considering a superWIMP dark matter scenario
\footnote{Going beyond MSSM, i)
Ref.\cite{Banerjee:2020zvi} is an analysis with
supersymmetric gauged ${U(1)}_{L_\mu-L_\tau}$ model, ii)
Ref.\cite{Crivellin:2018qmi} pointed
out the level of phenomenological stringency of the issue while citing
varieties of generic models including MSSM, leptoquarks and little Higgs
inspired  models and showed the usefulness of
abelian ${L_\mu-L_\tau}$ to explain the anomalies.
Ref.\cite{Crivellin:2018qmi} also
commented on the possibility of non-holomorphic trilinear terms to address the
issue.
iii) Ref.\cite{Frank:2021nkq} used
additional $U(1)^\prime$ SUSY model with family dependent (non-universal)
$U(1)^\prime$ charges, iv) Ref.\cite{Cao:2021lmj} used
inverse-seesaw extended NMSSM to address the anomalies. 
}.
As mentioned earlier, Ref.\cite{Endo:2019bcj} explored the
above magnetic moment constraints also in the MSSM context using a very 
large $\mu$ (up to $ \sim 500$~TeV) to generate large threshold corrections to 
the relevant Yukawa couplings in the analysis.
A large degree of
sensitivity of the above corrections to the slepton masses is used
to accommodate the $(g-2)$ constraints.
Selectron masses were considered
to be heavy (multi TeV) whereas smuon masses are less than a TeV
in order to satisfy both 
the $g-2$ constraints. The analysis required a very 
large $\tan\beta$($=70$) and
light wino (~$500$~GeV) and massive higgsinos.
The work that is associated with a large value
for the electroweak fine-tuning 
complied with the vacuum stability condition to find the valid
parameter space. On the theoretical side, both the analyses 
(\cite{Endo:2019bcj,Badziak:2019gaf}) discussed the flavor dependence of
the slepton masses 
by considering a Higgs mediation scenario\cite{Yamaguchi:2016oqz,Yin:2016shg} so as to have an
alignment of the slepton and lepton matrices.
Ref.\cite{Li:2021xmw} addressed the
 magnetic moment constraint pair by considering CP-violating phases and the
constraint from electron dipole moment (EDM). 

  In contrast to the above analyses that had to manage the flavor issues
  carefully and are strongly constrained by
  $\mu \rightarrow e \gamma$ or may have to live with light SUSY spectra or
  very large $\tan\beta$, 
  we focus on including nonstandard soft terms. We must emphasize here
  that as mentioned before, the MSSM soft terms and the nonstandard
  soft SUSY breaking terms do not
  have the same status. While the regular soft terms are
  highly supported by popular UV complete models, the NH terms are in general
  difficult for modelling. As mentioned 
  earlier, in a model independent standpoint, we
  analyze the effect of the nonstandard terms considering an
  approach of ``Minimal Effective Supersymmetry'' of 
  Ref.\cite{Hall:1990ac}. 
 An analysis with NH terms would not demand any large $\mu$,
  or light spectra and we will use a flavor friendly scenario of having   
  identical slepton masses for the first two 
  generations with equal coefficients for the left and right mass 
  parameters. However, we will use two different non-holomorphic trilinear
  parameters for the first two generations of leptons, both given as inputs
  at the low scale. In general, had these been given at a much higher
  scale like the grand unification scale we might run into flavor issues.
  Nevertheless, a detailed analysis of the exact degree of flavor violation,   
  is beyond the scope of this work.

\section{Results}
\label{resultsection}
\subsection{Muon magnetic Moment in NHSSM}
Our NHSSM analysis on lepton $g-2$ in its first part involves
studying the effects of the nonstandard soft terms on the
enhancement of the SUSY contributions to muon g-2.
We will explore at the beginning the constraints from $(g-2)_\mu$
data and analyze the
parameter space that would be consistent with dark matter (DM)
and other constraints.
We will demonstrate 
the mechanism that enhances $(g-2)_\mu$ in NHSSM.
For DM, we will consider the case where the lightest supersymmetric particle
(LSP) is expected to obey only 
the upper limit of the relic density bound from
PLANCK data\cite{Planck:2018vyg}. We will identify the parameter space
that would satisfy all the direct detection limits from XENON1T
namely both the spin-independent\cite{XENON:2018voc}
and spin-dependent\cite{XENON:2019rxp}
type of data. Once the above analysis delineates the parameter space, we
will explore 
the effect of 
imposing the constraint from the electron's magnetic moment $(g-2)_e$
due to fine-structure constant measurement.
We will see how the Yukawa threshold corrections due to NHSSM effects can
affect leptonic $g-2$, particularly $(g-2)_e$. Because of the above, 
we will see that it is indeed possible to 
find NHSSM parameter regions such that the scaled
magnetic moment ratio of interest $R_{e,\mu}$ as discussed before
may be consistent with Eq.\ref{ratio_eqn}. We use
{\tt SARAH-4.14.4}\cite{Staub:2013tta,Staub:2015kfa} and 
{\tt SPheno 4.0.4}\cite{Porod:2003um,Porod:2011nf} to implement the
model and for computing SUSY spectra and observables.

     In order to generate a sufficient amount of SUSY contributions to leptonic $g-2$ in NHSSM
     we take the help of chirality flip via the 
Left-Right scalar mixing due to the non-holomorphic trilinear SUSY breaking interaction parametrized by ${A}^\prime_l$. We use $T_l^\prime$ as an input
defined via $T^\prime_l =y_l {A}^\prime_l$
(with $y_l=\frac{m_l}{(v_d/\sqrt 2)}$).
This follows the input convention used in the codes SARAH-SPheno
(for example see Ref.\cite{Ross:2016pml}) for all the trilinear
soft parameters.  We should point out
that with this parametrization a value of a few GeV for
$T^\prime_l$ may mean a large value
for ${A}^\prime_l$. The largeness is most prominent 
for electron, whereas for top quark,
the associated values of $T_t$ and $A_t$ are not so different from each other.
 We must also remember that
because of the appearance of the fermionic mass $m_e$ in the off-diagonal
entry of Eq.\ref{slepton_mass},
any non-negligible L-R mixing of selectrons
would need a significantly large ${A}^\prime_e$ because of the
smallness of the electronic mass.
In one-loop vertex correction radiative diagrams (Figure~\ref{one-loop-diags}), this is achieved
via the NH trilinear interactions of scalars in the bino-slepton loops\footnote{We use vanishing 
coefficients $A_l$ for the trilinear SUSY breaking parameters throughout our study.}. Thus, both the above trilinear
parameters and the mass of bino $M_1$ will have significant roles in our analysis apart from 
the masses of sleptons for which we consider equal left and right mass
parameters $m_L,m_R$ for the first two generations.
We must emphasize that the above NH terms may potentially enhance $a_l$ significantly in comparison with
the MSSM contributions involving chargino or neutralino loops.

In order to probe NHSSM effects clearly we keep
the SU(2) gaugino mass to be sufficiently heavy, much above $\mu$ or $M_1$.
A large part of parameter space 
where $(g-2)_e$ can be accommodated in our work
involves i) direct effect of larger $A_e^\prime$ on the bino-seletron loop
of Fig.\ref{one-loop-diags} (i.e. through Eq.\ref{amu_bino_smuon_loop} 
with $\mu \rightarrow \mu+A_e^\prime$) and ii) via the effect of the 
enhancement factor $\frac{1}{1+\Delta_e}$ directly on $a_e$ (Eq.\ref{a_e_scaling}).   
Of course, $\frac{1}{1+\Delta_e}$ influences the 
Yukawa coupling for electron ($y_e$) directly via $A_e^\prime$ through
the same L-R mixing.
The effect of the said NH trilinear mixing may easily supersede
the contribution from the higgsino mixing superpotential term
characterized by $\mu$. On the other hand, 
the corrections may become smaller for larger slepton mass values.
Keeping our choice 
of scalar mass spectra in tact, we will rather  
limit the associated NH trilinear SUSY breaking parameters such as
$A^\prime_e$ so as to limit $y_e$.
We will label the above as the ``Limited Threshold Corrections of Yukawa
Coupling (LTCYC)'' zone. To specify explicitly, 
by LTCYC zone of $y_e$ we mean that the radiatively corrected value of 
$y_e$ can at most be twice the corresponding MSSM value, or
in other words $\frac{1}{1+\Delta_e} < 2$. However, with the
chosen SUSY mass parameters we will see that the required agreement with data 
for ${(g-2)}_e$ range is possible even with a factor quite smaller than two.
As we will point out in Section~\ref{LTCYCdetails} the chosen SUSY
parameter space that satisfies ${(g-2)}_\mu$ requires only moderately large 
$A_\mu^\prime$ in contrast to $A_e^\prime$ for the case of ${(g-2)}_e$.
Thus, LTCYC for $y_\mu$ is automatically satisfied for all the parts of our
analysis meant for ${(g-2)}_\mu$. We 
will explore explicitly how far LTCYC for $y_e$ as a condition 
becomes important when we incorporate 
$(g-2)_e$ as a constraint in Section~\ref{yukawathreshold}.

Toward the end, we will impose the constraints from 
ATLAS search for slepton-pair production\cite{ATLAS:2019lff}
in the LSP-slepton mass plane and explore a few benchmark points that would
satisfy all the constraints. 
While the neutralino-slepton loop dominated by bino type of neutralino
would enhance $(g-2)_l$, 
we will primarily consider higgsino as the LSP which would be suitable for
satisfying the DM relic density 
data from PLANCK (only the upper limit, in a multi-component DM scenario)
as well as the spin independent (SI) direct-detection (DD)
scattering cross section data of
LSP-nucleon scattering. The bino-higgsino mixed region
would obviously be disfavored or discarded via the SI DD constraints. 
Below, we focus on the minimal requirement for enhancing SUSY contributions to leptonic $g-2$ and the variables that
are closely connected to $g-2$ and dark matter. With the above in mind, we keep the squark masses decoupled to large values and we do the
 same for the tau-sleptons too. All the left and right handed scalars
including that of the first two generations of sleptons 
will assume equal SUSY breaking mass parameter 
input values ($m_L=m_R$). We keep the trilinear coefficient for
top-squark namely $T_t$ at a high negative value that would be
consistent with Higgs mass data.  The Table~\ref{table-parameters} refers to the input parameter values/ranges\footnote{The relevant SM parameters are:
$m_t^{\rm pole}=173.5,~m_b^{{\overline{\rm MS}}}=4.18, ~m_\tau=1.776$ all in GeV.}.

\begin{table}[!htb]
  	\centering
	\begin{tabular}{|c|c|c|c|}
		\hline\hline 
		Parameters  &  Value &  Parameters & Value   \\ [0.5ex]
		\hline
		$M_1$ & [100 to 1000] & $M_{2}$  &  1500\\
		$M_3$ & 2800 & $\mu$ & [150 to 1000] \\
		$m_A$ (as tree-level input toward ${B_\mu}$ ) & 2500 & $\tan\beta$ & 10 \& 40 \\
		$M_{\tilde{q}_{33}} / M_{\tilde{u}_{33}}$ & 4000 &  $M_{\tilde{d}_{33}}$  & 4000 \\
		$M_{\tilde{q}_{11,22}}/ M_{\tilde{u}_{11,22}}$ & 2500 & $M_{\tilde{d}_{11,22}}$  & 2500 \\
                $m_L \equiv M_{\tilde{L}_{11,22}}=M_{\tilde{e}_{11,22}} \equiv m_R$ & [200 to 1000] & $ M_{\tilde{L}_{33}}=M_{\tilde{e}_{33}}$  & 2000\\
		\hline
		$T_{t}$  & -3500 & $T_{t}^{\prime}$ & 0  \\
		\hline
                $T_{e,\mu,\tau}$  & 0 & $T^\prime_e$,$T^\prime_\mu$, $T^\prime_\tau$ & [-25 to 25], [0 to 1000],0\\  
                \hline
	\end{tabular}
	\caption{SUSY scale input ranges of Soft masses and Trilinear
        SUSY breaking parameters. Quantities labelled by $T$
        are scaled trilinear 
        parameters, e.g. $T_t =y_t {A}_t$ or
        $T^\prime_\mu =y_\mu {A}^\prime_\mu$. 
        Masses and trilinear couplings are in GeV.
        All the trilinear parameters not mentioned here are understood to 
       assume zero values.
        } 
	\label{table-parameters}
\end{table}
With the above parameter space been defined, some of our constraints are that from the Higgs data, $Br(b \rightarrow s \gamma)$, relic
density upper limit from PLANCK. This is apart from the $(g-2)_\mu$ and $(g-2)_e$ constraints mentioned earlier and the SI and SD direct
detection constraints from XENON1T that depend on the mass of the LSP. With
$M_{\rm SUSY}$ being large, we consider a 3 GeV theoretical uncertainty
in SUSY
Higgs mass leading to the following as the acceptable
range\cite{Bahl:2019hmm} for the SM-like Higgs in MSSM. 
\begin{equation}
122<m_h<128~{\rm GeV}. 
\label{higgsmassdata}
\end{equation} 
The flavor limits at 2$\sigma$ are $3.02 \times 10^{-4}< Br(b \rightarrow s \gamma)< 3.62 \times 10^{-4} $\cite{HFLAV:2019otj}
and $2.23 \times 10^{-9} <Br(B_s \rightarrow \mu^+ \mu^-)< 3.63
\times 10^{-9}$~\cite{Altmannshofer:2021qrr}.
Considering the PLANCK result for dark matter namely  
$\Omega_{{\tilde \chi}^0} h^2=0.120 \pm 0.001$\cite{Planck:2018vyg}, the
$2\sigma$ level limits are
${[\Omega_{{\tilde \chi}^0} h^2]}_{\rm min}=0.118$ and 
${[\Omega_{{\tilde \chi}^0} h^2]}_{\rm max}=0.122$.  
The DM relic density is expected to satisfy only the upper bound.
In our generally underabundant scenario, for the purpose of
direct detection cross-section,  
we define a scale factor $\xi$\cite{Chattopadhyay:2000qa} as given below.
\begin{equation}
\xi=\Omega_{{\tilde \chi}^0} h^2 /{[\Omega_{{\tilde \chi}^0} h^2]}_{\rm min}
~{\rm if}~ \Omega_{{\tilde \chi}^0} h^2
\leq {[\Omega_{{\tilde \chi}^0} h^2]}_{\rm min}, {\rm else}~ \xi=1.
\label{scalefactorequation}
\end{equation}

\begin{figure}[hbt] 
     \begin{center}
        \subfigure[]{%
           \label{amuvsM1}
           \includegraphics[width=0.40\textwidth]{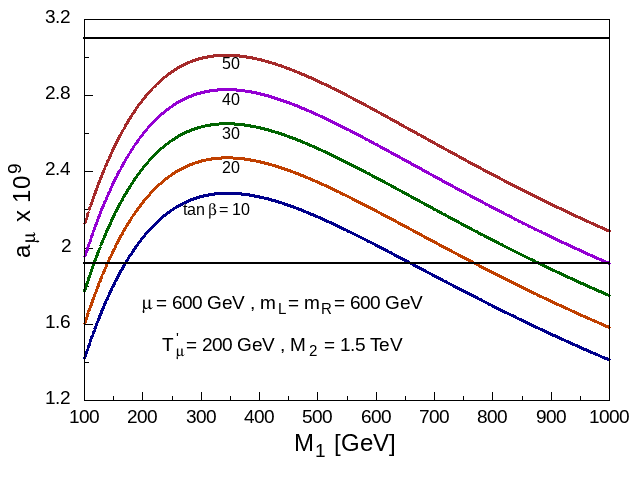}
                 }%
\hskip 30pt
        \subfigure[]{%
           \label{amuvsaprimevM1}
           \includegraphics[width=0.40\textwidth]{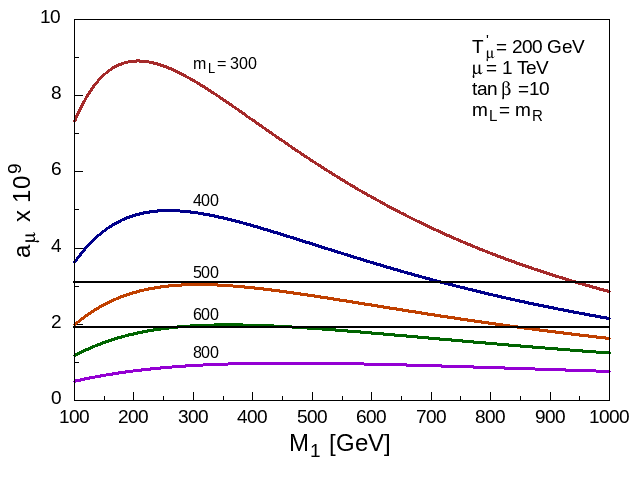}
        }%
        \caption{Fig.\ref{amuvsM1}: Variation of SUSY contributions to
        $a_{\mu}$ (referred as $a_{\mu}$ itself)
        with respect to the mass of bino ($M_1$) for a few values of
        $\tan\beta$. Fixed parameters are as mentioned 
        in the plots. The black horizontal lines are the 1$\sigma$ limits of
        $a_\mu$. For each $\tan\beta$, $a_{\mu}$ has an ascending
        and 
        a descending part over the range of variation of $M_1$.
	The location of the peaks depend on 
        $m_L(=m_R)$, and these are
        essentially unchanged with respect to $M_1$. 
        Fig.\ref{amuvsaprimevM1}: Same as Fig.\ref{amuvsM1} except for a given
        value of $\tan\beta$ and varying $m_L$.
	There are similar peaks in $a_{\mu}$, 
        but they shift with varying $m_L$.}
        \label{amu_vs_M1_fixed_tanb_mL}
\end{center}
\end{figure}

We plot Figure~\ref{amu_vs_M1_fixed_tanb_mL} that describes
our result of the SUSY contributions to 
$a_{\mu}$ (referred hereafter as $a_{\mu}$ itself)
in relation to the relevant SUSY
breaking parameter $M_1$, the mass of bino.
Fig.~\ref{amuvsM1} shows the 
variation of $a_{\mu}$ with the mass of bino
($M_1$) for $\tan\beta=$~10 to 50 in steps of 10. Here, the fixed parameters
used are $T^\prime_\mu=200$~GeV, $m_L(=m_R)=600$~GeV, $\mu=600$~GeV, and 
$M_2=1.5$~TeV. As we will see later (Figure~\ref{bothmagmom_withDM_M1_Amuprime}), the NHSSM contributions  
to $a_\mu$ are much larger than the corresponding MSSM ones (i.e. with 
vanishing $T^\prime_\mu$) for the chosen regions of values of mass and coupling
parameters of Table~\ref{table-parameters}.
The black horizontal lines are the 1$\sigma$ limits of
        $a_\mu$. 
For each $\tan\beta$, 
$a_{\mu}$ increases with $M_1$, and then it decreases. The peaks occur at around a given value of
        $M_1$ and 
        the corresponding locations are almost independent
        of $\tan\beta$.  Fig.\ref{amuvsaprimevM1} shows the variation of $a_{\mu}$ with $M_1$ for specific values of slepton mass parameters $m_L ~(=m_R)$
        corresponding to the first two generations. The fixed parameters
        chosen are $T^\prime_\mu=200$~GeV, $\mu=1$~TeV and $\tan\beta=10$ with
        $M_2$ taking an identically large value as before. The curves show 
       similar peaks and their locations depend on the slepton mass parameter
       $m_L$.

%

\begin{figure}[hbt] 
     \begin{center}
     \subfigure[]{%
           \label{amuvsM1values}
           \includegraphics[width=0.32\textwidth]{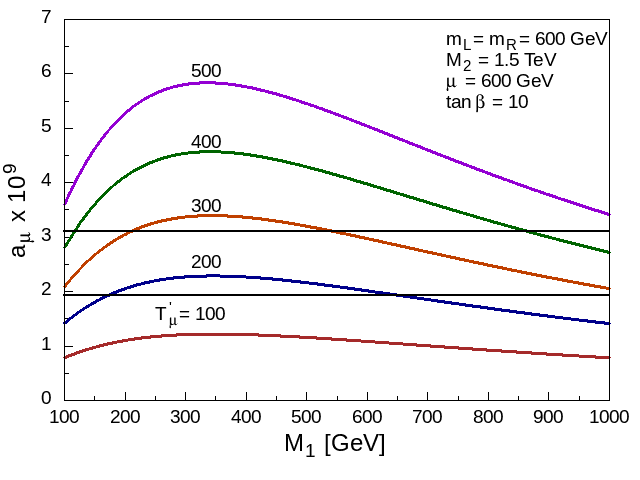}
        }%
        \subfigure[]{
           \label{amuvsAmuprimeline} 
           \includegraphics[width=0.32\textwidth]{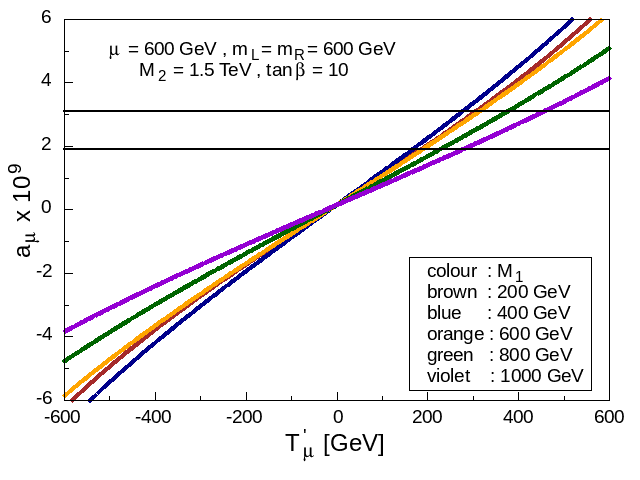}
        }%
    \subfigure[]{%
    	\label{muscanned}
    	\includegraphics[width=0.32\textwidth]{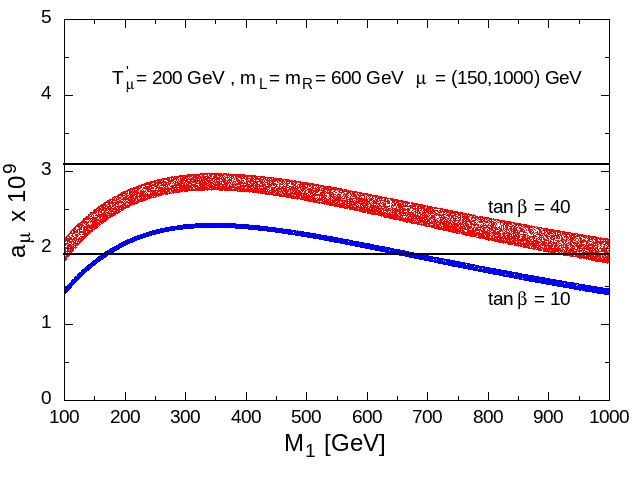}
    }%
\caption{
Fig.\ref{amuvsM1values}: Variation of $a_{\mu}$ with the mass of bino
($M_1$) for a few values of $T_\mu^\prime$ where $T_\mu^\prime$ relates to
${A}^\prime_\mu$ of Eq.\ref{nonstandardsoftterms} via
$T^\prime_\mu =y_\mu {A}^\prime_\mu$. 
Fig.\ref{amuvsAmuprimeline}:  
Variation of $a_{\mu}$ over $T_\mu^\prime$ for different values of $M_1$.
$|a_{\mu}|$
generally decreases with $M_1$ for values above 400 GeV. 
The exceptions are the cases of
$M_1=200$ and 400 GeV, 
which are flipped because they belong to the ascending and the descending
zones of the corresponding curve for $m_L=600$ GeV of Fig.\ref{amuvsaprimevM1}.
Fig.\ref{muscanned}: Variation of $a_{\mu}$ with 
$M_1$ for the scanned range of $\mu$ satisfying $150<\mu<1000$~GeV. The blue and red regions refer to $\tan\beta=10$ and 40 respectively.
The limited level of thickening 
of lines even for a large $\tan\beta$ demonstrates only a mild degree of
dependence of $a_{\mu}$ on $\mu$ over the range. The black horizontal lines are the 1$\sigma$ limits of $a_\mu$.
}
\label{further_amu_2d_plots}
\end{center}
\end{figure}

We continue to study the behavior of $a_\mu$ concerning
the relevant SUSY breaking parameters in Figure~\ref{further_amu_2d_plots}.
Undoubtedly,
$a_\mu$ is enhanced most prominently via the trilinear parameter $A_\mu^\prime$
via its strong effect through chirality flipping L-R scalar interaction
in the bino-smuon loop. For $\tan\beta=10$, $m_L=\mu=600$~GeV and
$M_2=1.5$~TeV, Fig.\ref{amuvsM1values} shows the  
variation of $a_\mu$ with respect to $M_1$ for a few different values
of $T_\mu^\prime$, 100~GeV to 500~GeV in steps of 100~GeV. A substantial
amount of enhancement of $a_\mu$ occurs due to a change in $T_\mu^\prime$. In
this analysis, we satisfy all the charge-breaking related constraints
for NHSSM (Eq.\ref{ccbconst})\cite{Beuria:2017gtf} 
and stay within the LTCYC zone for the 
threshold corrections to $y_l$ due to the variation over $T_l^\prime$. 
Fig.\ref{amuvsAmuprimeline} refers to the  
variation of $a_{\mu}$ over $T_\mu^\prime$ for different values of $M_1$.
$|a_{\mu}|$ generally decreases with $M_1$ for values above 400 GeV. 
The exceptions are the cases of $M_1=200$ and 400 GeV, 
which are flipped because they belong to the ascending and the descending
parts of the corresponding curve for $m_L=600$ GeV of Fig.\ref{amuvsaprimevM1}.
Furthermore, the sign of $a_\mu$ is approximately given by the
sign of $T^\prime_\mu$. 
In the NHSSM scenario of interest where non-vanishing
${A}^\prime_\mu$ enhances the bino-smuon loop contribution causing the    
same loop to dominate over all the other diagrams, at the lowest order,
$\mu\tan\beta$ gets replaced by $(\mu+{A}^\prime_\mu)\tan\beta$
in Eq.~\ref{amu_bino_smuon_loop}. 
With vanishing ${A}_\mu$, and 
${A}^\prime_\mu~(=T^\prime_\mu/y_\mu)$ large enough to offset $\mu$, 
the sign of $a_\mu$ is determined by the sign of $A^\prime_\mu$. Numerically
this is seen to be valid as long as $|T^\prime_\mu|\gsim 20$~GeV or so.
Consistency with the 1$\sigma$ band demands that $(g-2)_\mu$ data can be
satisfied with only positive $T^\prime_\mu$.
Fig.\ref{muscanned} displays the variation of $a_{\mu}$ with 
$M_1$ for $\mu$ satisfying $150<\mu<1000$~GeV.
The blue and red regions refer to $\tan\beta=10$ and 40 respectively. The limited level of thickening 
of lines even for a large $\tan\beta$ demonstrates that there is
only a mild degree of dependence of $a_{\mu}$ on $\mu$ over the range mentioned
above. This is consistent with the fact that a higgsino-smuon or a
charged higgsino-sneutrino loop for $a_\mu$ are hardly important in our 
analysis with a dominant effect due to a NH trilinear term.  
%

\begin{figure}[hbt] 
     \begin{center}     
     \subfigure[]{%
    	\includegraphics[width=0.45\textwidth]{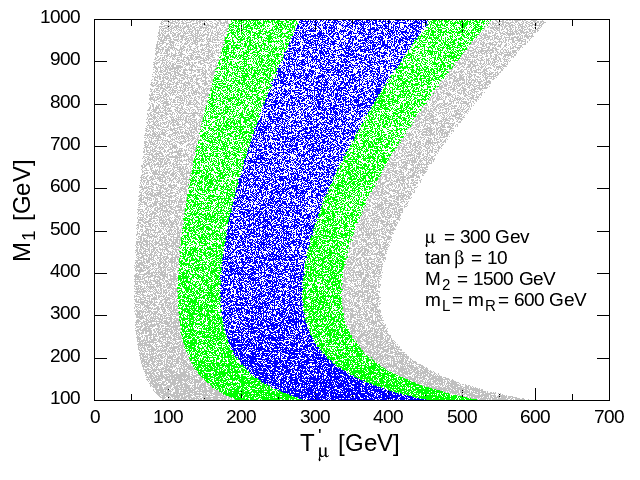}
        \label{M1vsAmuprimemu3t10sl6}
    }%
    \subfigure[]{%
    	\includegraphics[width=0.45\textwidth]{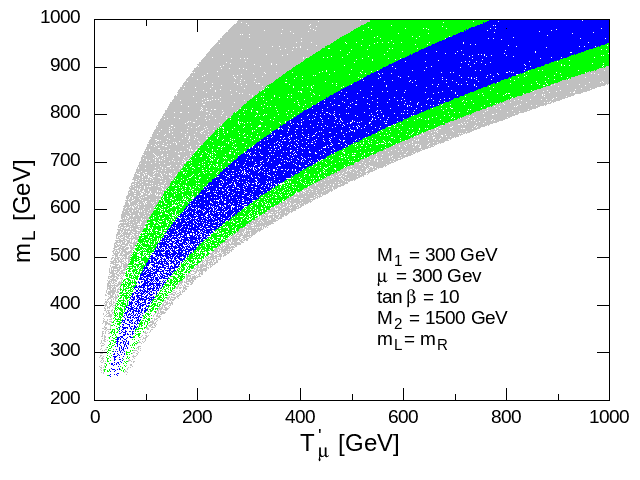}
        \label{mLvsAmuprimeM13mu3}
    }%
        \caption{Fig.\ref{M1vsAmuprimemu3t10sl6}:
        Display of 1$\sigma$ (blue),
        2$\sigma$ (green), and 3$\sigma$ (gray) regions with respect to the
        $(g-2)_\mu$ constraint in the $(T^{\prime}_\mu-M_1)$ plane for
        fixed parameters mentioned in the figure. 
        Fig.\ref{mLvsAmuprimeM13mu3}: Similar display in the $(T^{\prime}_\mu-m_L)$ plane.} 
\label{M1vsAprimefigs}        
\end{center}

\end{figure}

We now use the $(g-2)_\mu$ constraint in the most important
$M_1-T^\prime_\mu$ plane in Fig.\ref{M1vsAmuprimemu3t10sl6}.
The fixed parameter under the study are $\tan\beta=10$,
$\mu=300$~GeV, $m_L=600$~GeV and $M_2=1.5$~TeV.
We divide the region into 1$\sigma$ (blue),
        2$\sigma$ (green), and 3$\sigma$ (gray) zones with respect to the
        $(g-2)_\mu$ constraint. The 1$\sigma$ region extends from
        $T^\prime_\mu=175$~GeV to $450$~GeV while $M_1$ spans the entire
        space chosen in our analysis. The region bents toward the left
and this is indeed consistent with what follows from Fig.\ref{amuvsM1values}.
Fig.\ref{mLvsAmuprimeM13mu3} shows similarly constrained region with the same
color convention as above. A generic decrease of $a_\mu$ with increase
in $m_L$ is apparent since large $T^\prime_\mu$ is necessary to stay in a
given fixed colored zone for large $m_L$ values. 

%

\begin{figure}[hbt] 
     \begin{center}     
     \subfigure[]{%
    	\includegraphics[width=0.45\textwidth]{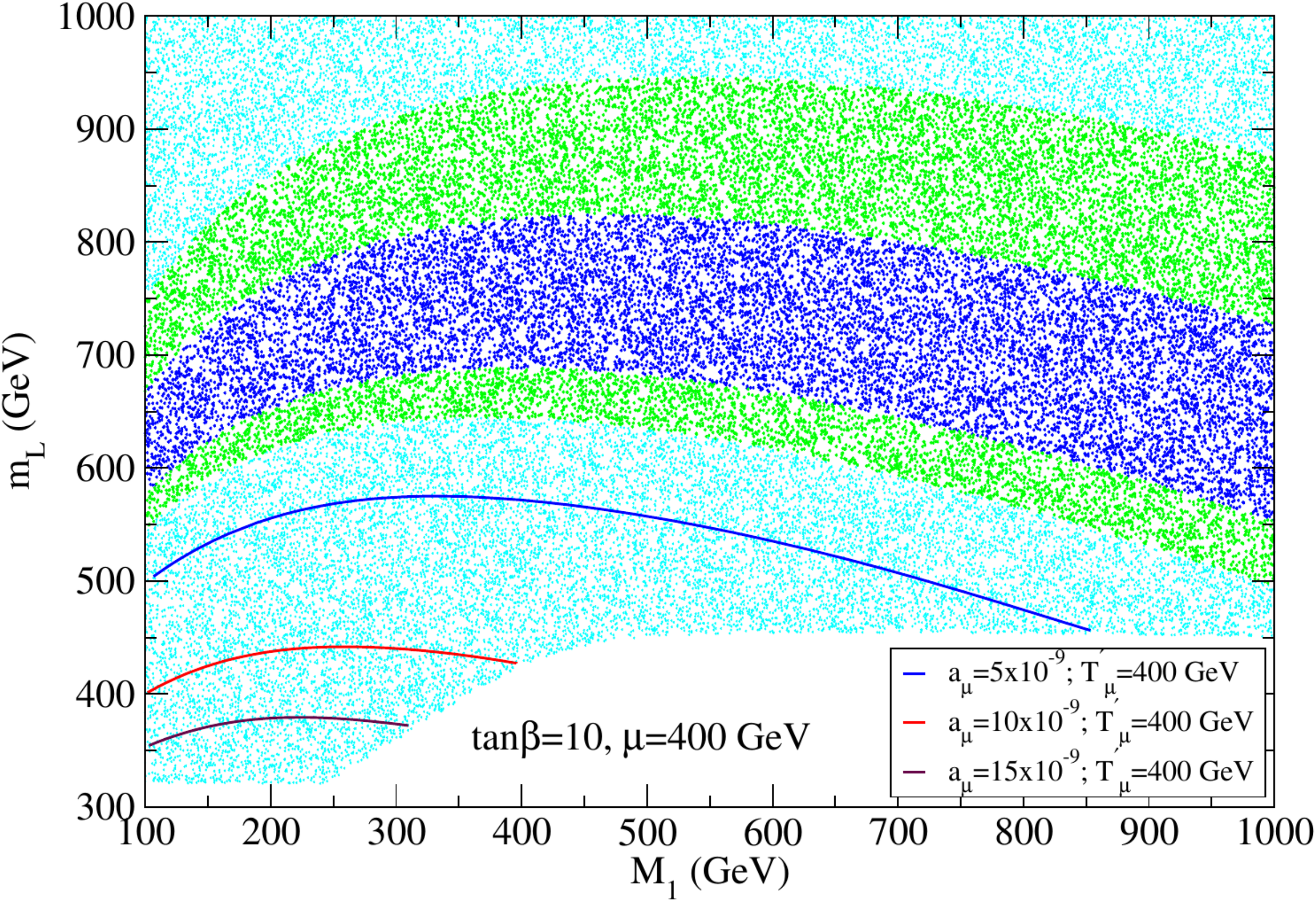}
        \label{tan10_amu_contour_M1-mL_amup400}
    }%
     \subfigure[]{%
    	\includegraphics[width=0.45\textwidth]{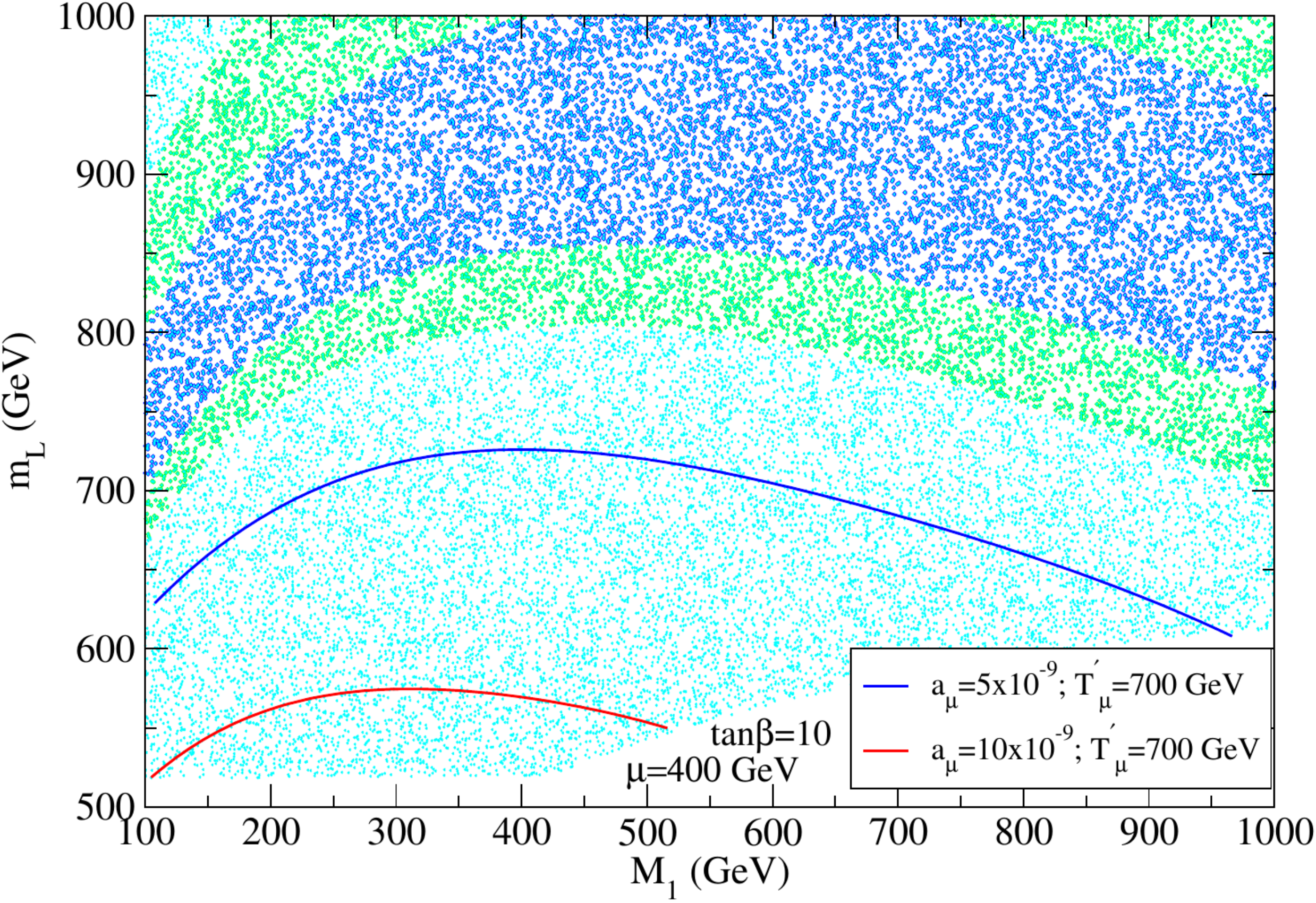}
        \label{tan10_amu_contour_M1-mL_amup700}
    }%
    
\subfigure[]{%
    	\includegraphics[width=0.45\textwidth]{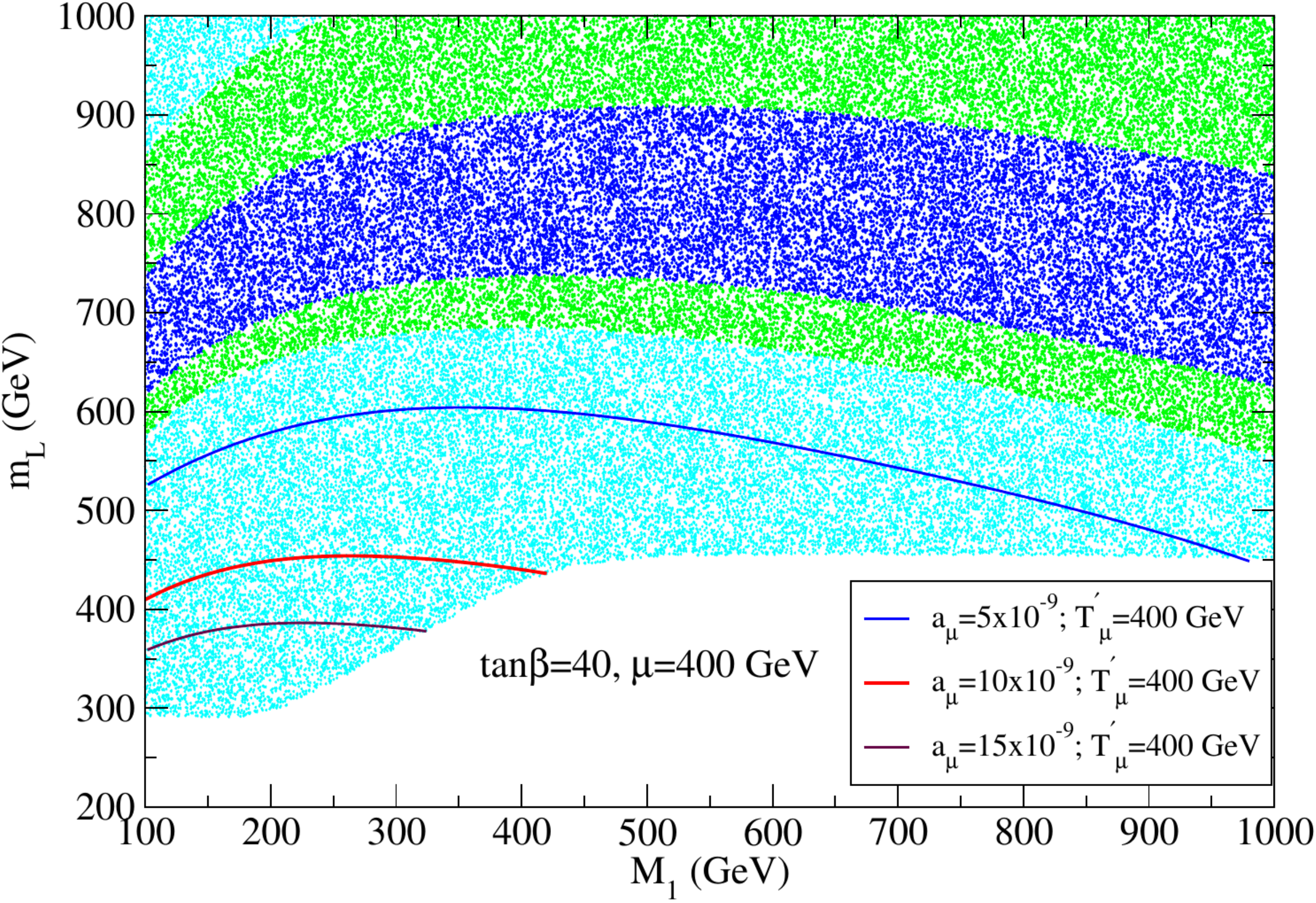}
        \label{tan40_amu_contour_M1-mL_amup400}
    }%
\subfigure[]{%
    	\includegraphics[width=0.45\textwidth]{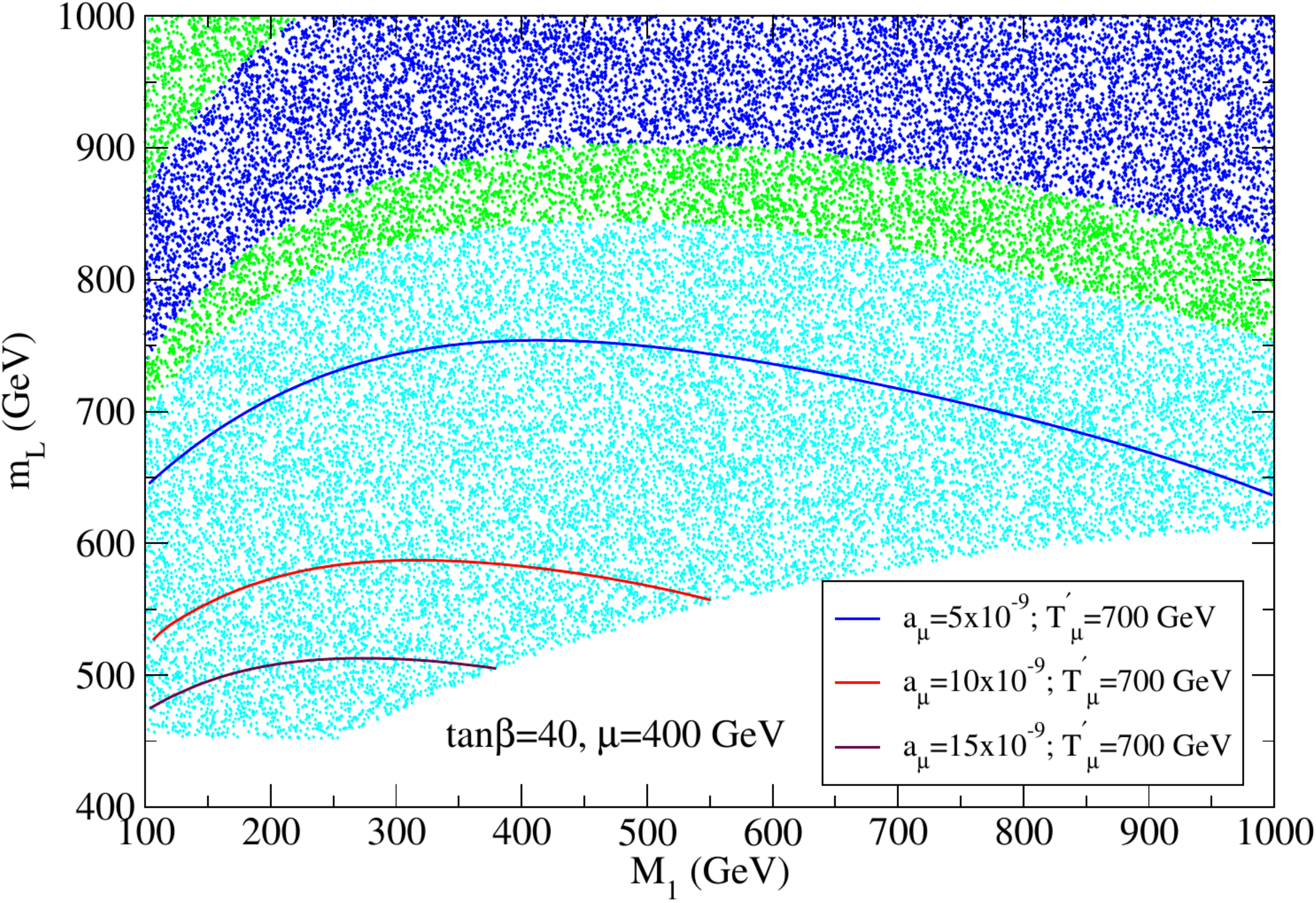}
        \label{tan40_amu_contour_M1-mL_amup700}
    }%
\caption{Fig.~\ref{tan10_amu_contour_M1-mL_amup400}: Plot in $M_1-m_L$ plane
for $\tan\beta=10$, $\mu=400$~GeV and $T^\prime_\mu=400$~GeV with all other fixed parameters same as mentioned in Table~\ref{table-parameters}.
The cyan-colored points represent
unconstrained parameter values that only satisfy the basic constraints like
Higgs data, $Br(b \rightarrow s \gamma)$, lightest neutralino to be
the LSP, and restriction from charge breaking minima.
The blue and
green shaded regions are $1\sigma$ and $2\sigma$ bands for
the ${(g-2)}_\mu$ constraint. 
Three lines are drawn for $a_\mu=5\times 10^{-9}$,
$a_\mu=10\times 10^{-9}$, and $a_\mu=15\times 10^{-9}$.  
Clearly, $a_\mu$ is
large for small $m_L$ or small smuon mass regions. The white region at the
bottom refers to a discarded parameter zone. This refers to not satisfying  
the requirements like LSP 
has to be the lightest neutralino ${\tilde \chi}^0_1$,
avoidance of tachyonic sleptons or 
any presence of charge-breaking minima.
Fig.~\ref{tan10_amu_contour_M1-mL_amup700}: Same as above except  
$T^\prime_\mu=700$~GeV. A larger $T^\prime_\mu$ enhances $a_\mu$,  
leading the possibility to accommodate a larger $m_L$ value for a given
$a_\mu$ at a given $M_1$ in comparison with
Fig.~\ref{tan10_amu_contour_M1-mL_amup400}. There is however no 
$a_\mu=15\times 10^{-9}$ line because of unavailability of valid
parameter space. 
Fig.~\ref{tan40_amu_contour_M1-mL_amup400} and
Fig.~\ref{tan40_amu_contour_M1-mL_amup700} are similar figures
as above for $\tan\beta=40$.
The lower discarded (white) regions extend with respect to
$\tan\beta=10$.}
\label{tan1040_amu_contour_M1-mL}        
\end{center}

\end{figure}

%
Figure~\ref{tan1040_amu_contour_M1-mL} shows a few $a_\mu$ contours in the
plane of $M_1-m_L$ for $\tan\beta=10$ and $40$ for $T^\prime_\mu=400$ and
700~GeV. The cyan-colored points represent
unconstrained parameter values that only satisfy the basic constraints like
Higgs data, $Br(b \rightarrow s \gamma)$, lightest neutralino to be
the LSP, and restriction from charge-breaking minima.
The blue and
green shaded regions are $1\sigma$ and $2\sigma$ bands for
the ${(g-2)}_\mu$ constraint. 
Three lines are drawn for $a_\mu=5\times 10^{-9}$,
$a_\mu=10\times 10^{-9}$, and $a_\mu=15\times 10^{-9}$.  
Clearly, $a_\mu$ is large for small $m_L$ or small smuon mass regions. We must
note that the reach of the smallness of $m_L$ to obtain an enhanced $a_\mu$
effectively may lead to picking up small $M_1$ regions. This is particularly
true for higher $\mu$ cases.  The requirement of
${\tilde \chi}^0_1$ to be the LSP, which in our case is either a bino
or a higgsino dominated state, to a reasonable degree of
approximation demands $m_L$ to be higher than
$M_1$ or $\mu$ whichever is the lowest among the last two.
There are other reasons why the 
bottom white regions of each of the figures are excluded. These may be
due to the requirements of avoidance of tachyonic sleptons or
charge-breaking minima. On the other hand, small $m_L$
regions find stringent constraints from the LHC data which we will discuss 
later. Comparing the figures,
we can further see that, as expected, 
larger $T^\prime_\mu$ values enhance $a_\mu$. 
For a given $a_\mu$, and a fixed $M_1$,
a larger $T^\prime_\mu$ creates a possibility to accommodate a
larger $m_L$. A large $\tan\beta$ like $40$
enhances $a_\mu$.

\subsection{Constraints from Dark Matter}
\label{gmuon_and_dm}
\begin{figure}[hbt] 
	\begin{center}
		\subfigure[]{%
			\includegraphics[width=0.45\textwidth]{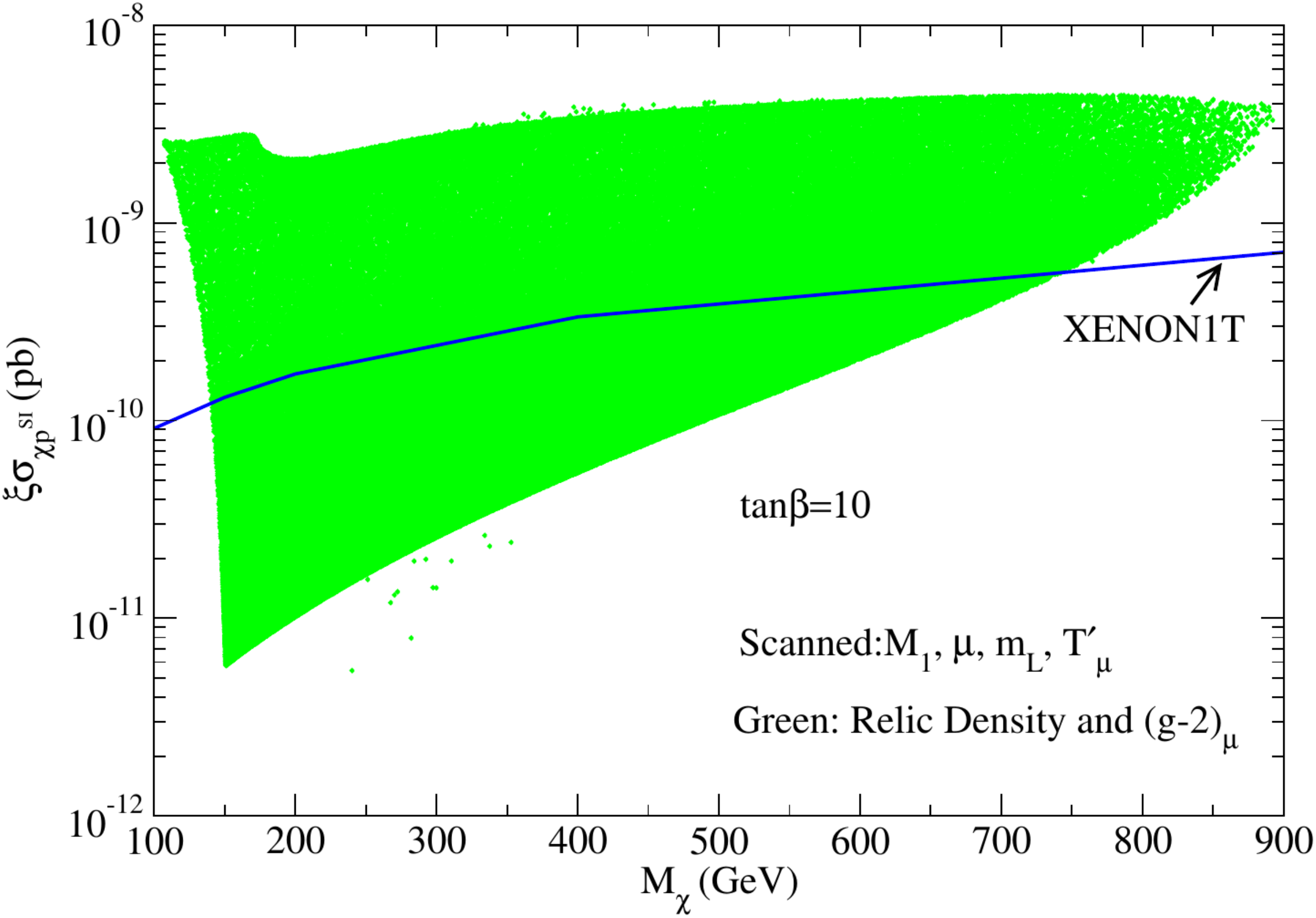}
           \label{tan10_all_scanned_all_constraints_LSP_vs_sigmaSI}        
		}%
		\hskip 30pt
		\subfigure[]{%
                 \includegraphics[width=0.45\textwidth]{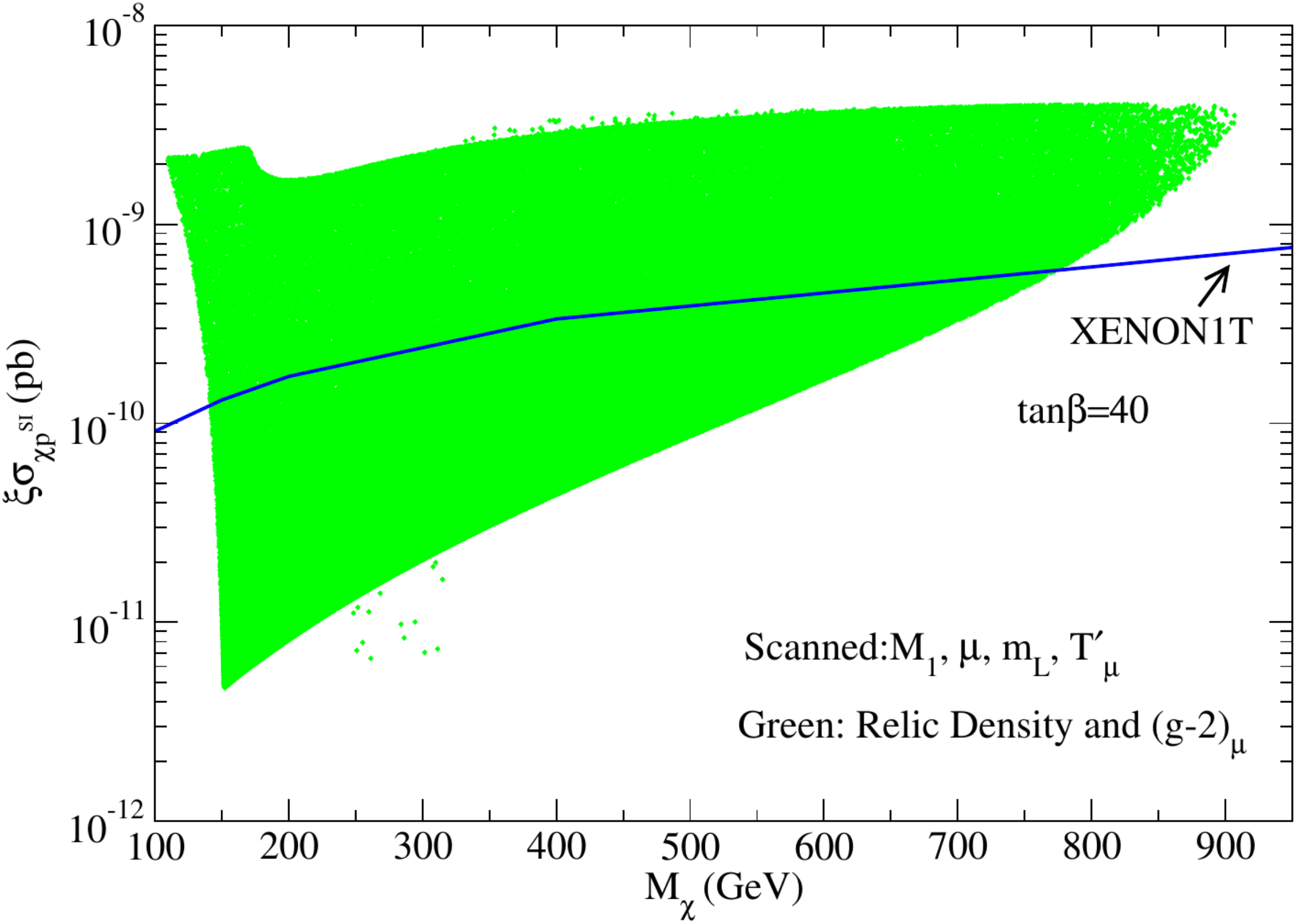}
  \label{tan40_all_scanned_all_constraints_LSP_vs_sigmaSI}               
		}%
		\caption{Fig.\ref{tan10_all_scanned_all_constraints_LSP_vs_sigmaSI}: Spin-independent (SI) scattering cross-section of the LSP
     with a proton $\sigma^{\rm SI}_{\chi p}$ as a function of the LSP mass for $\tan\beta = 10$. Considering the large number of points with underabundance of DM we multiply $\sigma^{\rm SI}_{\chi p}$ with a scale factor $\xi$ (Eq.\ref{scalefactorequation}).
The points satisfying the DM relic density upper 
bound and $(g-2)_\mu$ limits at 2$\sigma$ level are shown in green.
The blue line is the constraint from $\sigma^{\rm SI}_{\chi p}$, the spin-independent (SI) direct detection (DD) experiment 
of XENON1T indicating discarded regions above the line (at 90\%CL).  
     Fig.\ref{tan40_all_scanned_all_constraints_LSP_vs_sigmaSI}: Similar
     figure for $\tan\beta = 40$.
     \label{LSP-sigmaSI}
		}
	\end{center}
\end{figure}

We now include the constraints from dark matter in this NHSSM $(g-2)_\mu$
analysis.
Of course, with our choice of an underabundant scenario of a
higgsino dark matter, no NHSSM trilinear parameter would
directly affect the DM analysis. We will   
study DM for its effects on the combined parameter space.
In our analysis, wino mass $M_2$ is chosen to be quite heavy
(1.5 TeV), whereas the higgsino mass has a range of $150<\mu<1000$~GeV, and
for bino, we have $100<M_1<1000$~GeV.  
Our parameter space further consists of heavy tau-sleptons (2 TeV) along with
decoupled squarks, 
whereas we choose the range of the first two generations of sleptons to
vary within 200~GeV to 1~TeV. With $m_A=2.5$~TeV as an input   
the neutral CP-even or CP-odd Higgs masses are way above to encounter the so-called funnel
region that is 
characterized by bino-dominated LSPs undergoing self-annihilation
via s-channel processes involving $A$ or $H$-bosons.
Coming to the higgsinos, the chosen mass range is such that
the upper limit of 1 TeV is 
what is necessary for higgsino LSP to become a single component DM but
the upper limit of $M_1$ is also chosen to be the same.
A chance equality of $M_1$ and $\mu$
would produce a large bino-higgsino mixing, but this would not be
friendly with the SI direct detection. On the other hand,
regions with $M_1<\mu$ would produce overabundance. 
Hence, in most of the available parameter space, the LSP is of higgsino
type, except in a few occasions when there is a possibility of
$\lspone$-slepton coannihilations. With higgsino LSPs to have mass below a
TeV the associated relic density is supposed to satisfy only the upper limit of
DM relic density constraint from PLANCK data. One of the important
channels for DM production would be the coannihilation channel 
between the higgsno LSP with the lighter chargino state 
(${\tilde \chi}^0_1-{\tilde \chi}^\pm_1$) which is also higgsino dominated
in nature.
Fig.\ref{tan10_all_scanned_all_constraints_LSP_vs_sigmaSI} shows the
Spin-independent scattering cross-section of the LSP
     with a proton ($\sigma^{\rm SI}_{\chi p}$)
     as a function of the LSP mass
     for $\tan\beta = 10$ for the parameter space mentioned in
     Table~\ref{table-parameters}. We used {\tt micrOMEGAs 5.2.13} for
     dark matter-related computations\cite{Belanger:2013oya,Belanger:2010pz,Belanger:2008sj}.  
Considering the large number of points with underabundance of DM  
we multiply $\sigma^{\rm SI}_{\chi p}$ with a scale factor $\xi$ which was
defined in Eq.\ref{scalefactorequation}.
Only those points satisfying the DM relic density upper bound, Higgs mass
data, B-physics related limits and $(g-2)_\mu$ at 2$\sigma$ level
are plotted and these are shown in green.
The blue line is the constraint from 
the spin-independent (SI) direct detection (DD) experiment
of XENON1T indicating discarded regions above the line (at 90\%CL).
Using interpolation we will further use the
XEONON1T line to judge whether a parameter point with a given mass of
the LSP would survive the SI DD experiment limit. Fig.\ref{tan40_all_scanned_all_constraints_LSP_vs_sigmaSI} shows the results for $\tan\beta=40$.   


\begin{figure}[hbt] 
	\begin{center}
		\subfigure[]{%
			\includegraphics[width=0.45\textwidth]{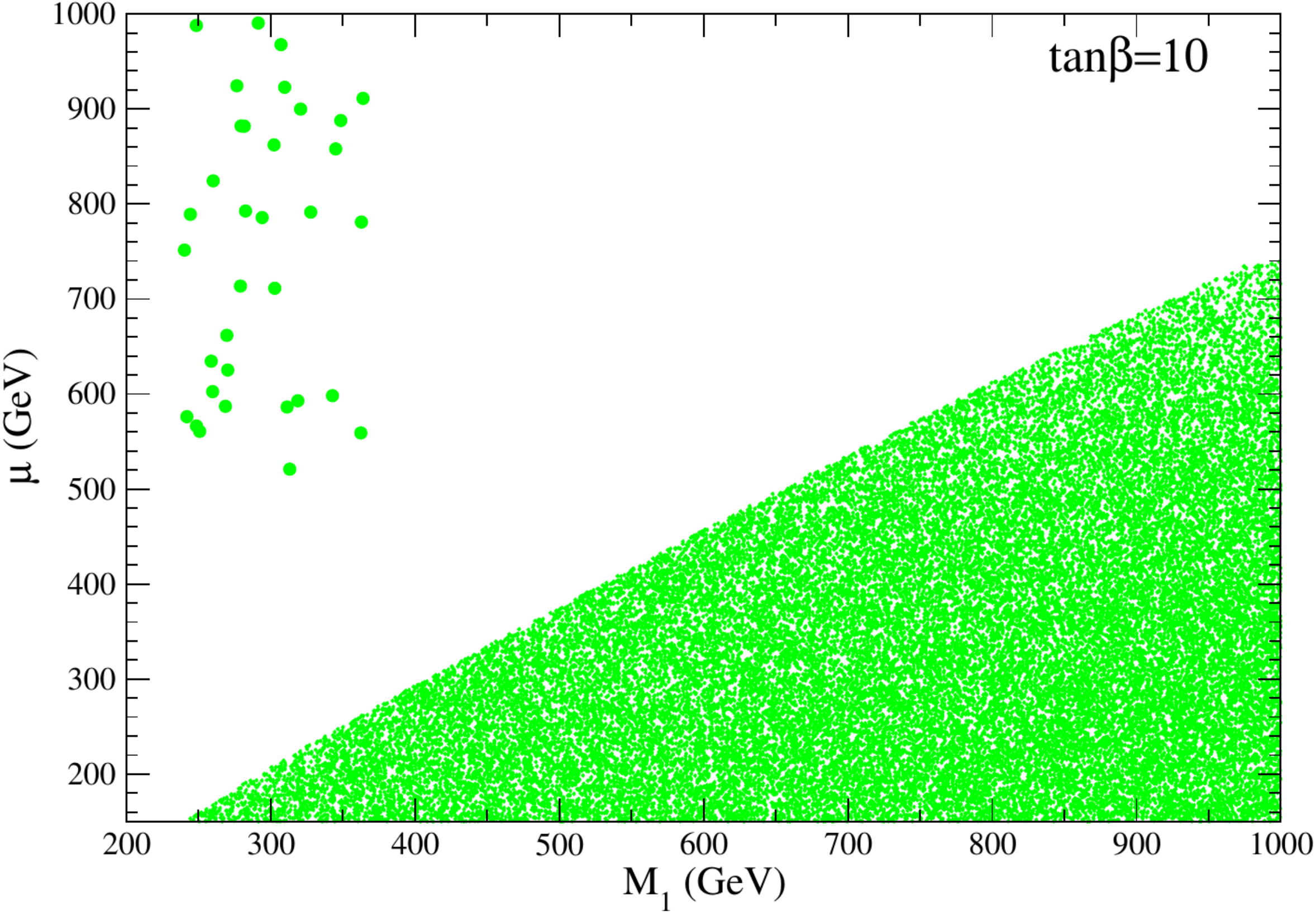}
           \label{tan10_M1_vs_Mu_muon}        
		}%
		\hskip 30pt
		\subfigure[]{%
                 \includegraphics[width=0.45\textwidth]{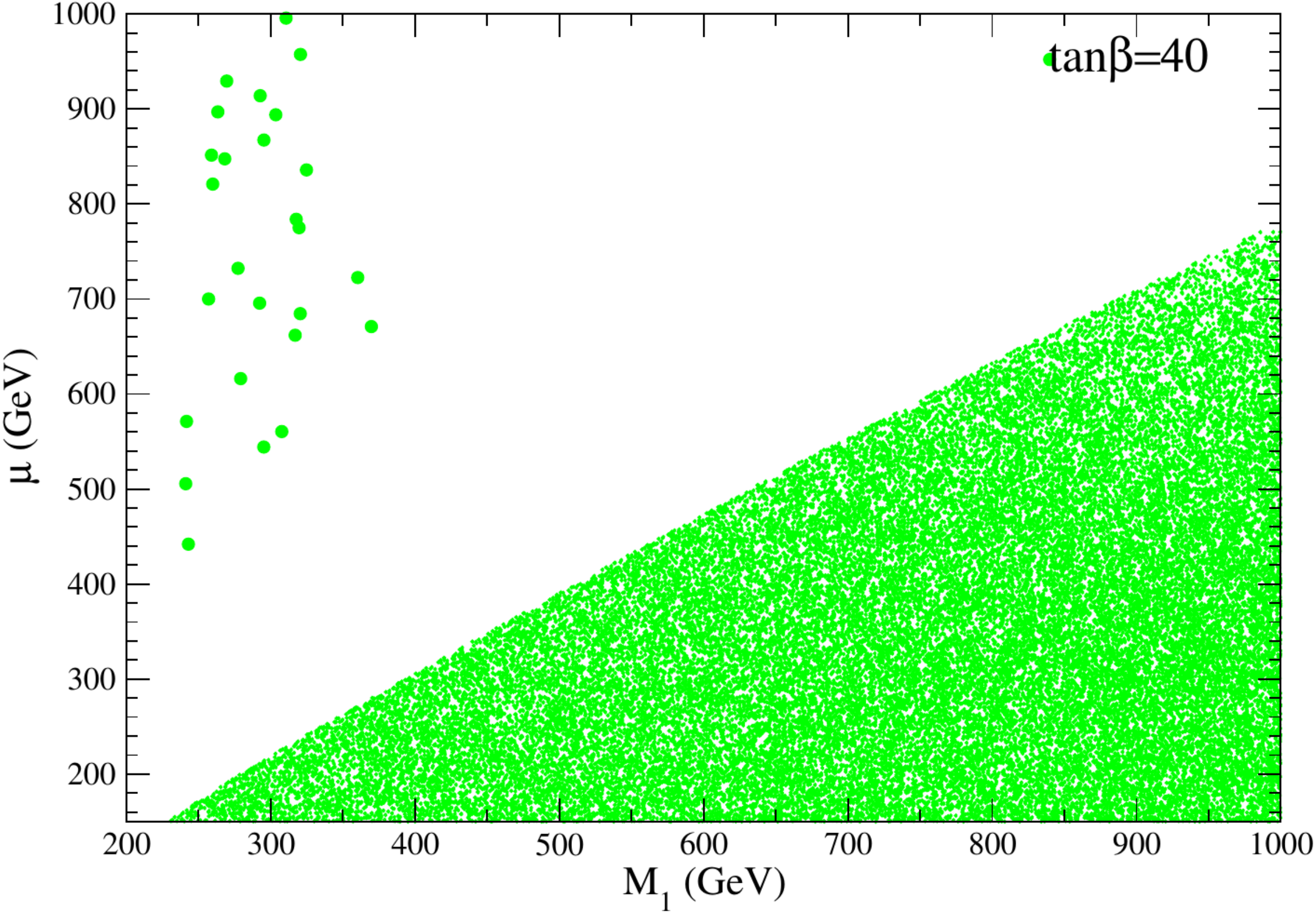}
            \label{tan40_M1_vs_Mu_muon}                       
		}%
		\caption{Fig.\ref{tan10_M1_vs_Mu_muon}: Scatter plot in
                $M_1-\mu$ plane for the analysis of
                Fig.\ref{LSP-sigmaSI}.
Points shown in green satisfy the $(g-2)_\mu$ and the DM relic
density limits both at $2\sigma$ level while also having 
$\sigma^{\rm SI}_{\chi p}$ value below
the XENON1T limit at 90\% CL. The LSP is generally of higgsino type.
There are a few isolated points in the small $M_1$($<\mu$)
zone with $\mlspone$ between 250 to 350 GeV that
correspond to bino-dominated LSPs satisfying the DM
relic density limits via slepton coannihilations. An
appropriately large choice of $m_{A/H}$  avoids pair-annihilations of
essentially binos via the s-channels Higgs process.
Fig.\ref{tan40_M1_vs_Mu_muon}: Similar figure for $\tan\beta=40$.   
		}
         \label{M1_vs_Mu_muon}         
	\end{center}
\end{figure}

    Fig.\ref{tan10_M1_vs_Mu_muon} shows the 
    projection of the analysis of 
                Fig.\ref{LSP-sigmaSI} on the 
                $M_1-\mu$ plane. Points shown in green satisfy the DM relic density upper bound, the XENON1T spin-independent direct detection cross-section limit at 90\%CL 
                as well as $(g-2)_\mu$ values within 2$\sigma$ level.
The parameter zone with bino-higgsino mixed type of LSPs
($M_1 \simeq \mu$) are 
typically discarded via the XENON1T SI-DD limits. 
Generally, LSPs are visibly higgsino-dominated in nature and satisfy
all the constraints within the mass range of 150
(the chosen lower limit of $\mu$) to 780~GeV.
Apart from higgsinos, additionally, there are a very few parameter points corresponding to 
bino-like LSPs undergoing occasional coannihilations with sleptons for
$\mlspone$ between 250 to 350 GeV.   
Fig.\ref{tan40_M1_vs_Mu_muon}, shows a similar plot
for $\tan\beta=40$.   
Here $\mlspone$ ranges 
from 150 GeV to 760 GeV. Both for $\tan\beta=10$ and $40$, the upper
limit of the mass of $\lspone$ is restricted via the XENON1T SI-DD data.
    
\begin{figure}[hbt] 
	\begin{center}
		\subfigure[]{%
\includegraphics[width=0.45\textwidth]{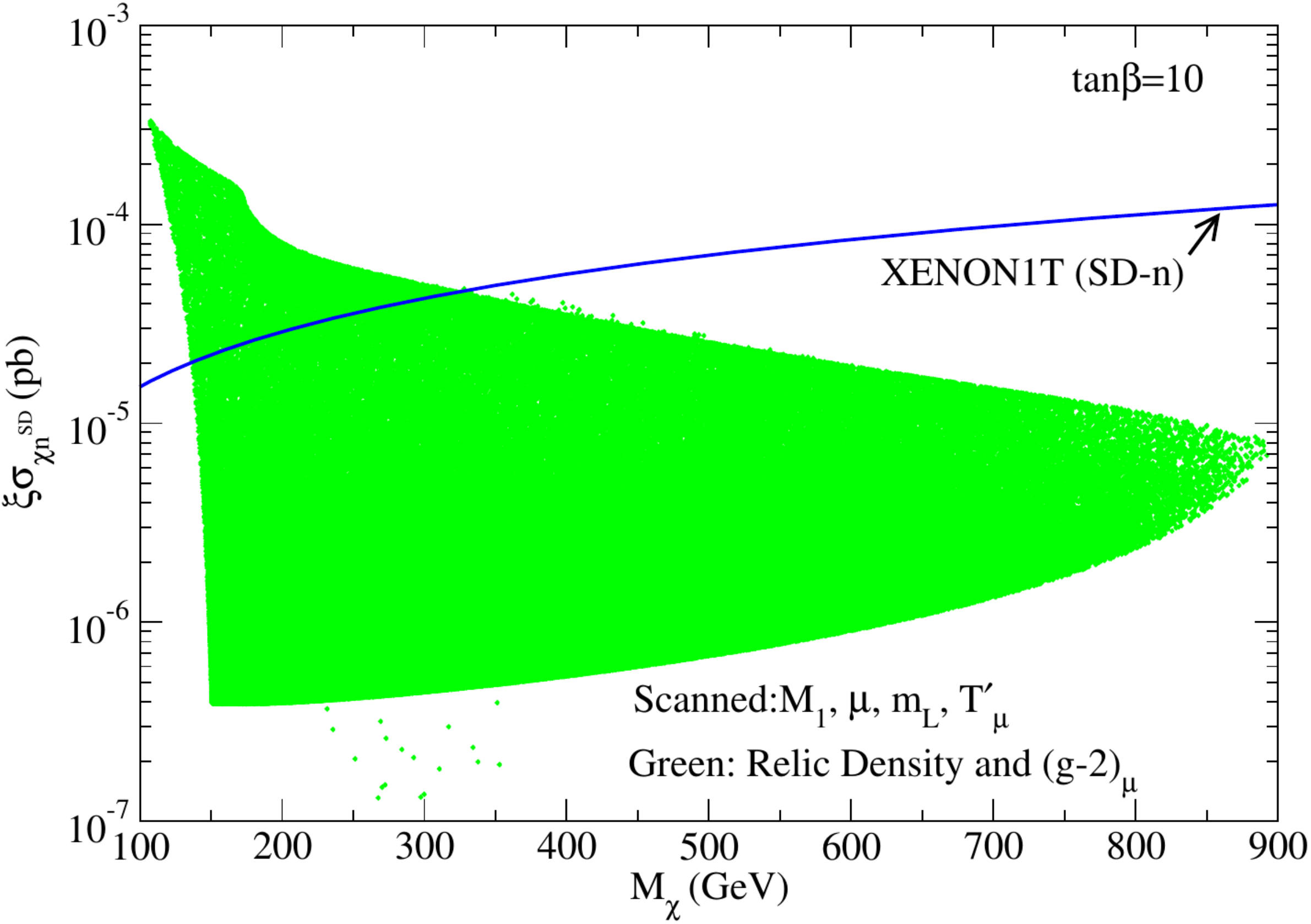}
\label{tan10_all_scanned_all_constraints_LSP_vs_sigmaSD}
		}%
		\hskip 30pt
		\subfigure[]{%
                \includegraphics[width=0.45\textwidth]{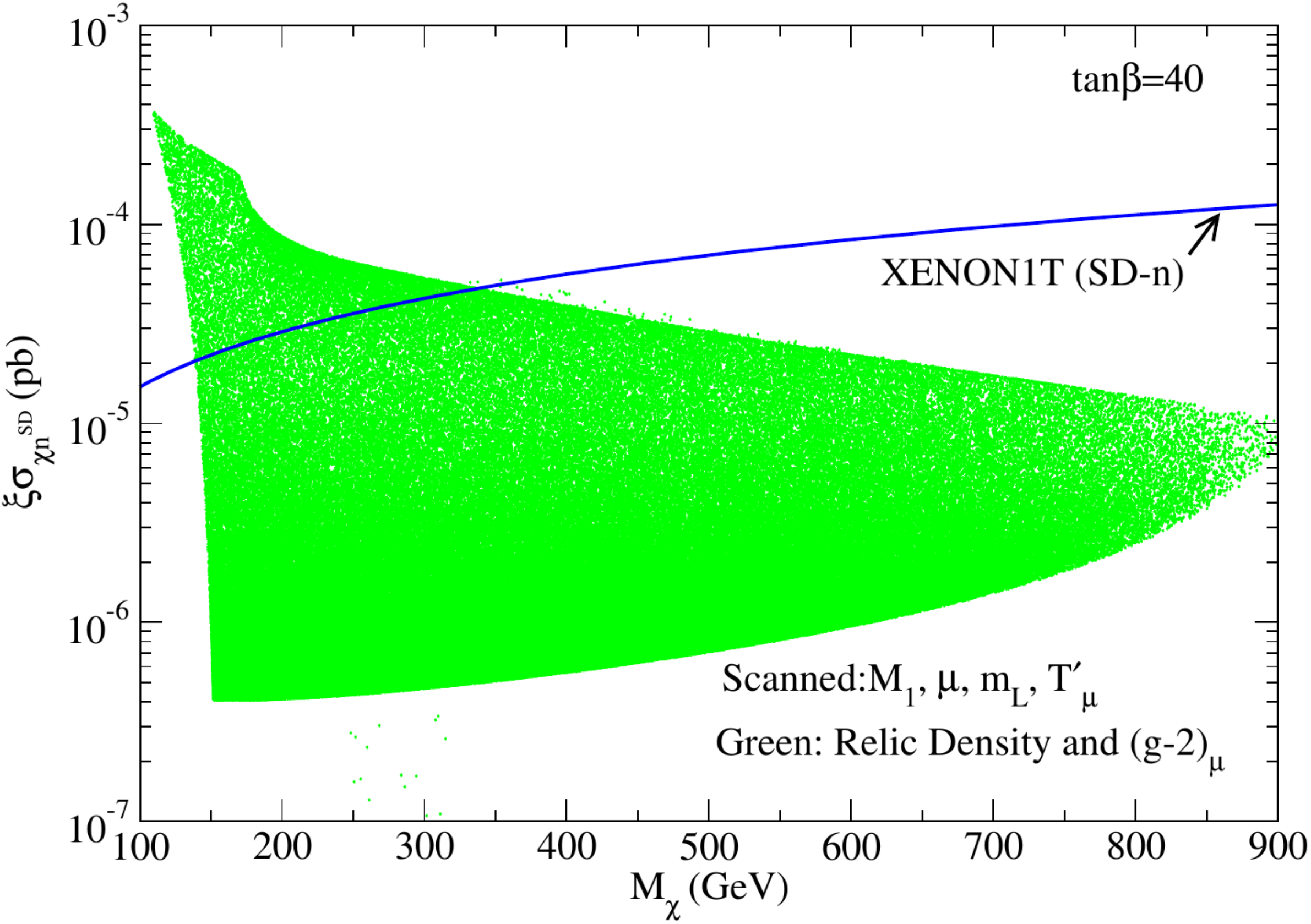}
        \label{tan40_all_scanned_all_constraints_LSP_vs_sigmaSD}        
		}%
		\caption{
Fig.\ref{tan10_all_scanned_all_constraints_LSP_vs_sigmaSD}: Spin-dependent (SD) scattering cross-section of the LSP 
     with a neutron $\sigma^{\rm SD}_{\chi n}$ multiplied
     by the previously mentioned 
scale factor $\xi$, as a function
     of the LSP mass, for $\tan\beta = 10$.
The points satisfying the DM relic density
upper bound and $(g-2)_\mu$ limits at 2$\sigma$ level are shown in green.
The blue line refers to the constraint from the SD direct-detection experiment
of XENON1T at 90\% CL. 
The constraint from 
$\sigma^{\rm SD}_{\chi p}$ is weaker than $\sigma^{\rm SD}_{\chi n}$.  
     Fig.\ref{tan40_all_scanned_all_constraints_LSP_vs_sigmaSD}: Similar
     figure for $\tan\beta = 40$.
}
\label{LSP-sigmaSD}
	\end{center}
\end{figure}

We now include the Spin-dependent direct detection of DM
study in our analysis. 
Fig.\ref{tan10_all_scanned_all_constraints_LSP_vs_sigmaSD} shows the
Spin-dependent (SD) scattering cross-section of the LSP 
     with a neutron ($\sigma^{\rm SD}_{\chi n}$) as a function
     of the LSP mass for $\tan\beta = 10$. As before,
we multiply $\sigma^{\rm SD}_{\chi n}$ 
with a scale factor $\xi$ (Eq.\ref{scalefactorequation}).
Only those points satisfying the DM relic density upper bound as well as
$(g-2)_\mu$ at 2$\sigma$ level are plotted and these
are shown in green.
The blue line is the constraint from the SD direct-detection experiment
of XENON1T indicating discarded regions above the line at 90\% CL.
We further checked that the constraint from 
$\sigma^{\rm SD}_{\chi p}$ is weaker than $\sigma^{\rm SD}_{\chi n}$. 
     Fig.\ref{tan40_all_scanned_all_constraints_LSP_vs_sigmaSD} is a similar
     plot for $\tan\beta = 40$. Compared to the SI DD
     cross-sections of 
     Fig.\ref{LSP-sigmaSI}, here 
     we do not find any new discarded region. This is
     true for both choices of $\tan\beta$.
     Hence, all our conclusion for the SI DD cross-section analysis remain
     valid and analyzing $\sigma^{\rm SI}_{\chi p}$ itself is 
     sufficient for a conclusion regarding the direct detection of DM.

\begin{figure}[hbt] 
	\begin{center}
		\subfigure[]{%
			\includegraphics[width=0.45\textwidth]{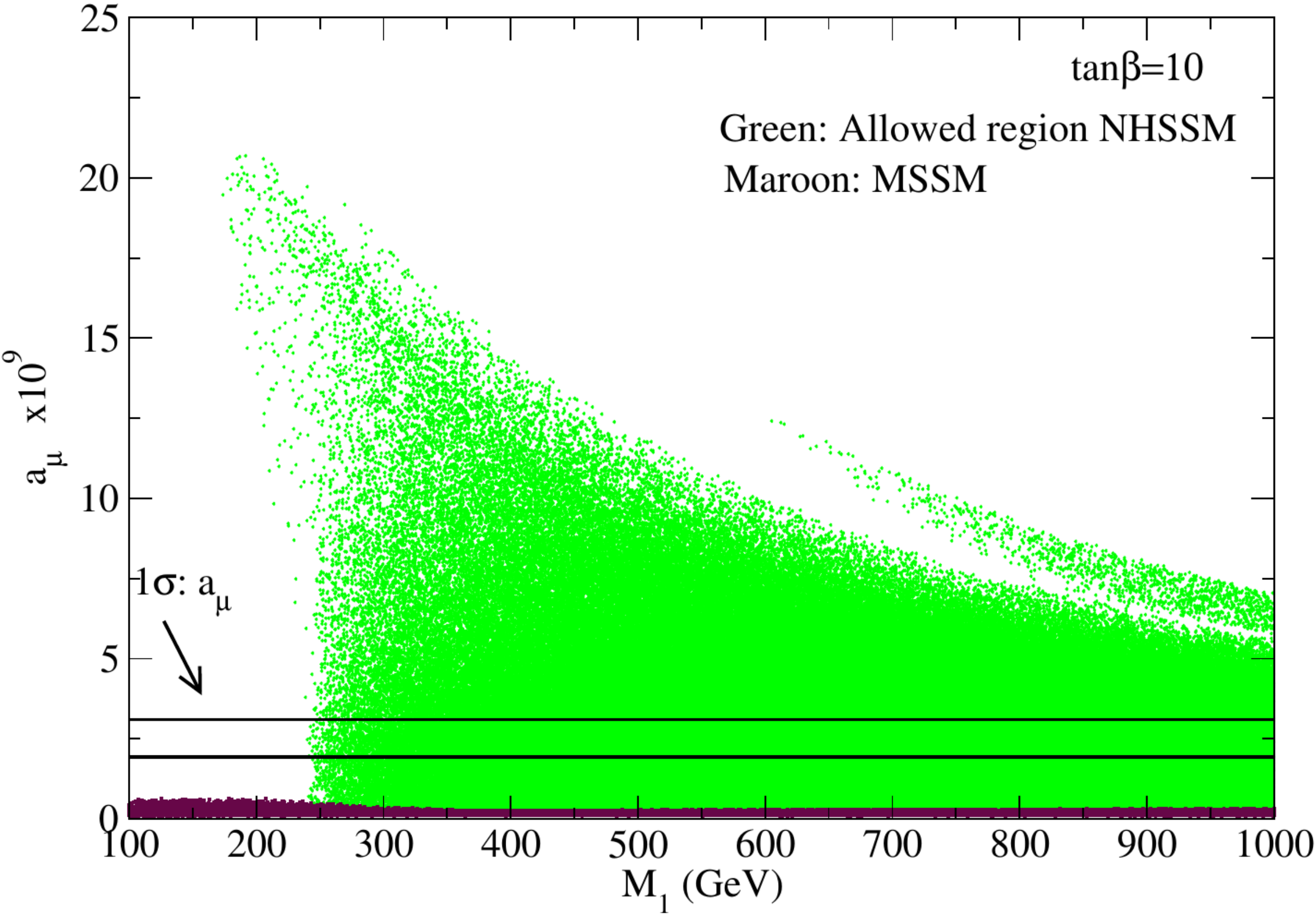}
        \label{tan10_allscanned_M1_vs_amu}                
		}%
		\subfigure[]{%
			\includegraphics[width=0.45\textwidth]{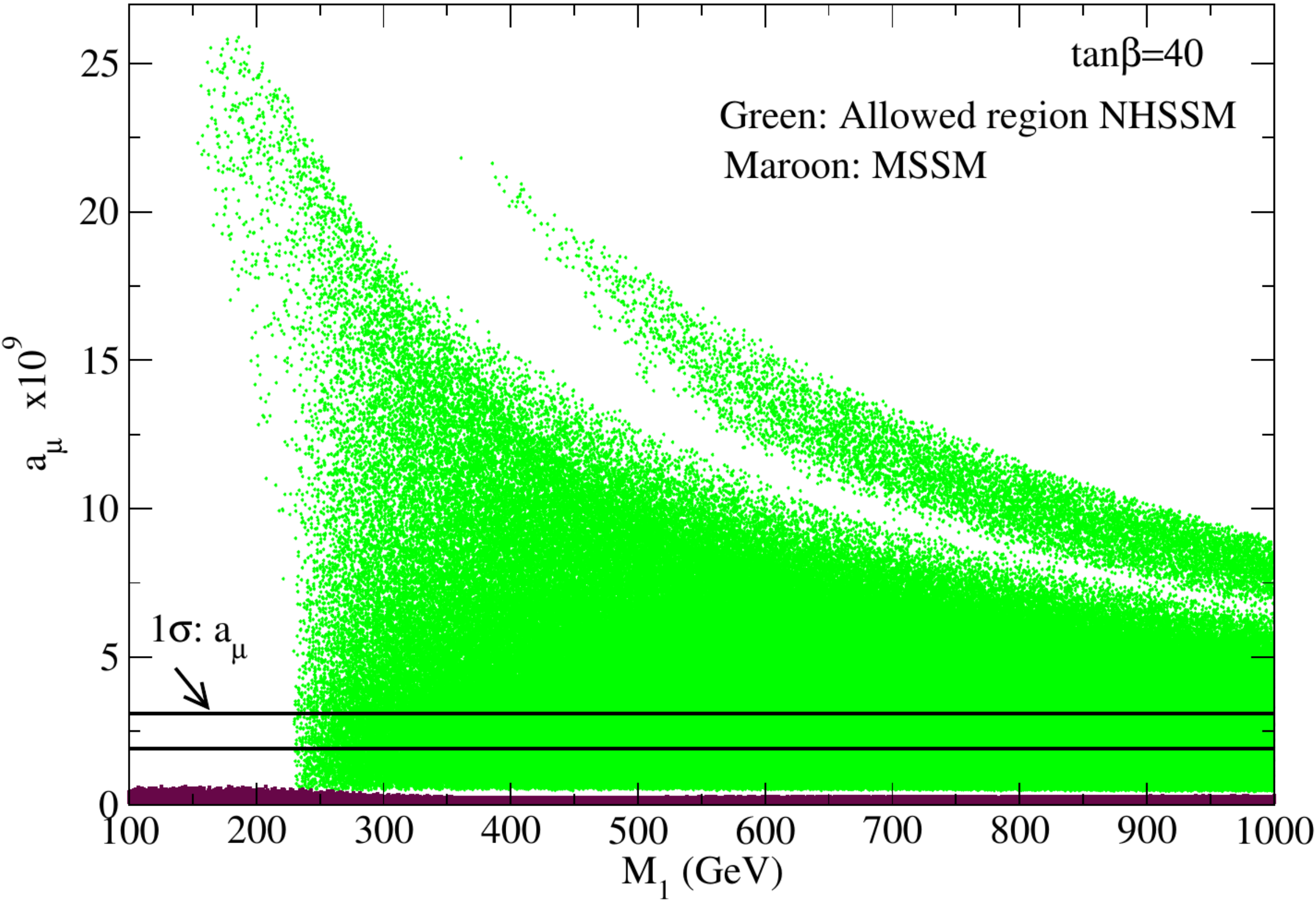}
                        \label{tan40_allscanned_M1_vs_amu}                
		}%
\caption{Fig.\ref{tan10_allscanned_M1_vs_amu}: Scatter plot of
        $a_\mu$ for varying $M_1$ and other parameters as mentioned in the
        text for $\tan\beta = 10$. The parameter points shown in green
        satisfy the dark matter
        constraints including the direct detection limits from XENON1T. 
In order to probe the extent to which NHSSM can enhance $\amususy$,
the maroon points refer to the result of a similar 
scanning in an MSSM parameter space (using $T^\prime_l=0$). None of the
dark matter constraints are applied here. 
Throughout the analyses, we confirm
that the points that satisfy $\sigma^{\rm SI}_{\chi p}$ also respect
both n and p SD DD cross-sections namely, $\sigma^{\rm SD}_{\chi n}$ and
$\sigma^{\rm SD}_{\chi p}$ respectively. The $1\sigma$ allowed band for $a_{\mu}$
is shown as black horizontal lines.  
Fig.\ref{tan40_allscanned_M1_vs_amu}: Similar figure for $\tan\beta=40$.
 }
 \label{M1_amu1040}        
	\end{center}
\end{figure}

Below, we probe how large $a_\mu$ can go depending on the variation of basic
MSSM parameters like $M_1$ and $m_L$ and NHSSM parameter $T^\prime_\mu$. 
Thus, Fig.\ref{tan10_allscanned_M1_vs_amu} shows a scatter plot of
        $a_\mu$ vs. $M_1$ for $\tan\beta = 10$.
The parameter points shown in green 
        satisfy the dark matter
        constraints including the direct detection limits from XENON1T. 
For
comparison purposes, we draw maroon points located near the
bottom of the $a_\mu$ axis. These points  
refer to the result of a similar
scanning in an MSSM parameter space (by using $T^\prime_\mu=0$). 
None of the dark matter constraints are however applied here in the MSSM case. 
The smallness of the 
MSSM $a_\mu$ values shows that the bino-smuon loop contribution with
effects from $T^\prime_\mu$ supersedes all the MSSM loop contributions. This
indeed helps us in explaining the sign-correlation of
$a_\mu$ with $T^\prime_\mu$ as mentioned earlier.
The regions with large $a_\mu$ typically arise in small $M_1$ zones and
this is likely to be case due to the associated smallness of
slepton masses (see Figure~\ref{tan1040_amu_contour_M1-mL}).         
The figure shows that the SI 
direct detection cross-section is below the XENON1T limit throughout the
domain of variation chosen for $M_1$. The results shown in
Fig.\ref{tan40_allscanned_M1_vs_amu}
confirm larger MSSM contributions to $a_\mu$ for an increase of 
$\tan\beta$ to 40, although the enhancement is much smaller than
the NHSSM effects toward $a_\mu$, hence not so visible. 

\begin{figure}[hbt] 
	\begin{center}
		\subfigure[]{%
			\includegraphics[width=0.45\textwidth]{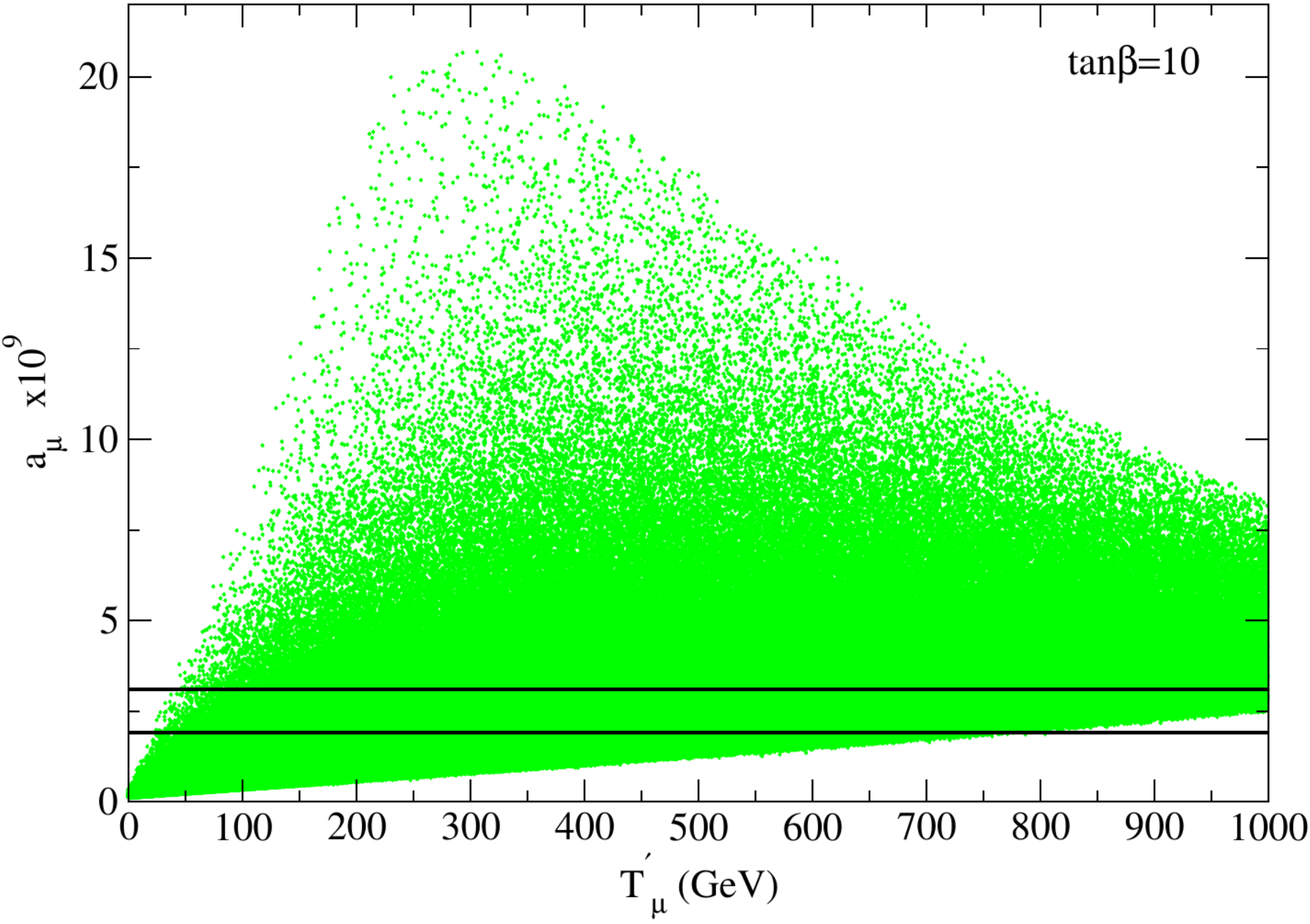}
        \label{tan10_allscanned_Amuprime_vs_amu}                
		}%
		\subfigure[]{%
			\includegraphics[width=0.45\textwidth]{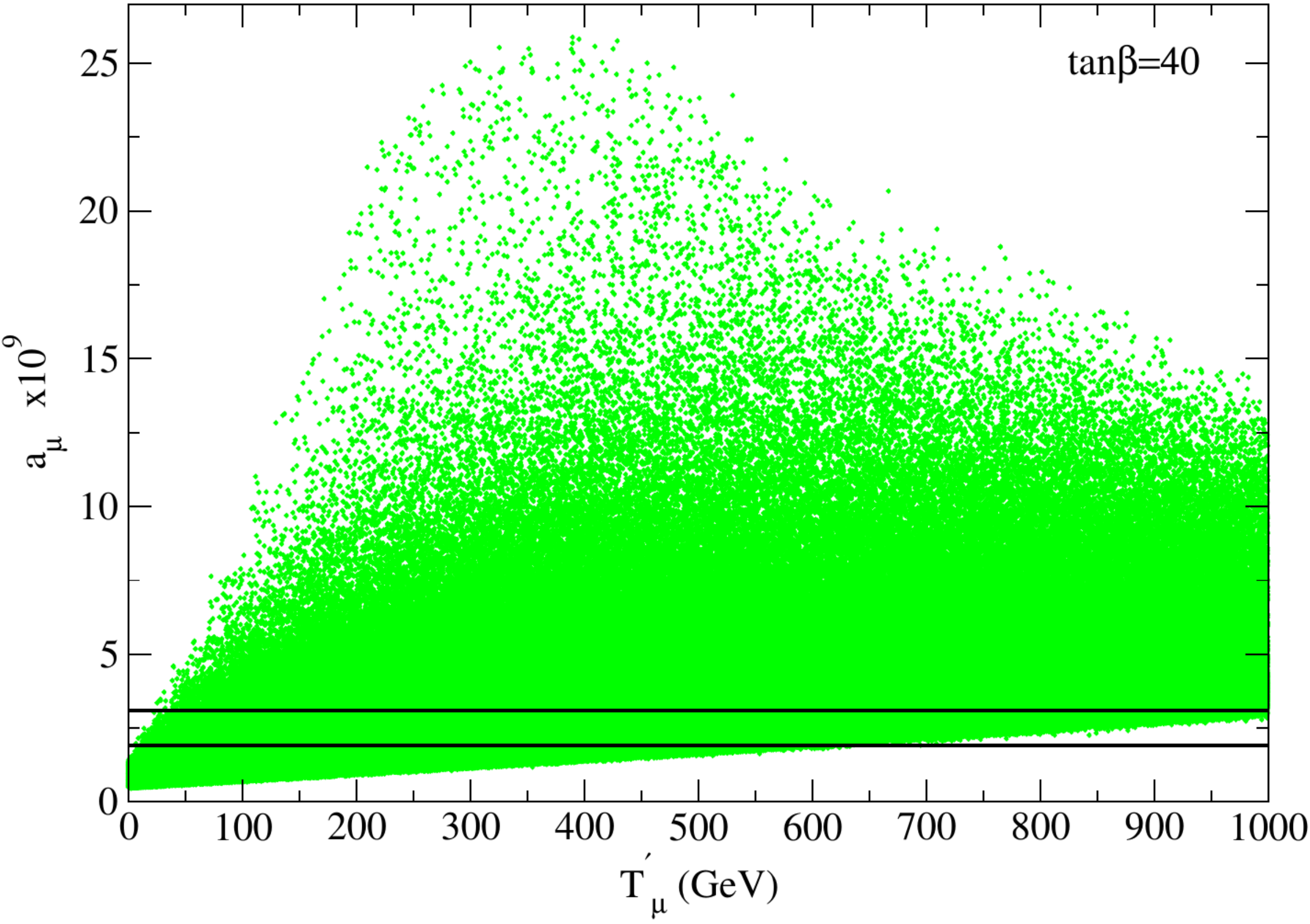}
                        \label{tan40_allscanned_Amuprime_vs_amu}                
		}%
\caption{
Fig.\ref{tan10_allscanned_Amuprime_vs_amu}: Scatter plot of
        $a_\mu$ vs. $T^\prime_\mu$ for $\tan\beta = 10$ for varying
parameters as mentioned in the text. All the green points 
satisfy the same constraints as mentioned in Fig.\ref{M1_amu1040}.	
Fig.\ref{tan40_allscanned_Amuprime_vs_amu}: Similar figure for $\tan\beta=40$.
 }
 \label{Amuprime_amu1040}        
	\end{center}
\end{figure}

\begin{figure}[hbt] 
	\begin{center}
		\subfigure[]{%
			\includegraphics[width=0.45\textwidth]{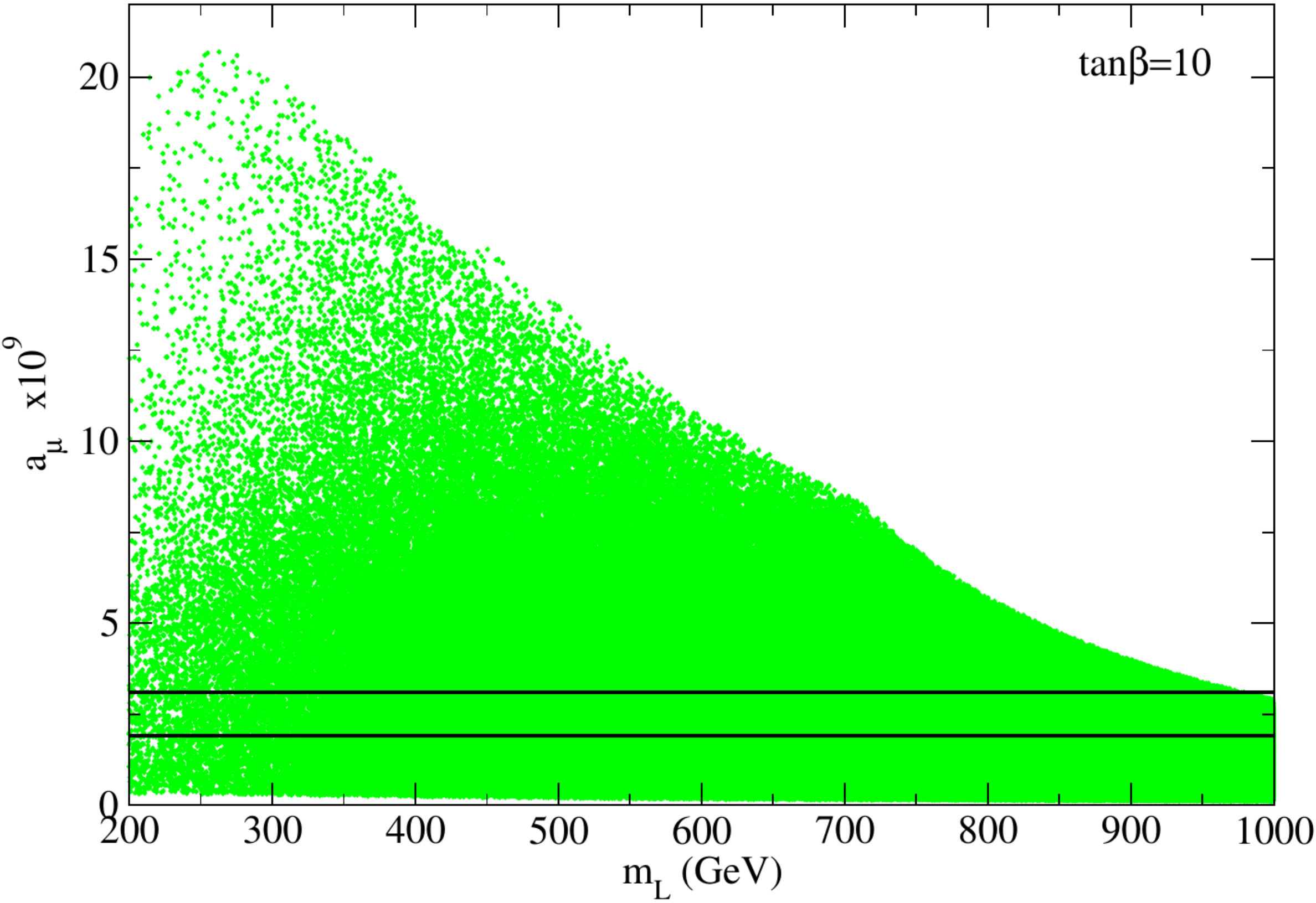}
            \label{tan10_allscanned_mL_vs_amu}            
		}%
		\subfigure[]{%
			\includegraphics[width=0.45\textwidth]{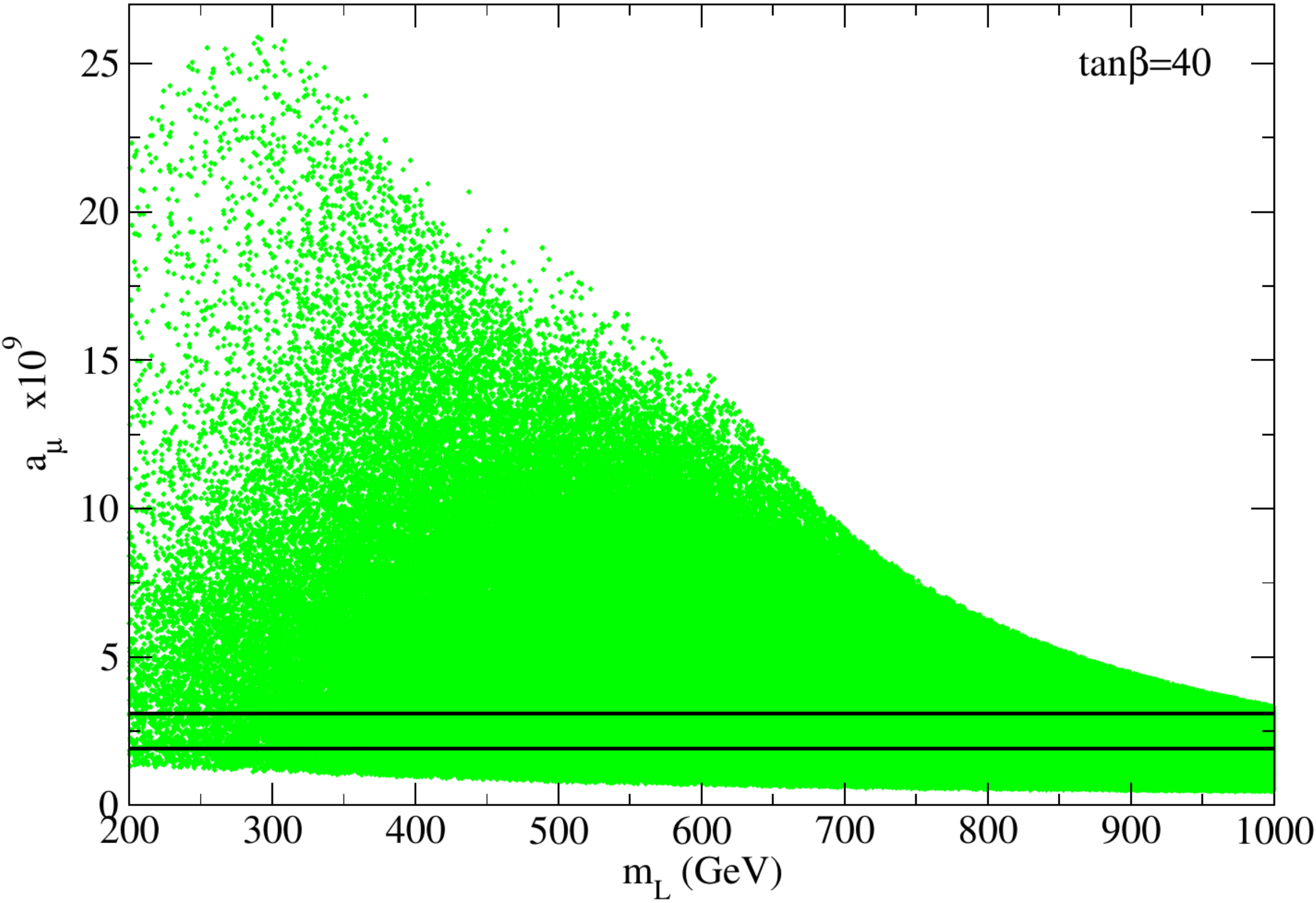}
                        \label{tan40_allscanned_mL_vs_amu}            
		}%
	\caption{
Fig.\ref{tan10_allscanned_mL_vs_amu}: Same as 
Fig.\ref{tan10_allscanned_Amuprime_vs_amu} for $\tan\beta = 10$ 
except for varying $m_L$ instead of $T^\prime_\mu$.   
Fig.\ref{tan40_allscanned_mL_vs_amu}: Similar plot for $\tan\beta=40$. 
}
\label{mL-amu1040}
	\end{center}
\end{figure}

Figure~\ref{Amuprime_amu1040} and Figure~\ref{mL-amu1040} are the
scatter plots for the dependence of $a_\mu$ on $T^\prime_\mu$ and $m_L$ when
other parameters are varied. Fig.\ref{tan10_allscanned_Amuprime_vs_amu}
displays a 
scatter plot of
        $a_\mu$ for varying $T^\prime_\mu$ for $\tan\beta = 10$.
The points shown in green satisfy the dark matter relic density upper
bound 
and their SI direct detection (DD)
cross-section $\sigma^{\rm SI}_{\chi p}$ values fall 
below the XENON1T limit.
The large $a_\mu$ region of Figure~\ref{Amuprime_amu1040} refers to 
small $m_L$ values. This in turn correspond to small $M_1$ zones 
with enhanced $a_\mu$. 
The blanck
(white) strip near the $T^\prime_\mu$ axis i.e., near the region
with very small $a_\mu$ denotes the absence of valid parameter 
points, and this is related to the blanck region for 
smaller $m_L$ values (near the $m_L$ axis) 
of Fig.\ref{tan10_allscanned_mL_vs_amu}. In
combination, the resulting off-diagonal slepton mass values for large
$T^\prime_\mu$ and small $m_L$ are likely to generate excluded regions 
because of the appearance of tachyonic slepton states
or vacuum instability.
With $T^\prime_\mu=y_\mu {A}_\mu$, the effect becomes more prominent
for a larger $\tan\beta$ as may be seen in
Fig.\ref{tan40_allscanned_Amuprime_vs_amu} and
Fig.\ref{tan40_allscanned_mL_vs_amu}. We further note that the
large $a_\mu$ regions corresponding to small $m_L$ values in each of the parts of
Figure~\ref{mL-amu1040} refer to small
$M_1$ and large $T^\prime_\mu$ zones. The LSP that satisfies
the DM constraints is of higgsino type and associated with underabundance. 

\subsection{Inclusion of Electron Magnetic Moment limits; Yukawa coupling
enhancement in NHSSM}
\label{yukawathreshold}
\begin{figure}[hbt] 
	\begin{center}
		\subfigure[]{%
		\includegraphics[width=0.45\textwidth]{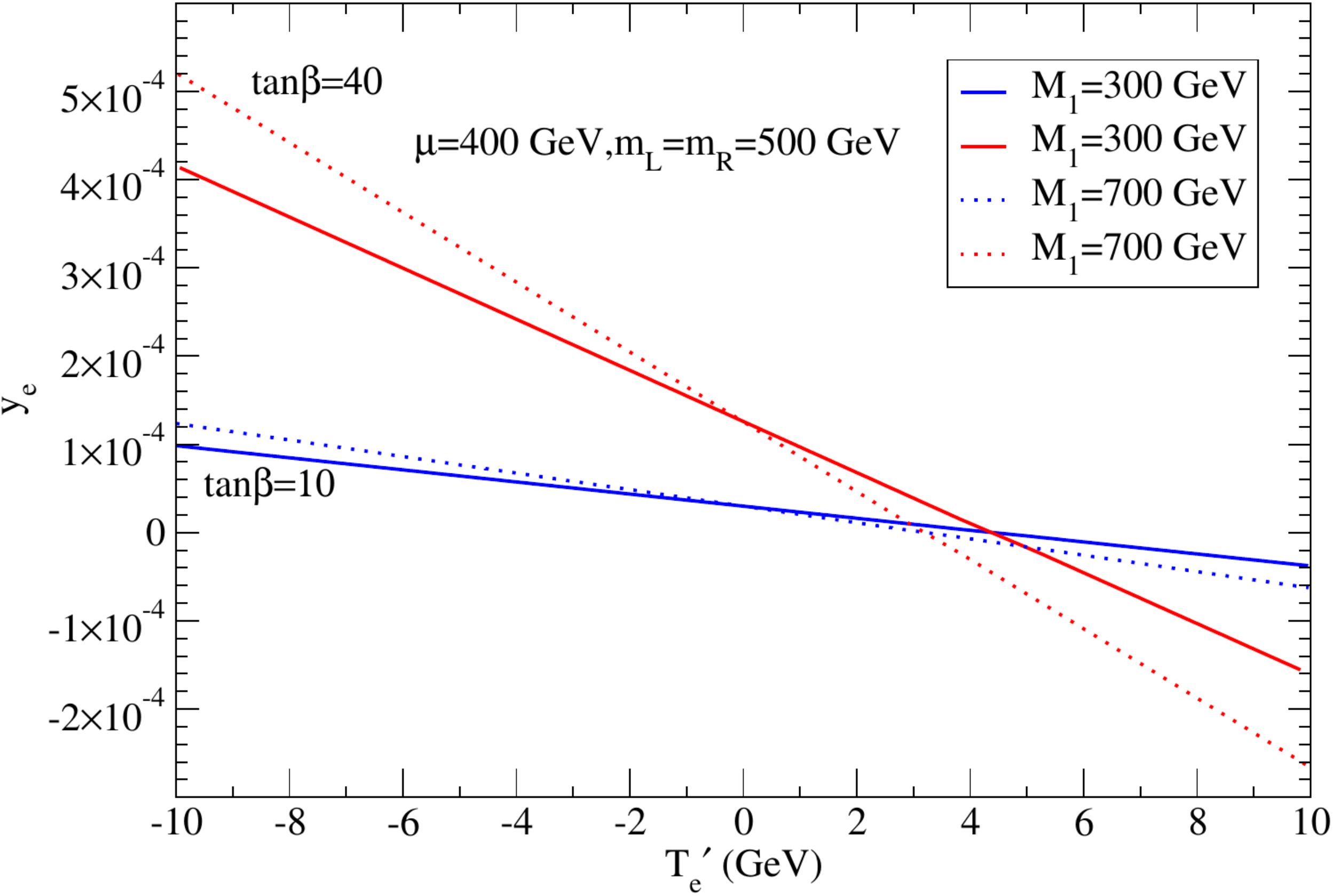}
                \label{yevsAeprime}
		}%
		\subfigure[]{%
			\includegraphics[width=0.45\textwidth]{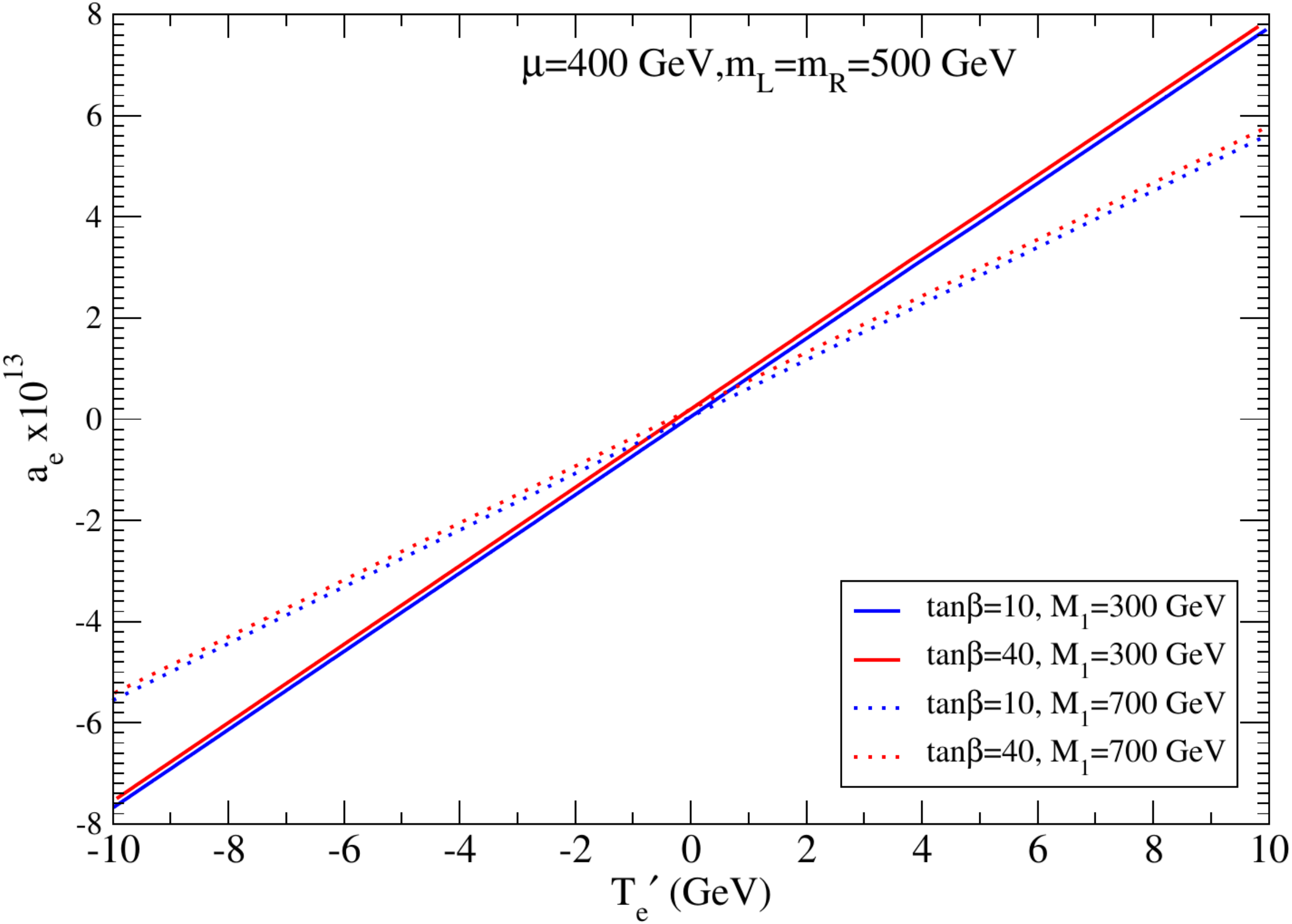}
                        \label{aevsAeprime}
		}%

		\subfigure[]{%
		\includegraphics[width=0.45\textwidth]{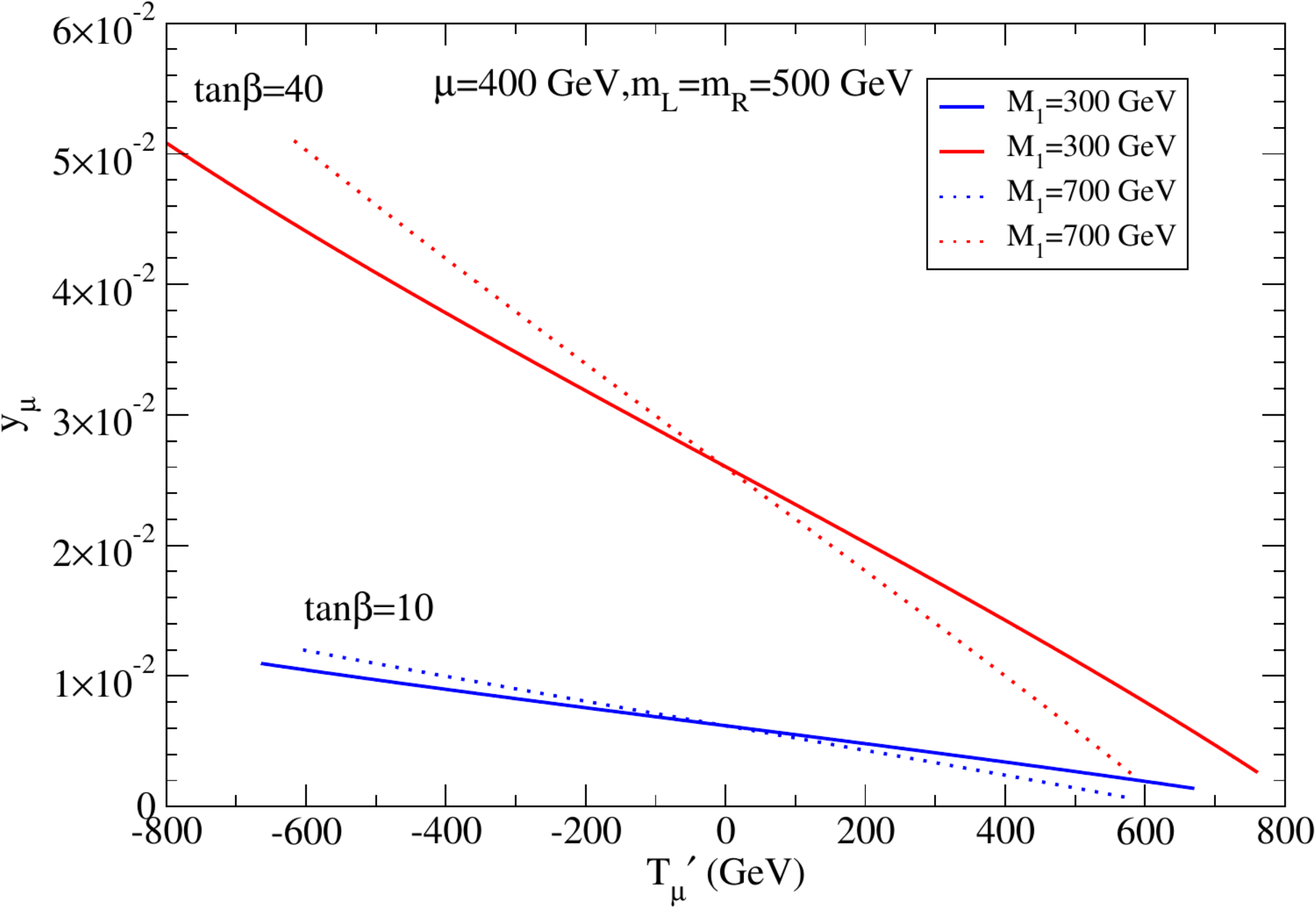}
                \label{ymuvsAmuprime}
		}%
		\subfigure[]{%
			\includegraphics[width=0.45\textwidth]{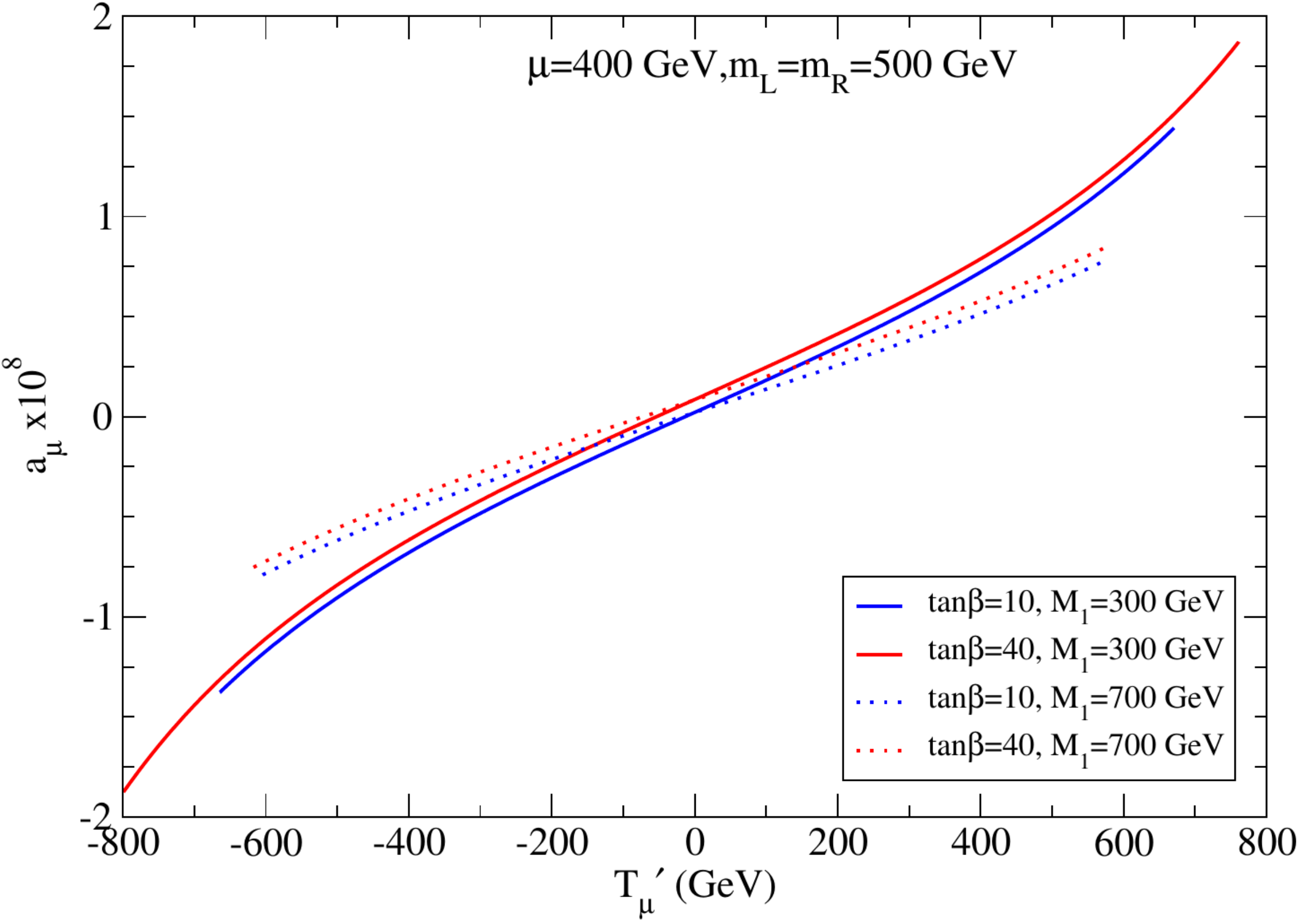}
                        \label{amuvsAmuprime}
		}%
\caption{Fig.\ref{yevsAeprime}: Display of the effect of
                $T^\prime_e$ on the electron Yukawa coupling $y_e$ for the shown values of the SUSY parameters. 
For each $\tan\beta$, 
the solid lines refer to $M_1=300$ GeV whereas the dotted lines are drawn for
$M_1=700$~GeV. A large negative value for $T^\prime_e$ gives rise to
a larger $|y_e|$ compared to the same for a similar positive value
for $T^\prime_e$.
$y_e$ is also seen to flip its sign as $T^\prime_e$ becomes positive. A larger
$M_1$ also enhances $|y_e|$.  No LTCYC condition for $y_e$ is applied here in order to display the extent of the Yukawa threshold correction of $y_e$.
Fig.\ref{aevsAeprime}: Display of 
                the electron magnetic moment $a_e$ for two different values of
                $\tan\beta$ as well as $M_1$. The other SUSY parameters
                are shown within the figure. The two
                solid lines refer to $\tan\beta=10,40$ for $M_1=300$~GeV, and 
                the two dotted lines stand for $\tan\beta=10,40$
                where $M_1=700$~GeV. Unless $T^\prime_e$ is tiny, 
                negative values of $a_e$ that is required via the experimental
                limits follow from negative values of $T^\prime_e$.
Fig.\ref{ymuvsAmuprime}: A similar plot like Fig.\ref{yevsAeprime}
for electron.
Fig.\ref{amuvsAmuprime}: A similar plot like Fig.\ref{aevsAeprime} for muon. 
}
         \label{ye_and_ae_vsAeprime_also_muon}        
	\end{center}
\end{figure}

We will now probe the effect of including the ${}^{133}{\rm Cs}$-based
fine-structure constant
measurement derived 
electron $g-2$ data in our analysis on top of the muon $g-2$ constraint.
We will only consider the necessary amount for $T^\prime_e$ that would 
generate the required value for $a_e$. We will also highlight the threshold 
corrections to the electron Yukawa coupling $y_e$ due to $T^\prime_e$ 
and constrain the NHSSM parameter space accordingly. This will involve both
cases namely a free enhancement of $y_e$ as well as limiting $y_e$ by requiring to
stay within the LTCYC zone as mentioned earlier.
Fig.\ref{yevsAeprime} shows the effect of
                $T^\prime_e$ on the electron Yukawa coupling $y_e$ for
  $\mu=400$~GeV and $m_L(=m_R)=500$~GeV. The blue lines are
  for $\tan\beta=10$, and the 
  red lines are for $\tan\beta=40$.  For each $\tan\beta$, 
the solid lines refer to $M_1=300$ GeV whereas the dotted lines are drawn for
$M_1=700$~GeV. 
No LTCYC condition for $y_e$ is applied here
in order to display the extent of the Yukawa threshold correction of $y_e$.
  A large negative value for $T^\prime_e$ gives rise to
a larger $|y_e|$ compared to the same for an identical positive value of 
$T^\prime_e$. 
$y_e$ is also seen to flip its sign as $T^\prime_e$ becomes positive. A larger
$M_1$ also enhances $|y_e|$. We point out that no LTCYC condition is
applied here in order to show the extent how $y_e$ is affected due to
an increased $T^\prime_e$.
Depending on the value of $m_L$ and $M_1$, the above
may in turn cause appearance of tachyonic selectron states.
We now try to understand how $a_e$ behaves as $T^\prime_e$ is varied. 
Fig.\ref{aevsAeprime} displays $a_e$ for two different values of
$\tan\beta$, and two different values
of $M_1$ identical with those of Fig.\ref{yevsAeprime}. The other SUSY
parameters are $\mu=400$~GeV and $m_L(=m_R)=500$~GeV as used before.
Unless $T^\prime_e$ is tiny, 
                negative value of $a_e$ that is required due to the
                experimental data is correlated with negative value
                of $T^\prime_e$. We note that the solid lines of $a_e$ for
                $\tan\beta=10$ and $40$ for $M_1=300$~GeV are very close to
                each other. The same is true for the two dotted lines
                corresponding to $M_1=700$~GeV. The reason for the approximate
                independence of $a_e$ on $\tan\beta$ would be clear in 
                Section~\ref{yukawathreshold}.
The display of $y_\mu$ for a variation of $T^\prime_\mu$ as shown in
Fig.~\ref{ymuvsAmuprime} indicates a larger available zone of variation
of $T^\prime_\mu$ compared to that of
$T^\prime_e$ of Fig.~\ref{yevsAeprime}. This arises because of
the difference of mass values of electron and muon. No LTCYC for
$y_\mu$ is an issue here since the threshold
correction of $y_\mu$ is moderate corresponding to the given
span of variation of $T^\prime_\mu$ and the chosen SUSY
mass parameters. 
A large negative value for $T^\prime_\mu$ gives rise to
a larger $|y_\mu|$ compared to the same for a similar positive value for
$T^\prime_\mu$. 
Unlike the case of $y_e$ of Fig.\ref{yevsAeprime}, here $y_\mu$ does not
change sign when
$T^\prime_\mu$ becomes positive, albeit within its given range of interest  
for our study. A larger $M_1$ also enhances $|y_\mu|$.     
Fig.~\ref{amuvsAmuprime} shows the variation of $a_\mu$ over
$T^\prime_\mu$. We note that with a given value $T^\prime_{(e,\mu)}$, varying $M_1$
will alter $y_{(e,\mu)}$ which in turn would affect $a_{e,\mu}$.

\FloatBarrier
\subsubsection{Absence of $\tan\beta$ scaling in $a_l$ within NHSSM
because of large ${A}^\prime_l$}
\label{tanbeta_scaling_behav_text}                 

\begin{figure}[hbt] 
\begin{center}
\includegraphics[width=0.65\textwidth]{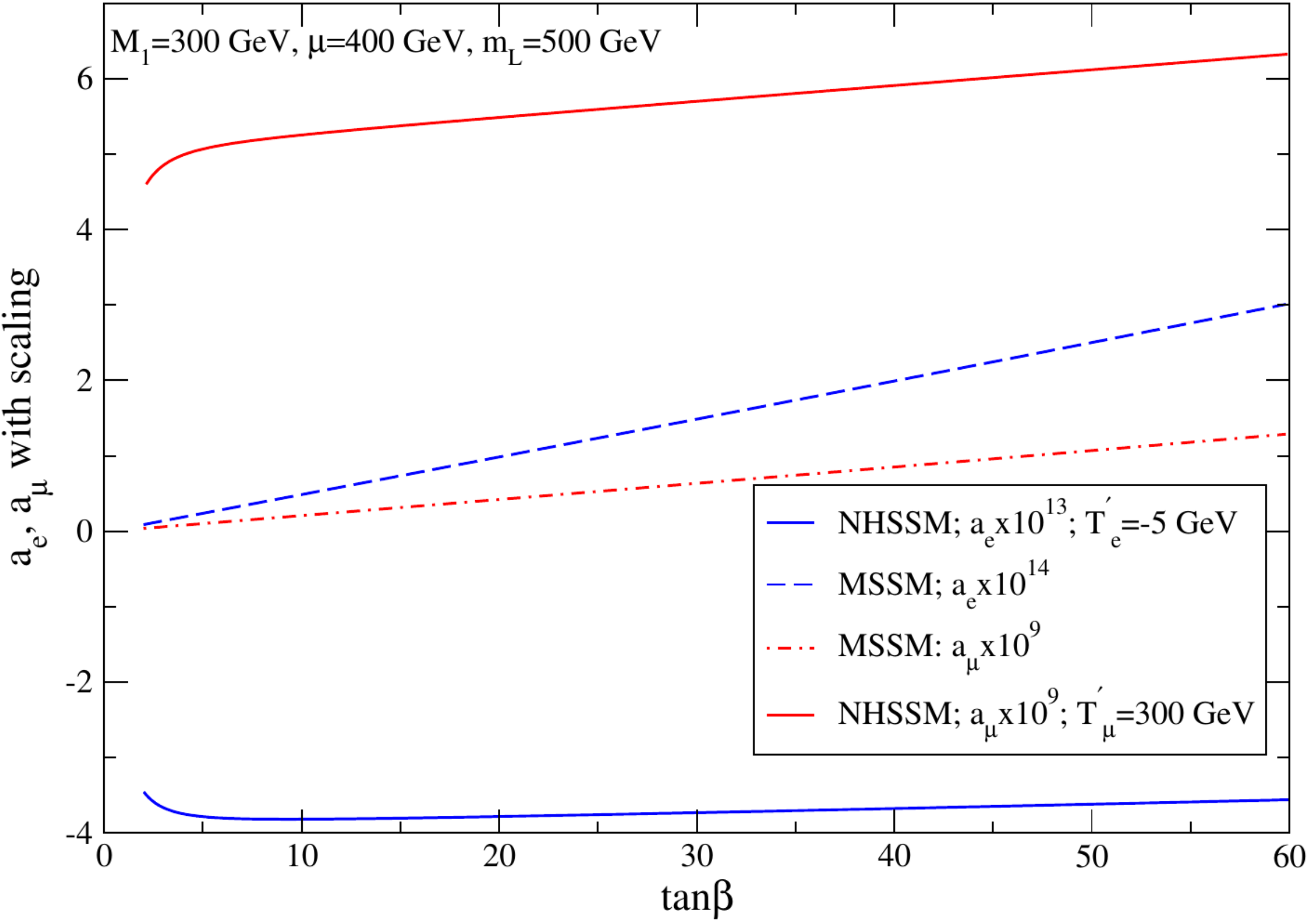}
\caption{Display of scaling related behavior of $a_e$,$a_\mu$ with respect to
$\tan\beta$ for the MSSM and the NHSSM cases. No LTCYC conditions are
used here. Here, $a_e$,$a_\mu$ are 
appropriately multiplied by appropriate powers of 10.  
The solid blue line refers to
$a_e \times 10^{13}$ for $T^\prime_e=-5$~GeV, whereas the dashed blue line
corresponds to $a_e \times 10^{14}$ for the MSSM case (i.e.$T^\prime_e=0$).
Clearly, $a_e$ satisfies proportional relationship with $\tan\beta$ in MSSM,
but $a_e$ is merely a slowly increasing function of $\tan\beta$ with
large intercept. Showing identical behavior, the solid and dashed red lines are
similar results for $a_\mu$.
} 
\label{tanbeta_scaling_behav}
\end{center}
\end{figure}

In Section\ref{yukawathreshold}, we discussed the effects on the $\tan\beta$ related behavior of 
magnetic moments of leptons in scenarios that may induce
large threshold corrections to the associated Yukawa couplings.
In the above context, an NHSSM study of $a_l$ becomes quite relevant.
With no LTCYC conditions applied Figure~\ref{tanbeta_scaling_behav} shows 
the scaling-related 
behaviors of $a_e$,$a_\mu$ with respect to a variation over 
$\tan\beta$ for the MSSM and the NHSSM cases. Here, $a_e$,$a_\mu$ are
appropriately multiplied by powers of 10 so that all of them
may be plotted in a single graph. The solid blue line refers to
$a_e \times 10^{13}$ for $T^\prime_e=-5$~GeV, whereas the dashed blue line
corresponds to $a_e \times 10^{14}$ for the MSSM case (i.e. $T^\prime_e=0$).
Clearly, $a_e$ satisfies proportional relationship with $\tan\beta$ in MSSM,
but this is no longer true in NHSSM, where $a_e$ is only 
a slowly increasing function of $\tan\beta$ with a large intercept
\footnote{This is consistent with what was first commented on 
non-holomorphic interactions in relation to $a_l$ in radiative generation
of fermion mass context in Ref.\cite{Borzumati:1999sp} (see their discussion
followed by Eq.38).} .
Showing identical behavior, the solid and dashed red lines are
similar results for $a_\mu$.
\FloatBarrier
\subsubsection{Limiting the degree of threshold corrections of Yukawa couplings for leptons: LTCYC criterion vs the $a_e$ constraint.}
\label{LTCYCdetails}
\begin{figure}[hbt] 
	\begin{center}
		\subfigure[]{%
		\includegraphics[width=0.45\textwidth]{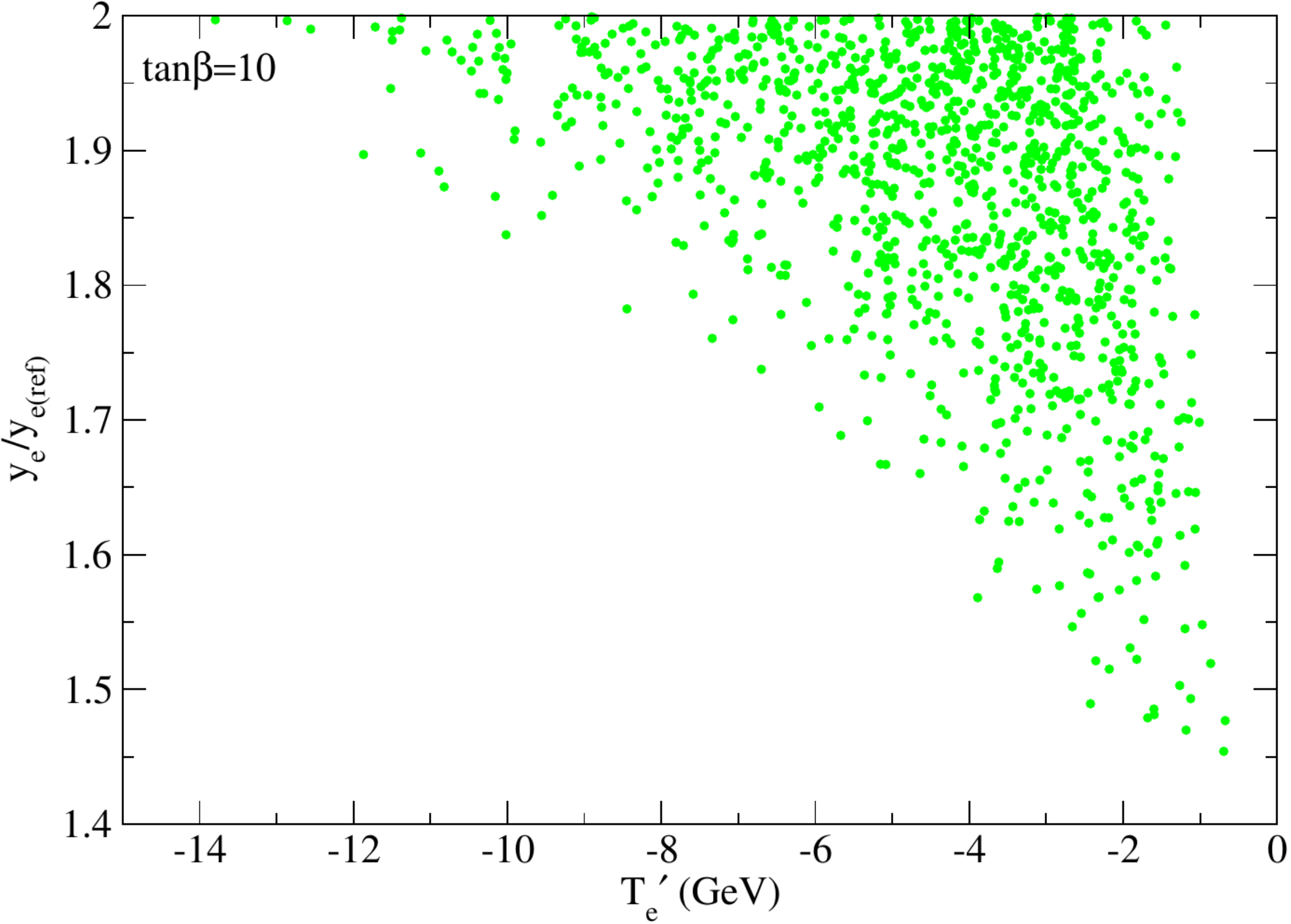}
                \label{Aeprime_yratio_tan10}
		}%
		\subfigure[]{%
			\includegraphics[width=0.45\textwidth]{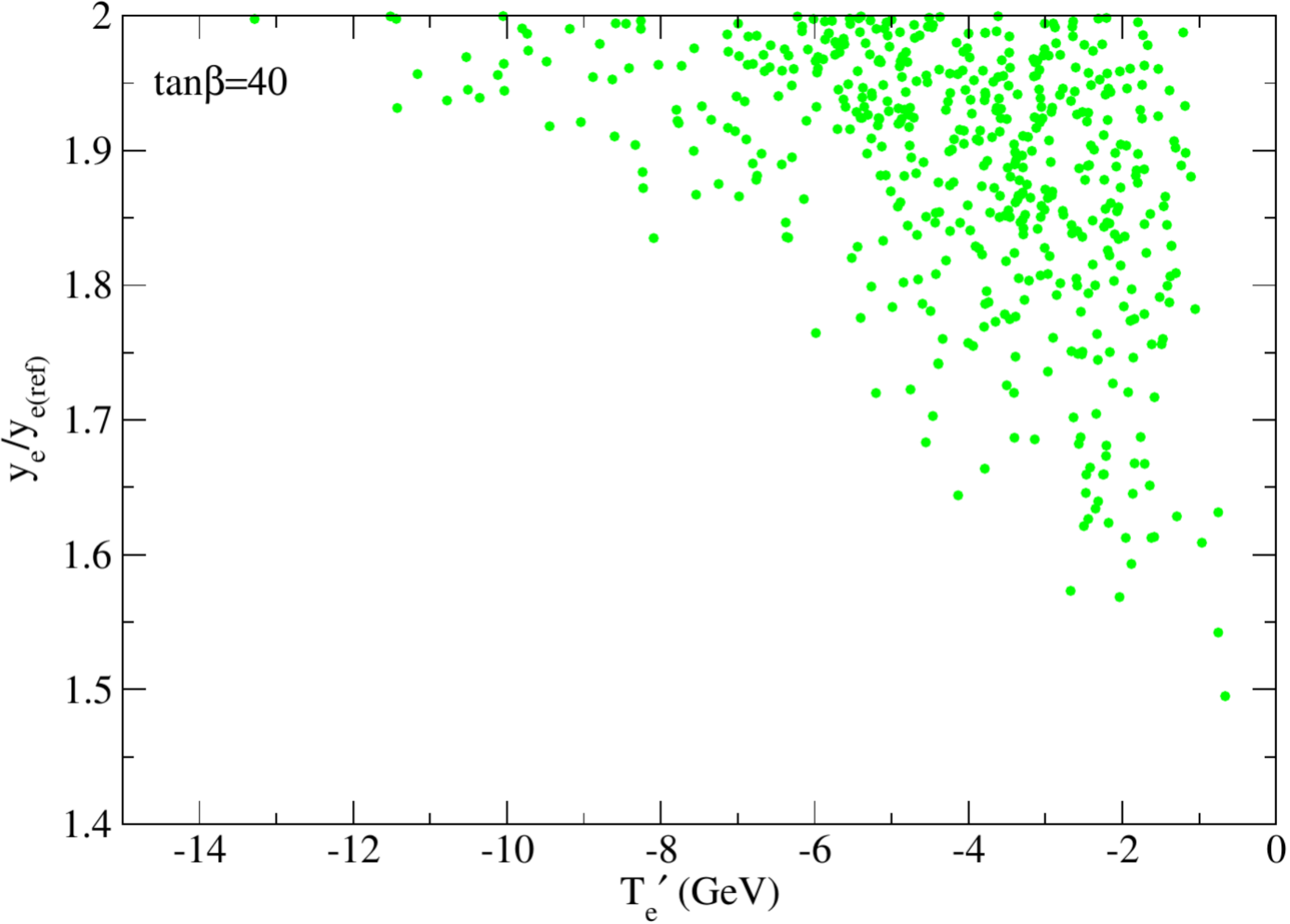}
                        \label{Aeprime_yratio_tan40}
		}%

		\subfigure[]{%
		\includegraphics[width=0.45\textwidth]{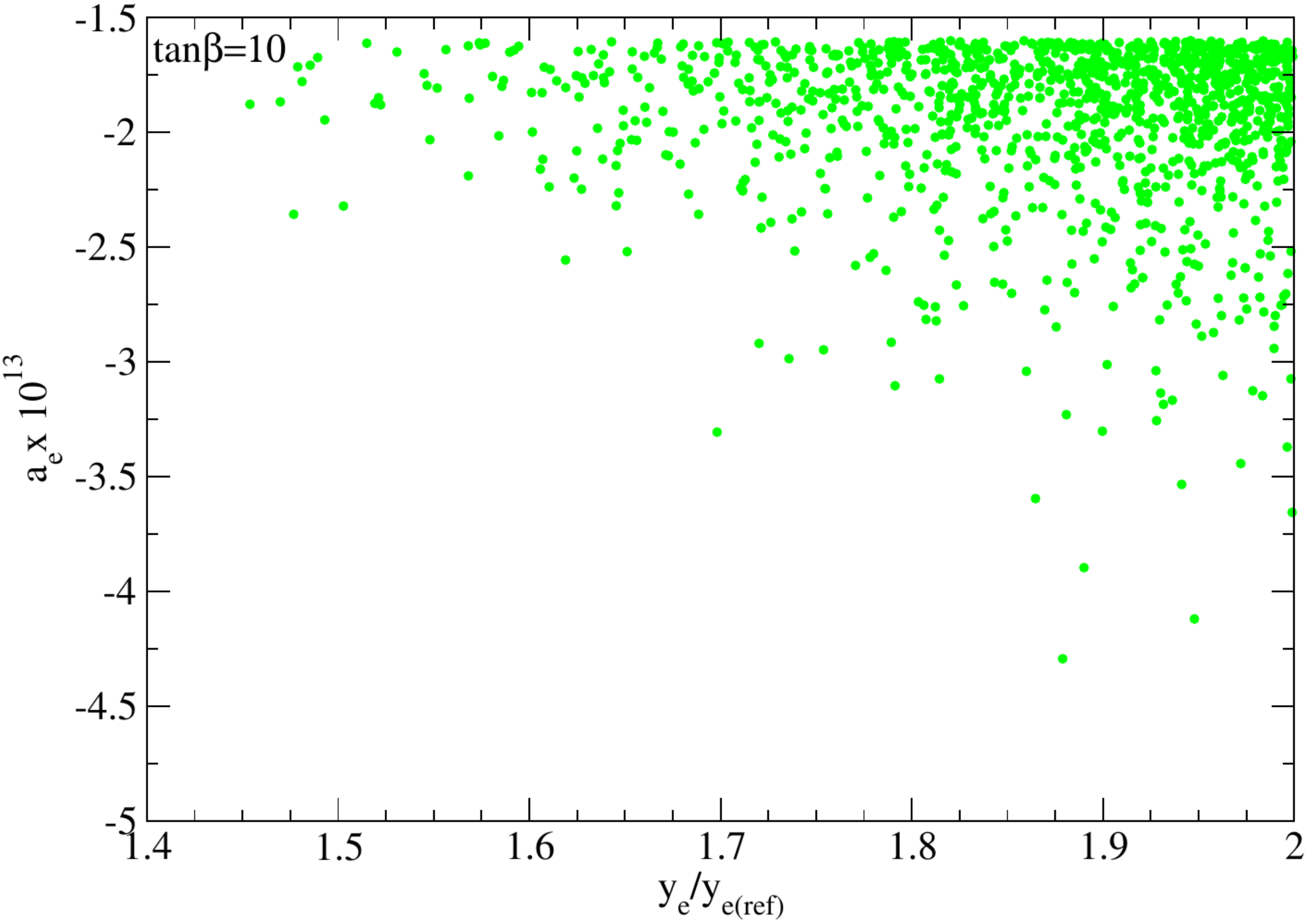}
                \label{tan10_yuk_ratio_ae}
		}%
		\subfigure[]{%
			\includegraphics[width=0.45\textwidth]{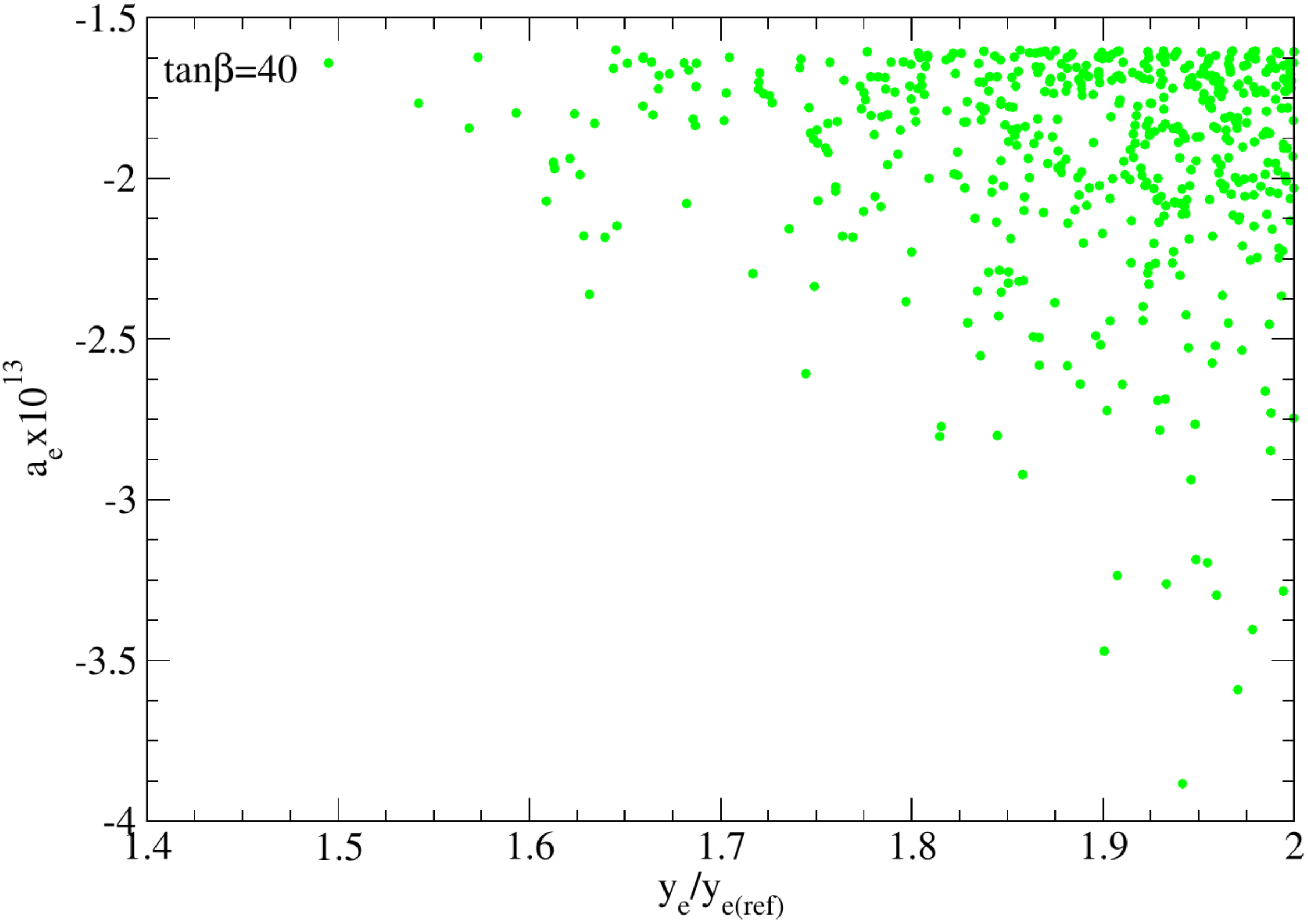}
                        \label{tan40_yuk_ratio_ae}
		}%
\caption{Limiting the level of leptonic Yukawa 
radiative corrections in NHSSM at the weak scale:
$y_l$ refers to output Yukawa coupling
value that is obtained after radiative corrections whereas
$y_{l(ref)}$ is the tree level MSSM input value out of leptonic mass $m_l$.
We choose to discard any parameter point that leads
to $y_l/y_{l(ref)}>2$ and denote it as {\it Limited Threshold Corrections of
Yukawa Coupling (LTCYC)} criterion.  
Fig.\ref{Aeprime_yratio_tan10}: Scatter plot in
$(y_e/y_{e(ref)}- T^\prime_e)$ plane satisfying the LTCYC criterion in
addition to all other constraints like DM relic density upper bound and
direct detection limits, Higgs mass
data, B-physics related limits and $(g-2)_{e,\mu}$ at 2$\sigma$ level for
$\tan\beta=10$.
A larger $|T^\prime_e|$ above 14 is seen not to satisfy the said limit. 
Fig.\ref{Aeprime_yratio_tan40}: Same as (a) for $\tan\beta=40$.
Fig.\ref{tan10_yuk_ratio_ae}: $(a_e-y_e/y_{e(ref)})$ plot showing a
possibility of finding the desired $a_e$ for radiative corrections 
above 45\%. 
Fig.\ref{tan40_yuk_ratio_ae}: Same as (c) for $\tan\beta=40$.
}
         \label{LTCYCdisplay}        
	\end{center}
\end{figure}

The non-holomorphic soft terms may be able to cause
large Yukawa corrections for fermions, particularly the leptons so that
the radiatively corrected values may be a few times the MSSM specified
value namely $y_{l,(ref)} \equiv y_l=\frac{m_e}{(v_d/\sqrt 2)}$. 
This was already discussed for phenomenological analyses 
 in Refs.\cite{Borzumati:1999sp,Marchetti:2008hw,Crivellin:2010ty,Crivellin:2011jt, Thalapillil:2014kya,
Bach:2015doa,Baker:2021yli} where lepton masses were generated radiatively,
obviously requiring a large radiative corrections. 
We should also point out that the present Higgs decay to $e^+ e^-$ from the 
LHC can only give an upper bound for $y_e$ which is about a  
few hundred times of its SM
value\cite{dEnterria:2021xij,Altmannshofer:2015qra,Das:2020ozo}\footnote{
As described in Ref.\cite{Altmannshofer:2015qra} Yukawa couplings may
not be proportional to fermion mass in presence of higher dimensional
operators (6 or more )that respect SM gauge symmetry. Thus a very
different Higgs-electron-electron coupling may be possible to arise from   
new physics effects from these operators.
There may be a significant amount of cancellation between the contributions 
to the mass of electron coming from the SM Yukawa coupling and that obtained from
higher dimensional operators. Similarly, the Higgs-electron-electron 
coupling will have two components, one from SM and the other from
the higher dimensional operators. See also Ref.\cite{Thalapillil:2014kya} for a SUSY
context}..  
In our analysis, a negative $a_e$ may be accommodated by a relatively
large radiative corrections to $y_e$.
Of course a choice of a very large slepton mass would be able to suppress the
radiative corrections, but here we want to explore the minimal zone of
values of $|T^\prime_e|$ that with the chosen SUSY mass spectra of our
analysis would be consistent with limits from $a_e$.
Since $a_e$ is driven by $y_e$, in general, both for muon and electron, as mentioned
before we choose to discard any parameter point that leads 
to ${y_l/y_{l(ref)}}>2$ with $l \equiv e,\mu$ and denote it as 
{\it Limited Threshold Corrections of Yukawa Coupling} (LTCYC) criterion.
Fig.\ref{Aeprime_yratio_tan10} shows a scatter plot in the 
$(y_e/y_{e(ref)}- T^\prime_e)$ plane satisfying the LTCYC criterion in
addition to all other constraints like DM relic density upper bound and
direct detection limits, Higgs mass 
data, B-physics related limits and $(g-2)_{e,\mu}$ at 2$\sigma$ level for
$\tan\beta=10$. The result is 
a bound of 14~GeV for $|T^\prime_e|$ and there are significant amount
of parameter space with much smaller values like a range of 1.5 to 1.8
for ${y_e/y_{e(ref)}}$. 
Fig.\ref{Aeprime_yratio_tan40} shows a similar result for $\tan\beta=40$ showing the same cut-off at 13~GeV. We observe that for
both the values of $\tan\beta$   there are appreciable amount of parameter
space where the ratio ${y_e/y_{e(ref)}}$ stays 
between 1.5 to 1.8 (much below the maximum of 2).
This corresponds to $-4 {\rm ~GeV} <T^\prime_e<-2 {\rm ~GeV}$.   
Fig.\ref{tan10_yuk_ratio_ae} and Fig.\ref{tan40_yuk_ratio_ae} display
the required values of the above Yukawa coupling ratio for correct $a_e$.
LTCYC issue is important for $a_e$ only. For muon, the experimental
limits of the magnetic moment is hardly stringent. Unlike the
case of electron, the
chosen region of variation of $T^\prime_\mu$ does not produce any appreciably 
large amount  of radiative corrections to $y_\mu$. Hence, LTCYC for $y_\mu$
is not an issue for $a_\mu$, so that the condition gets automatically
satisfied for the full parameter space under consideration. Indeed it may be seen that a positive
$T^\prime_\mu$ as required for positive $a_\mu$  leads to 
values below unity for $y_\mu/y_{\mu(ref)}$.
\FloatBarrier

\begin{figure}[hbt] 
	\begin{center}
		\subfigure[]{%
		\includegraphics[width=0.45\textwidth]{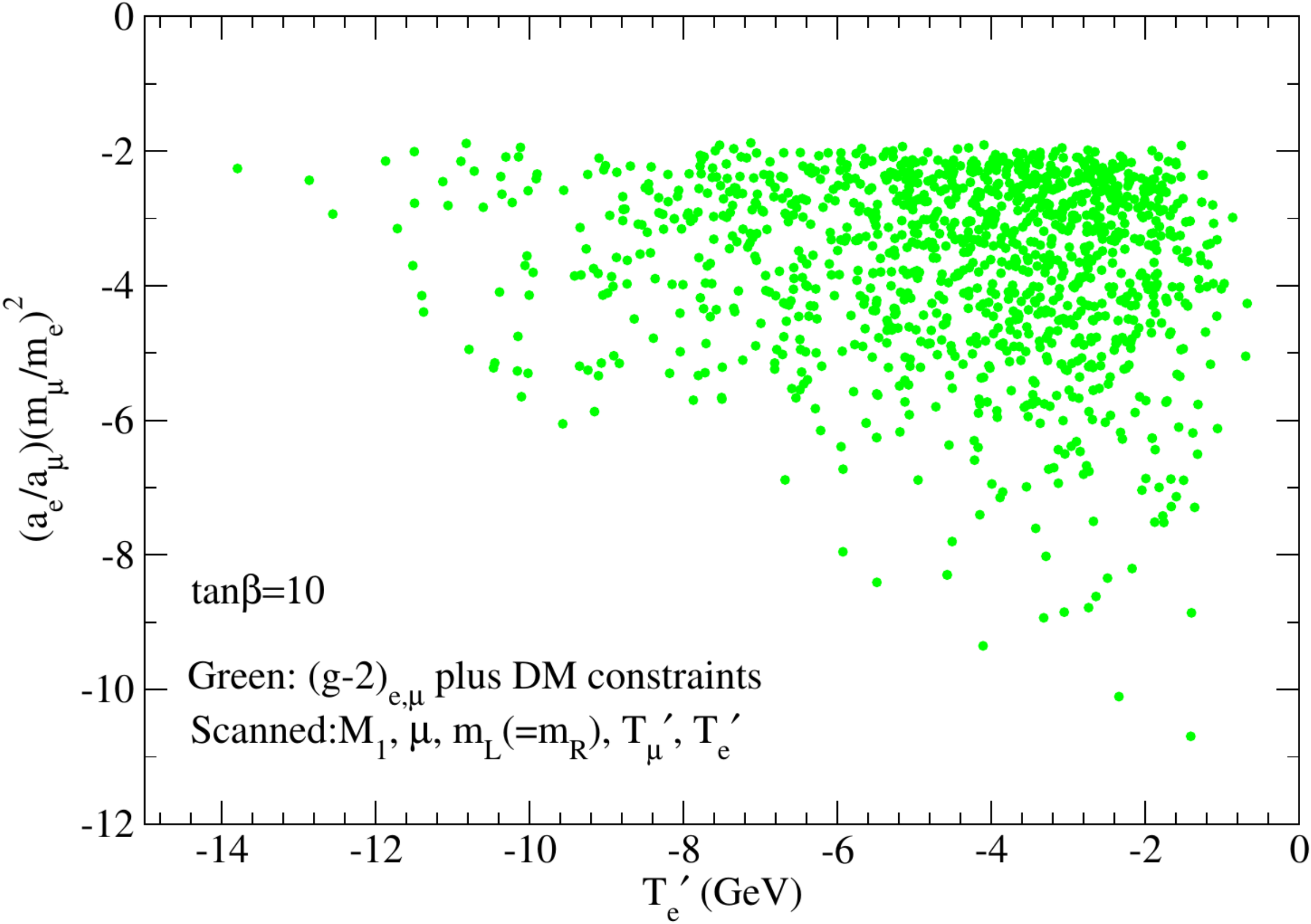}
                \label{magratioAeprime}
		}%
		\subfigure[]{%
			\includegraphics[width=0.45\textwidth]{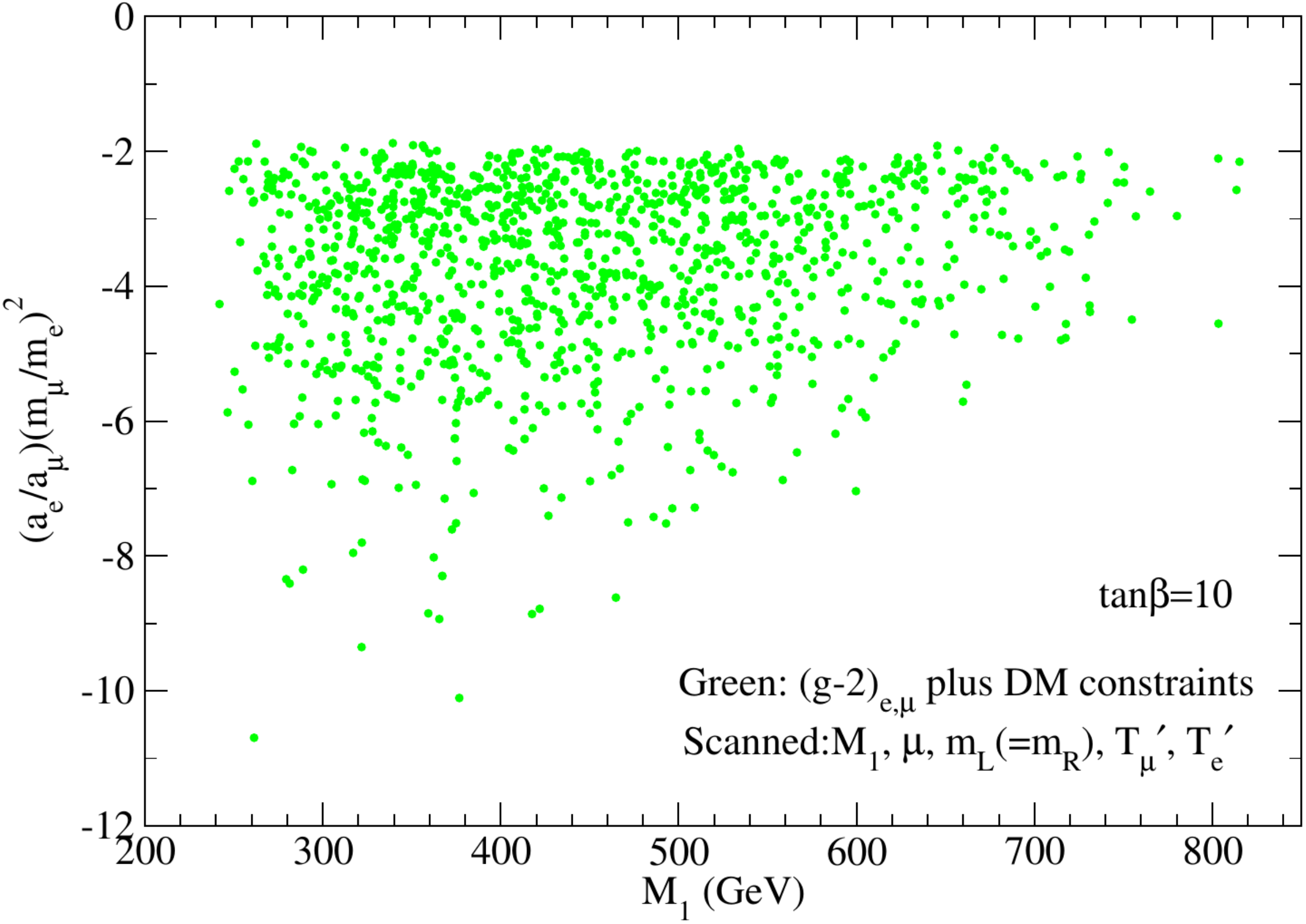}
                        \label{magratioM1}
		}%
		\caption{Fig.\ref{magratioAeprime}: Scatter plot
                of magnetic moment 
                ratio $a_e/a_\mu$ scaled with ${m_\mu}^2/{m_e}^2$ 
                against $T_e^\prime$ for $\tan\beta=10$.
All the points (shown in green) 
satisfy the DM relic density constraint for its upper limit, the XENON1T data for SI direct detection, $a_e$ and $a_\mu$ limits. LTCYC conditions are
applied in addition. 
Fig.\ref{magratioM1}: Similar scatter plot for a 
                variation of $M_1$. 
                 }
        \label{magmomratios}         
	\end{center}
\end{figure}

The fact that NHSSM parameter $T^\prime_e$ 
is able to generate a significant amount of correction to $y_e$,
while on the other hand there is an approximate sign
correlation between $a_{e,\mu}$ with $T^\prime_{e,\mu}$ gives one a
reasonable expectation for the scaled magnetic
moment ratio $R_{e,\mu}=\frac{(a_e/m_e^2)}{(a_\mu/m_\mu^2)}$ to be within the desired zone, namely
$R_{e,\mu} \simeq -15$ (see Eq.\ref{ratio_eqn}).
Figure~\ref{magmomratios} is a scatter plot of the above ratio with
parameters in the x-axis, namely $T^\prime_e$ and $M_1$
that are important for the radiative
corrections to Yukawa coupling $y_e$. All other parameters including also 
$T^\prime_\mu$ are varied 
according to the mentioned respective ranges of 
Table~\ref{table-parameters} considering $\tan\beta=10$ while also applying
the LTCYC conditions for the two leptons as mentioned earlier. 
All the points (shown in green) 
satisfy the DM relic density constraint for its upper limit, the XENON1T data for SI direct detection, $a_e$ and $a_\mu$ limits showing the desired $R_{e,\mu}$ values well within its uncertainty limits.
Fig.\ref{magratioM1} for $\tan\beta=10$ shows scatter points in the $M_1-R_{e,\mu}$ plane. The points with 
large $R_{e,\mu}$ 
correspond to smaller $M_1$ values. The DM satisfied (green) region satisfy 
$250 \lsim M_1 \lsim 800$~GeV. 

\FloatBarrier

\begin{figure}[hbt] 
	\begin{center}
		\subfigure[]{%
			\includegraphics[width=0.45\textwidth]{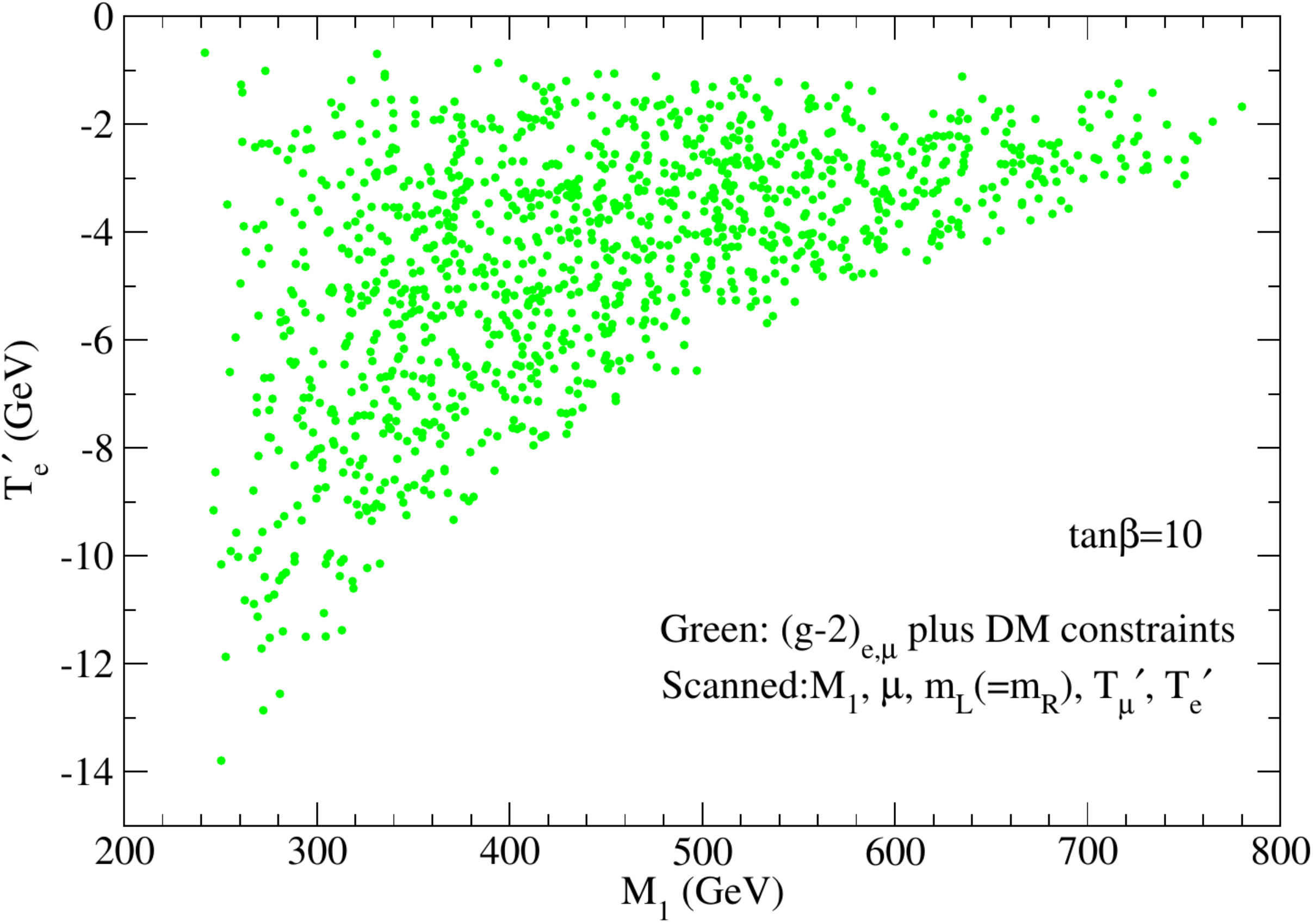}
\label{tanb10_bothmagmom_withDM_M1_Aeprime}
}%
		\subfigure[]{%
			\includegraphics[width=0.45\textwidth]{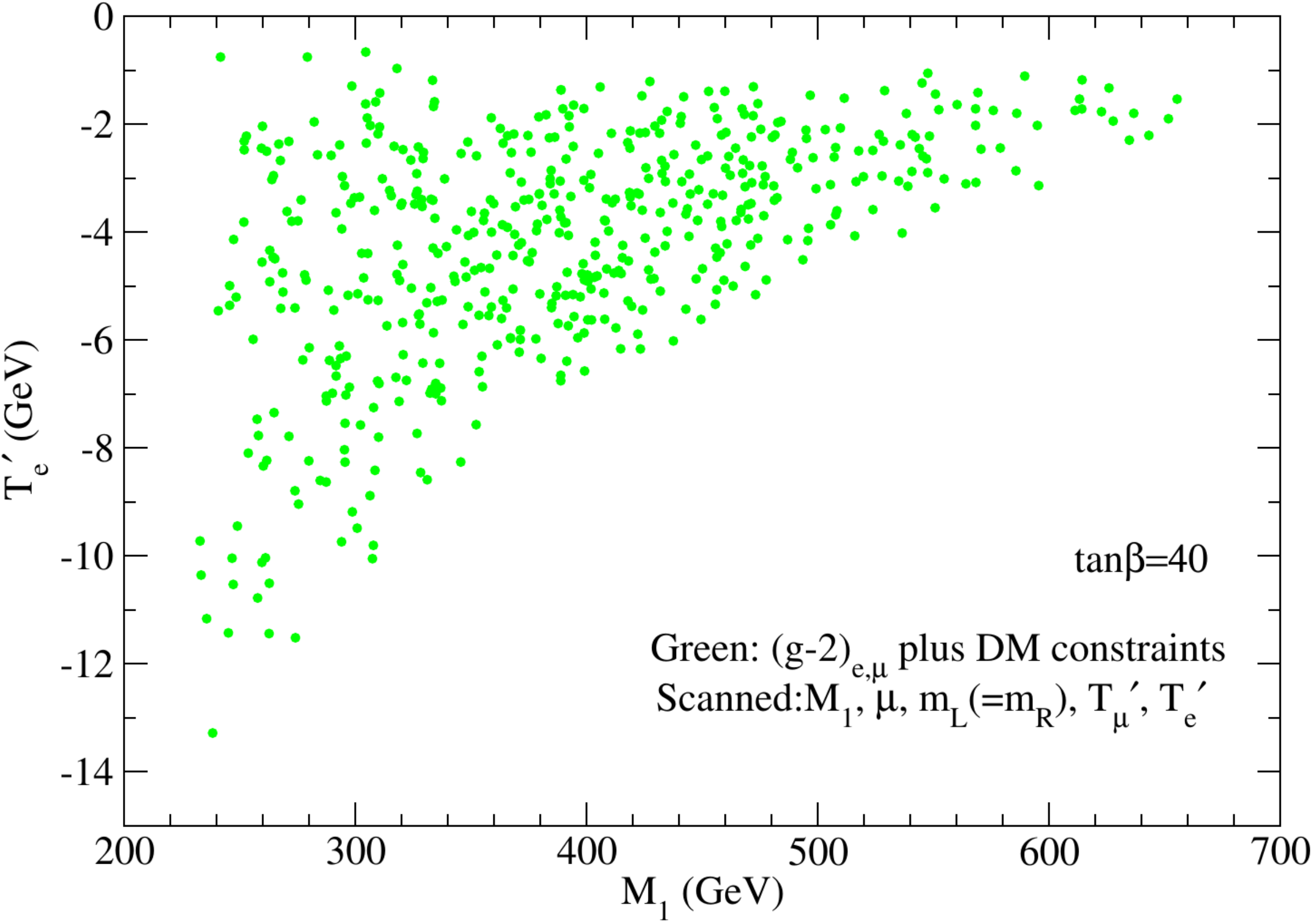}
            \label{tanb40_bothmagmom_withDM_M1_Aeprime}            
		}%
		\caption{Fig.\ref{tanb10_bothmagmom_withDM_M1_Aeprime}:
Scatter plots in the $M_1 - T^{\prime}_{e}$ plane 
                for $\tan\beta = 10$. The points shown in green
                arise from imposing all the constraints
like Higgs mass data, B-physics
limits, DM constraints for relic density and SI direct detection in addition
to $a_\mu$ and $a_e$ limits. Here and
henceforth in all the subsequent figures of this work
LTCYC limits for $y_e$ and $y_\mu$ are understood to have been applied by
default. 
Fig.\ref{tanb40_bothmagmom_withDM_M1_Aeprime}: Similar plot for $\tan\beta=40$.
}
\label{bothmagmom_withDM_M1_Aeprime}
	\end{center}
\end{figure}

We will now investigate how the conclusion of Section~\ref{gmuon_and_dm}
gets modified when 
we further restrict the NHSSM parameter space toward 
satisfying the $(g-2)_e$ constraint.
First, since we demand $y_e$ to be within the
LTCYC zone, $|T_e^\prime|$ should be small and this comes to around
14~GeV or less. Considering the fact that
SUSY contributions to $a_e$ needs to be negative,
the limits on $T^\prime_e$ 
becomes $-14 \lsim T^\prime_e< 0$.
Fig.~\ref{tanb10_bothmagmom_withDM_M1_Aeprime} shows the combined
results (green points) of imposing all the constraints
like Higgs mass data, B-physics
limits, DM constraints for relic density and SI direct detection in addition
to $a_\mu$ and $a_e$ limits drawn in the plane of $M_1-T_e^\prime$. Here and
henceforth in all the subsequent figures of this work
LTCYC limits for $y_e$ and $y_\mu$ are understood to have been applied by
default. 
The valid region for $M_1$ that satisfy all the constraints
comes out to be $230<M_1<800$~GeV for $\tan\beta=10$. 
Fig.~\ref{tanb40_bothmagmom_withDM_M1_Aeprime}, a similar plot
for $\tan\beta=40$ has a valid $M_1$ zone satisfying
$230<M_1<660$~GeV.

\begin{figure}[hbt]  
	\begin{center}
		\subfigure[]{%
			\includegraphics[width=0.45\textwidth]{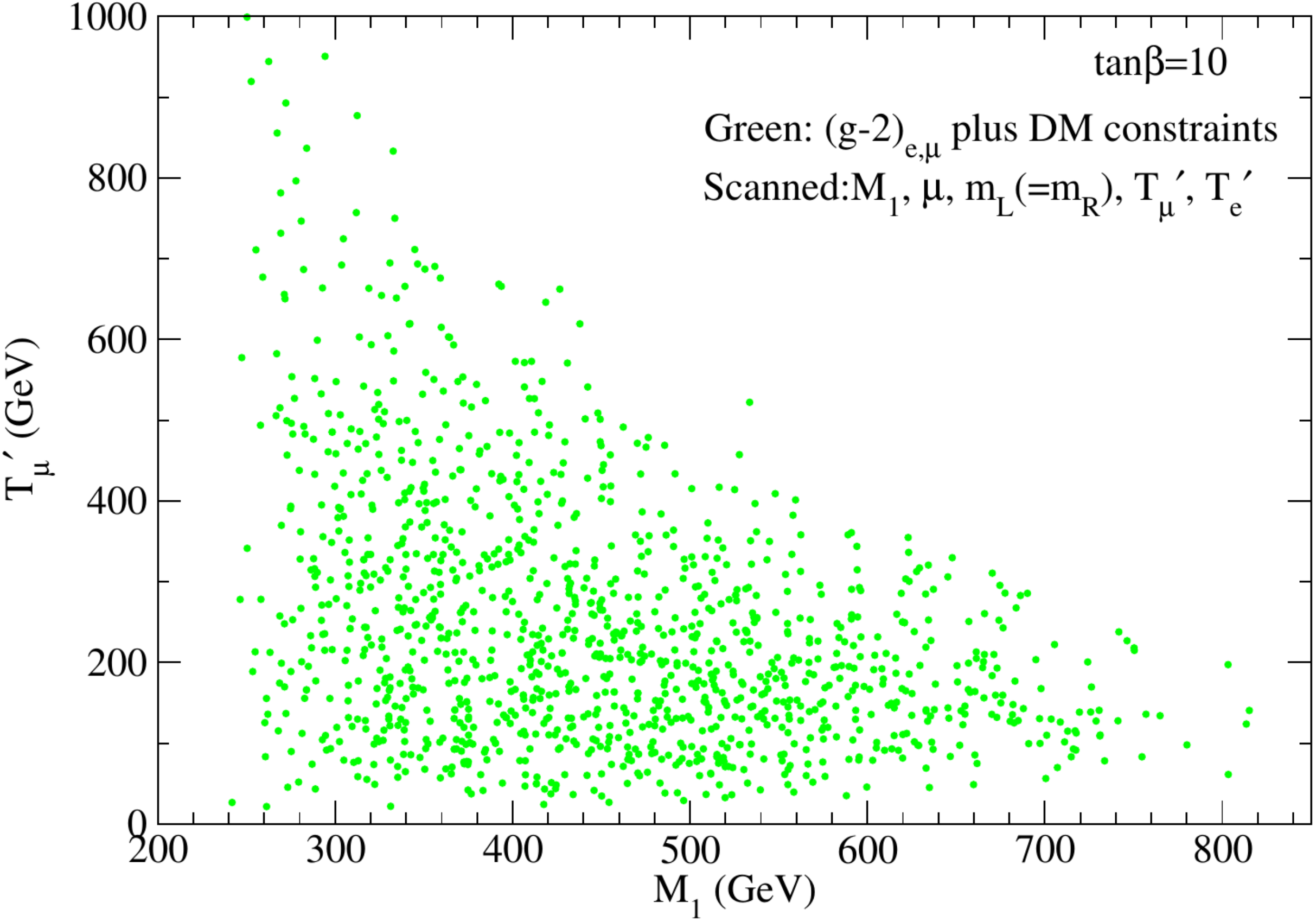}
       \label{tanb10_bothmagmom_withDM_M1_Amuprime}                 
		}%
		\subfigure[]{%
			\includegraphics[width=0.45\textwidth]{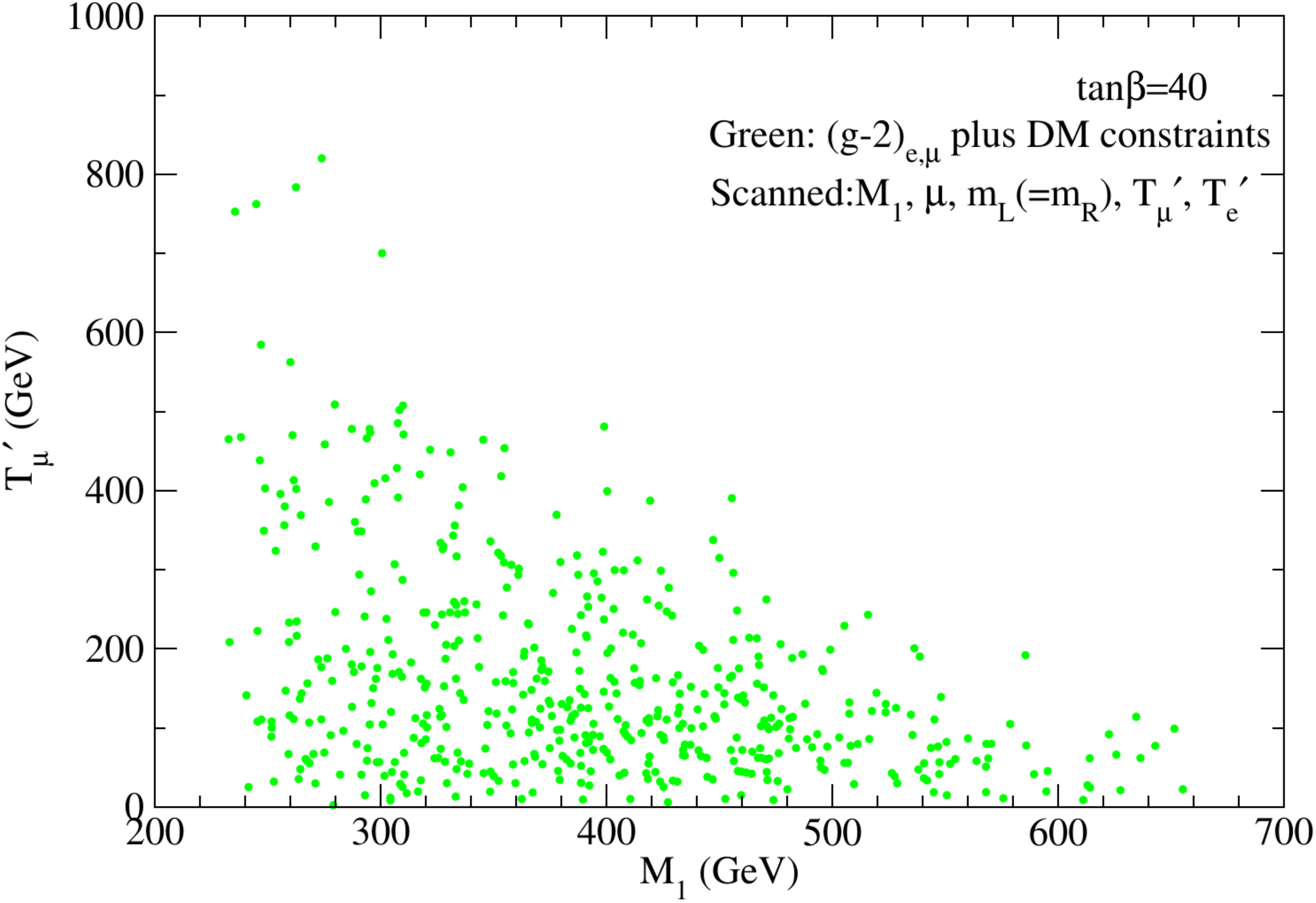}
        \label{tanb40_bothmagmom_withDM_M1_Amuprime}                
		}%
		\caption{Fig.\ref{tanb10_bothmagmom_withDM_M1_Amuprime}:                 Scatter plots in the $M_1 - T^{\prime}_{\mu}$ plane 
                for $\tan\beta = 10$.
The color scheme is same as that of Figure\ref{bothmagmom_withDM_M1_Aeprime}.  Fig.\ref{tanb40_bothmagmom_withDM_M1_Amuprime}:
Similar plot for $\tan\beta=40$.
                }
       \label{bothmagmom_withDM_M1_Amuprime}
	\end{center}
\end{figure}

Figure~\ref{bothmagmom_withDM_M1_Amuprime}
shows similar scatter plots in the $M_1-T^\prime_\mu$ plane.
The larger range of variation of $T^\prime_\mu$
compared to $T^\prime_e$ arises from the way $T^\prime_{e,\mu}$ are defined, 
namely these are scaled with the respective Yukawa couplings. 
The conclusion for 
valid upper limit of $M_1$ remains the same as before for both values of
$\tan\beta$.    

\begin{figure}[hbt] 
	\begin{center}
		\subfigure[]{%
           \includegraphics[width=0.45\textwidth]{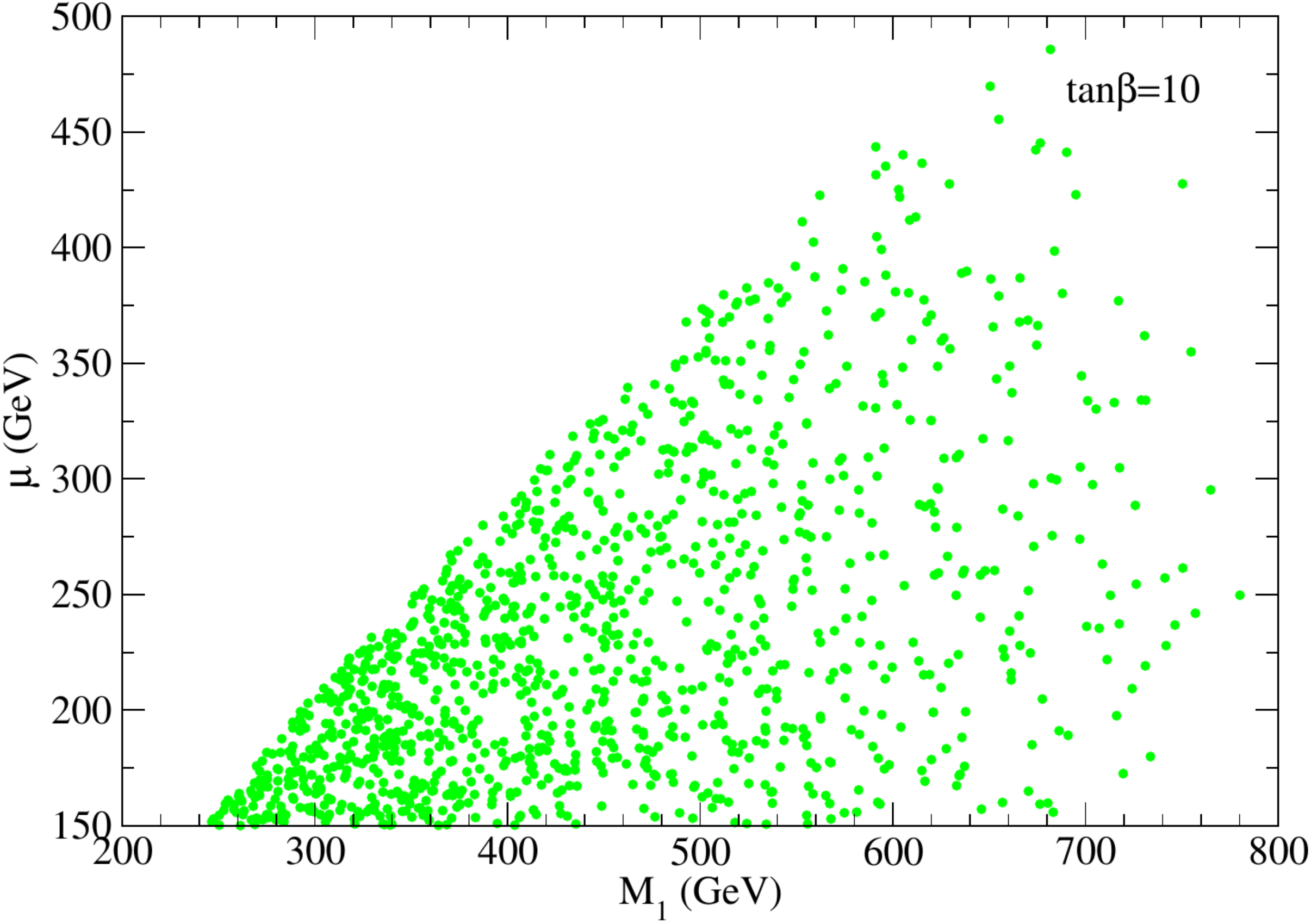}
         \label{tan10_allscanned_all_constraints_M1_vs_mu_both} 
		}%
		\hskip 30pt
		\subfigure[]{%
          \includegraphics[width=0.45\textwidth]{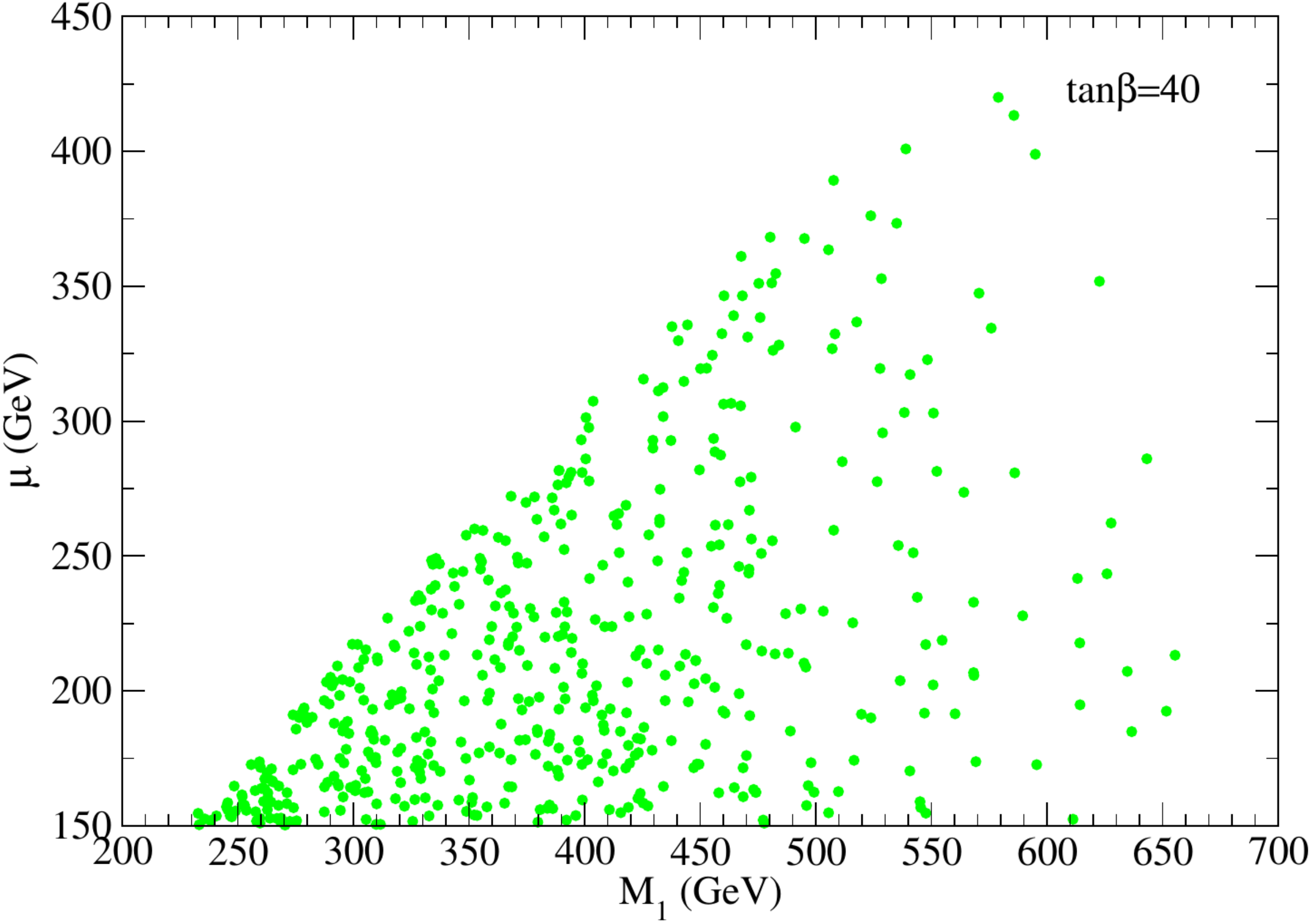}
        \label{tan40_allscanned_all_constraints_M1_vs_mu_both}                
		}%
		\caption{
Fig.\ref{tan10_allscanned_all_constraints_M1_vs_mu_both}: 
Scatter plot in $M_1-\mu$ plane for $\tan\beta = 10$. All the points shown in 
green satisfy $(g-2)_{e,\mu}$, dark matter relic density upper bound and
XENON1T provided $\sigma^{\rm SI}_{\chi p}$ bounds.
Fig.\ref{tan40_allscanned_all_constraints_M1_vs_mu_both}:
Similar plot for $\tan\beta=40$.
}
       \label{all_constraints_M1_vs_mu_both}         
	\end{center}
\end{figure}

We are to explore now the effect of imposing ${(g-2)}_e$ constraint on the
$M_1-\mu$ plane of Figure~\ref{all_constraints_M1_vs_mu_both}. We will compare
the above with the corresponding result of Figure~\ref{M1_vs_Mu_muon}
drawn for ${(g-2)}_\mu$. The availability of NHSSM parameter space gets
reduced significantly because the threshold corrections to
$y_e$ should not become too large. We remember
that the above corrections at least involve masses of bino,
sleptons, as well as the trilinear NH coupling $T^\prime_e$.
Hence, a given set of $M_1$ and $m_L$ that is 
available in the muon $(g-2)$ analysis, may not be 
consistent when combined with non-vanishing 
$T^\prime_e$ to satisfy ${(g-2)}_e$. Thus,
when one imposes the ${(g-2)}_e$ constraint, that in turn is
sensitive on the corrections to $y_e$, the reduction of parameter space is
rather unavoidable.
The green points of Fig.~\ref{tan10_allscanned_all_constraints_M1_vs_mu_both}
satisfy $(g-2)_{e,\mu}$ constraints at 2$\sigma$ level in addition to obeying
the dark matter relic density bound. The associated 
$\sigma^{\rm SI}_{\chi p}$ values also fall below the XENON1T limit.
Fig.~\ref{tan40_allscanned_all_constraints_M1_vs_mu_both} shows a
similar result for $\tan\beta=40$ where a largeness of $\tanbeta$ further
takes away a significant amount of parameter space so as to have $y_e$
within LTCYC zone.

\begin{figure}[hbt]
	\begin{center}
		\subfigure[]{%
			\includegraphics[width=0.45\textwidth]{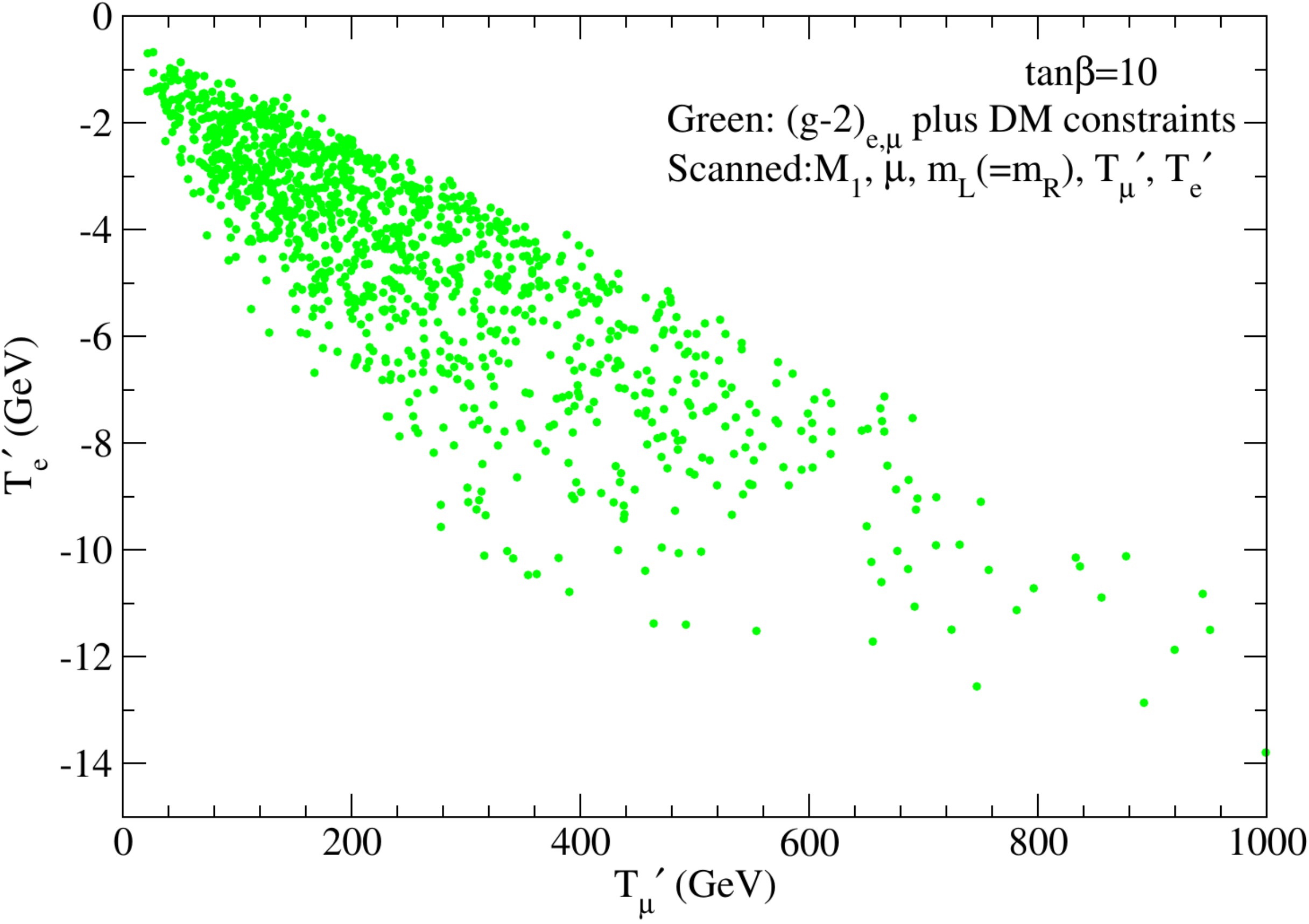}
          \label{tanb10_bothmagmom_withDM_Amuprime-Aeprime_new}              
		}%
		\hskip 30pt
		\subfigure[]{%
			\includegraphics[width=0.45\textwidth]{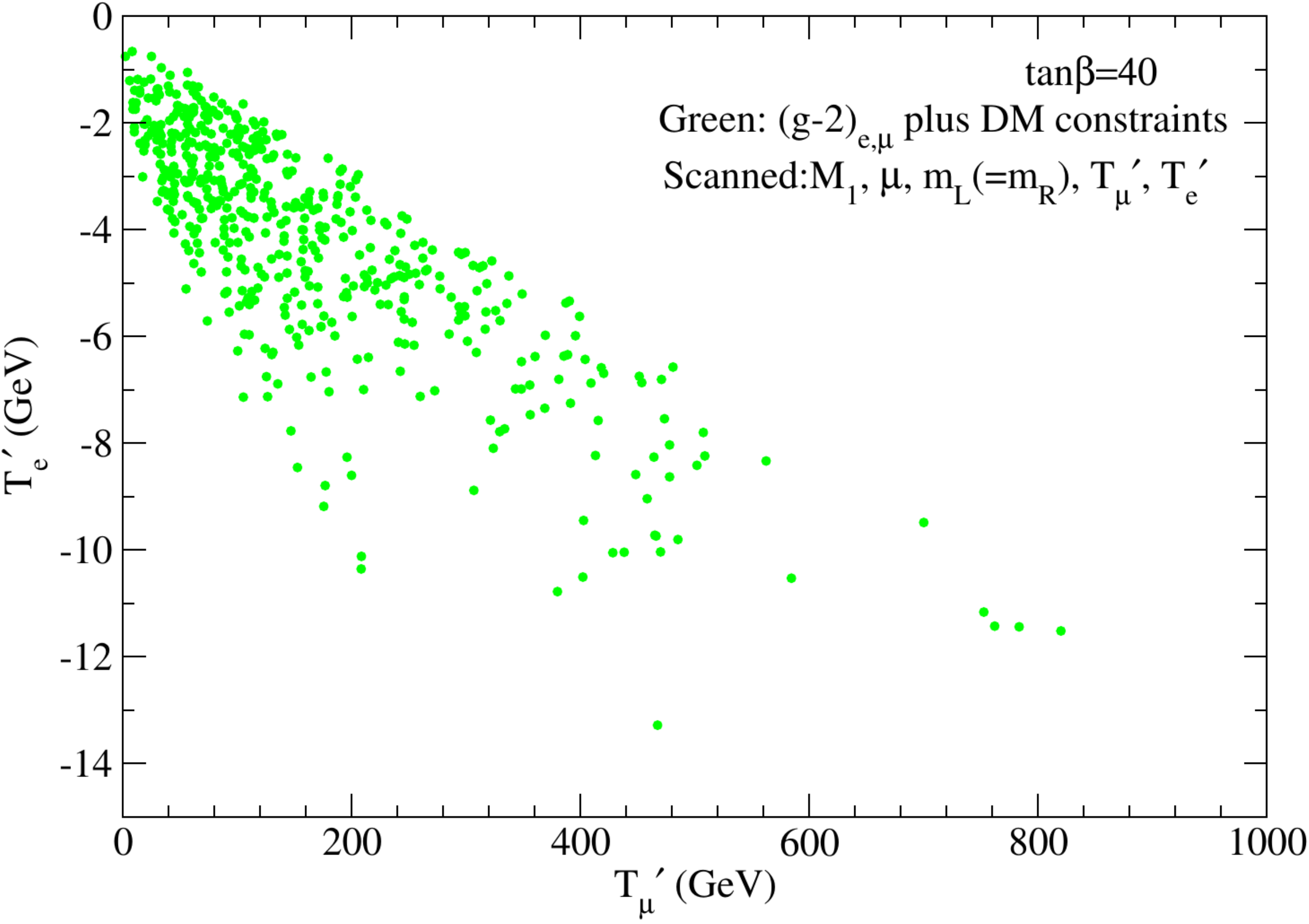}
          \label{tanb40_bothmagmom_withDM_Amuprime-Aeprime_new}     
		}%
\caption{Fig.\ref{tanb10_bothmagmom_withDM_Amuprime-Aeprime_new}:
Scatter plots in $T^{\prime}_e - T^{\prime}_{\mu}$ plane for $\tan\beta=10$.
The color scheme is same as that of
Figure\ref{bothmagmom_withDM_M1_Aeprime}.  Fig.\ref{tanb40_bothmagmom_withDM_Amuprime-Aeprime_new}: Similar plot for $\tan\beta=40$.
}
      \label{bothmagmom_withDM_Amuprime-Aeprime_new}          
	\end{center}
\end{figure}

Fig.\ref{tanb10_bothmagmom_withDM_Amuprime-Aeprime_new} is a scatter
plot in the plane of trilinear NH parameters for electron and muon for
$\tan\beta=10$. The  points shown in green  
satisfy the DM relic density constraint, the XENON1T $\sigma^{\rm SI}_{\chi p}$
data along with
$(g-2)_\mu$ and $(g-2)_e$ limits at $2\sigma$ level. The points also
have $y_e$ in the LTCYC zone.  
Clearly, $T^\prime_e$ should be 
appreciably large so as to satisfy $(g-2)_e$ data. We find the valid region
to have $-14~{\rm GeV} <T^\prime_e<0$.

\subsection{LHC Constraints from slepton pair production and constraint
from compressed higgsino LSP scenario}
\label{LHCconstraintsSection}
\subsubsection{Both $(g-2)_\mu$ and $(g-2)_e$}
With no difference in SUSY sparticle content 
between NHSSM and MSSM, we directly apply the SUSY constraints
from LHC data on our analysis. We intend to identify the exclusion
region of NHSSM parameter space from the ATLAS data
for slepton pair production that 
considered selectron and smuon in the analysis\cite{ATLAS:2019lff}. The
later gives an exclusion region in the $(m_{\tilde{\chi}^0_1}-m_L)$ plane.
We note here that by directly applying the ATLAS bound on our parameter
space we are taking a conservative approach as the 
exclusion can potentially get somewhat weaker
for the higgsino LSP scenario. The reason is as follows. The ATLAS limit is derived for
a simplified model assuming $BR(\tilde l \rightarrow  l \lspone) = 100 \%$,
This criterion is not strictly satisfied in the higgsino LSP region, 
where the proximity of $\mlspone$ and $\mchonepm$ allows for a 
significant branching ratio of the sleptons to final states involving $\nu_l$
and $\chonepm$. This may lead to a reduction in the  number of signal leptons
and thus to the weakening of the exclusion limit.

We will further analyze the constraint from a compressed scenario
with higgsino as LSP that is associated with closely spaced values for
the masses $\mlspone$, $\mchonepm$ and $\mlsptwo$. This
is based on the ATLAS result given in
Fig.14a of Ref.\cite{ATLAS:2019lng}.

\begin{figure}[hbt] 
	\begin{center}
\includegraphics[width=0.45\textwidth]{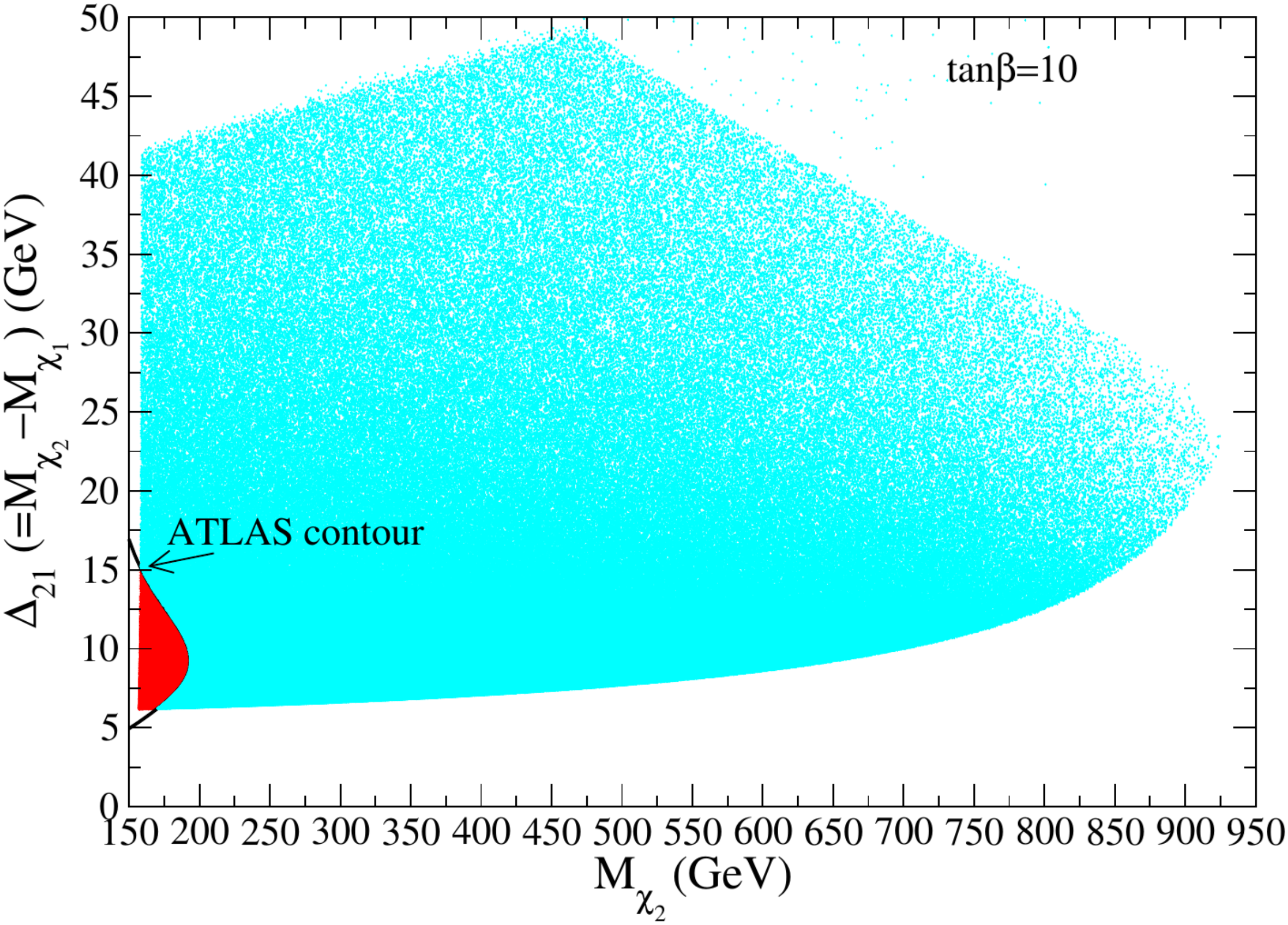}
\end{center}
\caption{Allowed (cyan) and disallowed (red) regions in the plane of
$\mlsptwo-\Delta_{21}$ with respect to
the 1$\sigma$-level contour (black line) as given by the
ATLAS data namely Fig.14a of Ref.\cite{ATLAS:2019lng}. The corresponding
lower limit of $\mlspone$ satisfying the above constraint is about 185 GeV in this higgsino dominated $\lspone$-$\lsptwo$-$\chonepm$ scenario.}
\label{atlas_compressed_higgsino_fig}
\end{figure}
Focusing on higgsino type of LSPs, we
show Figure~\ref{atlas_compressed_higgsino_fig} drawn in 
the plane of $\mlsptwo-\Delta_{21}$ where $\Delta_{21}=\mlsptwo-\mlspone$. 
The above-mentioned 1$\sigma$ contour of the ATLAS result
is shown in black. The red colored points appearing to the left of the   
contour form the discarded zone, whereas the cyan colored points survive
the constraint. Clearly, a compressed scenario with $\Delta_{21}$
having values up to 10 to 15 GeV are eliminated for $\mlsptwo \lsim 200$~GeV.
Larger values of $\Delta_{21}$ in the same zone of $\mlsptwo$ are alright.
The corresponding lower limit of $\mlspone$ is approximately
185 GeV. Thus, we may conclude that not all values of LSP masses 
below 185 GeV are valid. Hence, with a conservative standpoint we henceforth
consider the lowest value for a higgsino type of LSP to be 185 GeV.
\footnote{On the other hand,
regarding a compressed LSP-slepton
scenario there is hardly any constraint from LHC
(Fig. 16a of Ref.\cite{ATLAS:2019lff}) in our analysis.}

\begin{figure}[hbt] 
	\begin{center}
		\subfigure[]{%
                \includegraphics[width=0.45\textwidth]{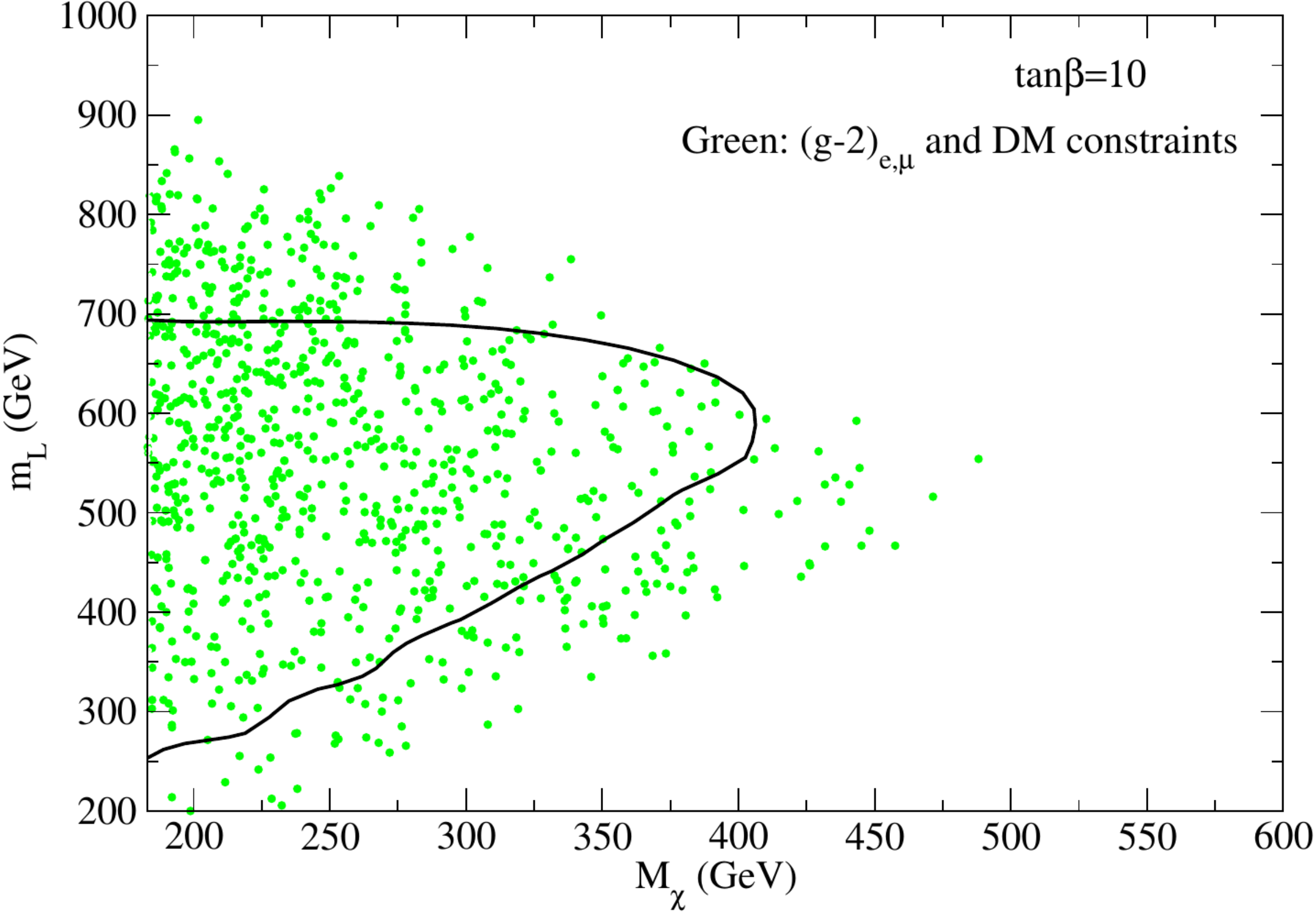}
       \label{tan10_allscanned_all_constraints_lsp_vs_ml_both}                 
		}%
		\hskip 30pt
		\subfigure[]{%
			\includegraphics[width=0.45\textwidth]{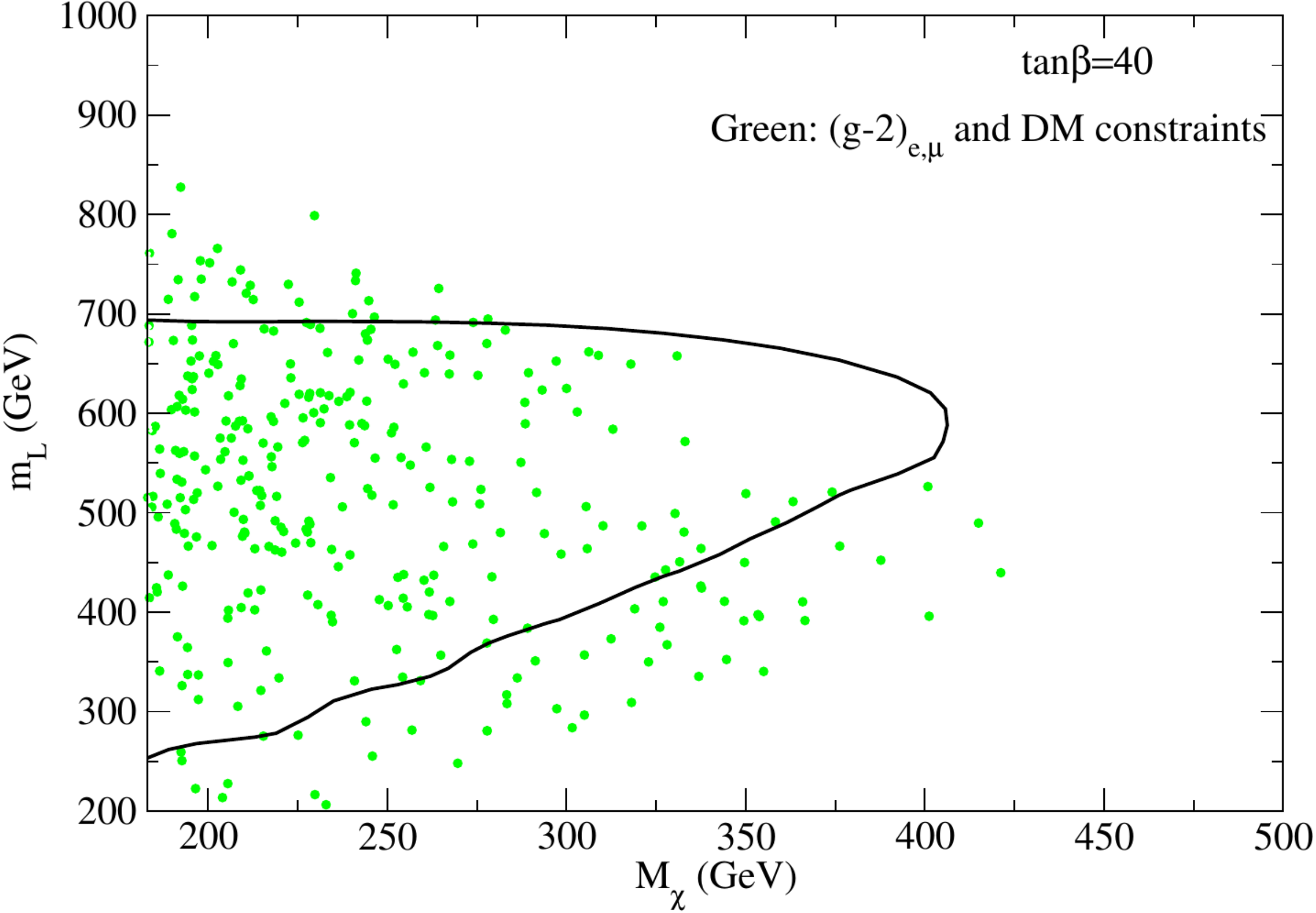}
        \label{tan40_allscanned_all_constraints_lsp_vs_ml_both}
		}%
	\caption{Fig.\ref{tan10_allscanned_all_constraints_lsp_vs_ml_both}
        Scatter plot in $m_{\tilde{\chi}^0_1}-m_L$ 
        plane for $\tan\beta=10$.
The  points shown in green  
satisfy the DM relic density constraint, the XENON1T $\sigma^{\rm SI}_{\chi p}$
data along with
$(g-2)_\mu$ and $(g-2)_e$ limits at $2\sigma$ level. The points also
have $y_e$ in the LTCYC zone.
 The black line indicates the current exclusion bounds in the
 slepton-neutralino plane
as obtained from slepton pair production data from ATLAS,
LHC\cite{ATLAS:2019lff}.
All the points pass the compressed higgsino
constraint as mentioned in the text.
Fig.\ref{tan40_allscanned_all_constraints_lsp_vs_ml_both}:
Similar plot for $\tan\beta=40$.
}
\label{tan1040_allscanned_all_constraints_lsp_vs_ml_both}
	\end{center}
\end{figure}

Fig.\ref{tan10_allscanned_all_constraints_lsp_vs_ml_both}
        shows a scatter plot in $m_L - m_{\tilde{\chi}^0_1}$
        plane for $\tan\beta=10$. All the points
        satisfy the
        perturbativity of $y_e$.
        Considering the constraint for compressed
        higgsino states as mentioned above, $m_{\tilde{\chi}^0_1}$ is allowed
        to have values above 185 GeV.
The  points shown in green  
satisfy the DM relic density constraint, the XENON1T $\sigma^{\rm SI}_{\chi p}$
data along with
$(g-2)_\mu$ and $(g-2)_e$ limits at $2\sigma$ level. The points also
have $y_e$ in the LTCYC zone.
On the top of the figure we draw the    
black line that indicates
the current exclusion bounds in the $m_L - m_{\tilde{\chi}^0_1}$ plane at the
LHC. The green points that stay outside the contour are the
residual parameter points that would also satisfy the LHC limit. On the
higher LSP mass side, irrespective of slepton masses, for $\tan\beta=10$
a value of $m_{{\tilde \chi}^0_1}$ between 400 to
500 GeV would satisfy all the constraints considered in this analysis. For slepton masses above 700 GeV, valid LSP
mass would lie below 350 GeV. 
Fig.\ref{tan40_allscanned_all_constraints_lsp_vs_ml_both}
shows a similar result for $\tan\beta=40$. Here, the LHC contour engulfs
a lot of region of the parameter space with a larger mass of the
LSP as shown in green. The region with slepton
masses above 700~GeV allows valid LSP mass up to 275~GeV. 
Apart from the above, the LSP mass values for valid parameter points outside
and near the 
black contour in its lower side are correlated with $m_L$ in an approximately
linear fashion.  One finds that with $\tan\beta=10$,
$m_{{\tilde \chi}^0_1}$ can be as large as 
400~GeV when $m_L=550$~GeV whereas the later numbers are 400 GeV
and 500 GeV respectively for $\tan\beta=40$.

\noindent
\subsubsection{Only $(g-2)_\mu$}
\begin{figure}[hbt] 
	\begin{center}
		\subfigure[]{%
                \includegraphics[width=0.45\textwidth]{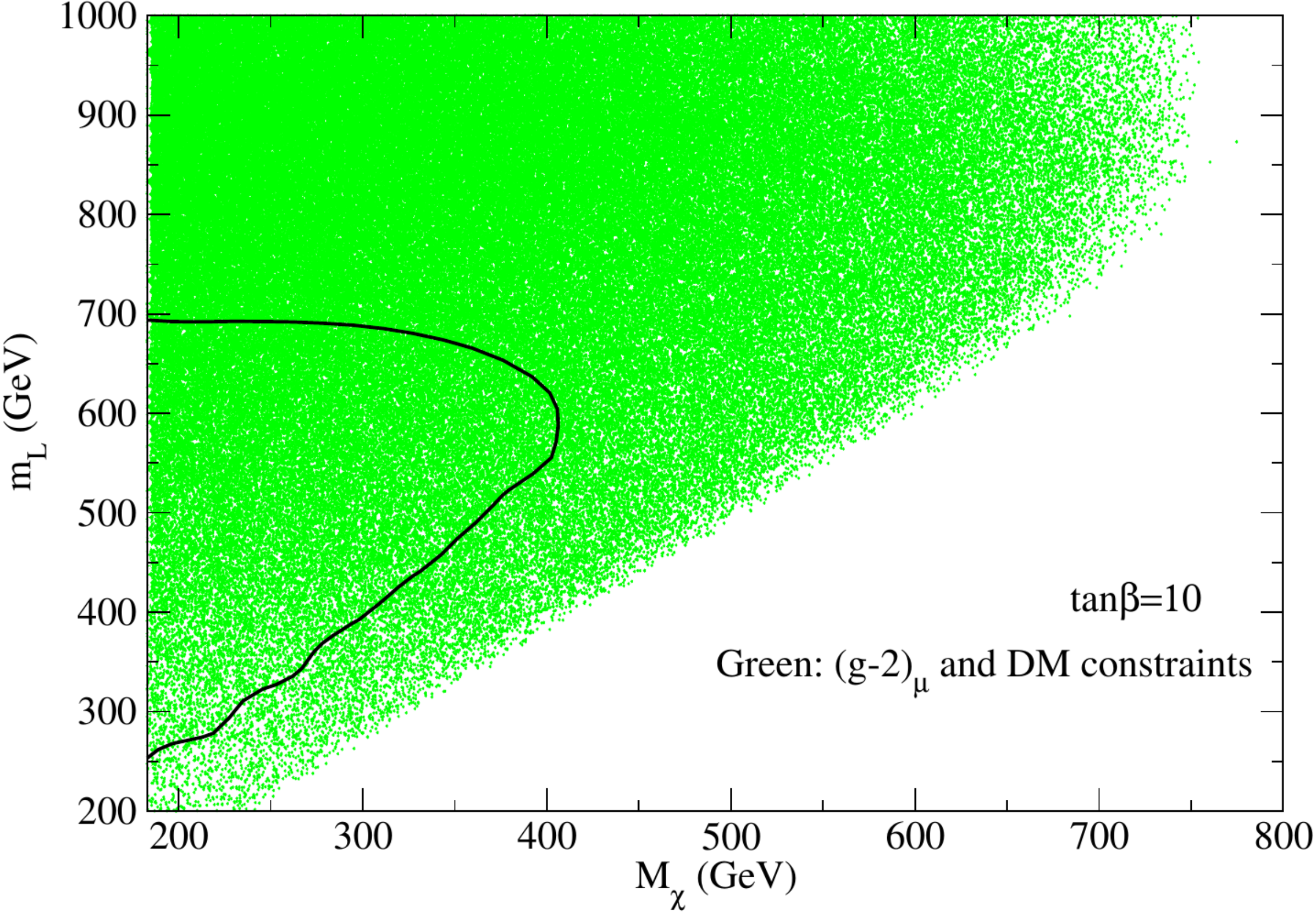}
       \label{tan10_allscanned_all_constraints_lsp_vs_ml_muon}                 
		}%
		\hskip 30pt
		\subfigure[]{%
			\includegraphics[width=0.45\textwidth]{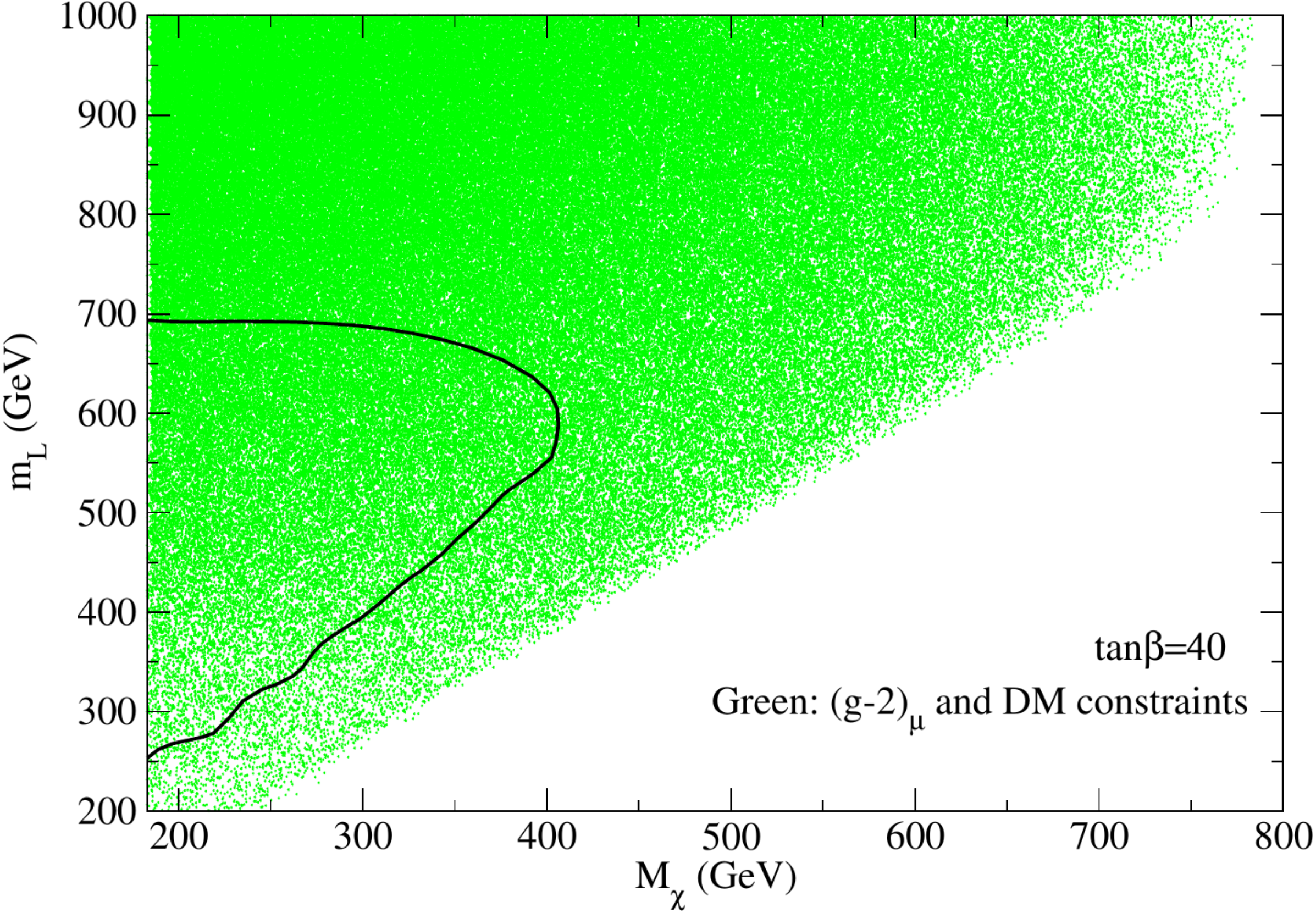}
        \label{tan40_allscanned_all_constraints_lsp_vs_ml_muon}
		}%
	\caption{Fig.\ref{tan10_allscanned_all_constraints_lsp_vs_ml_muon}
        Scatter plot in $m_{\tilde{\chi}^0_1}-m_L$ 
        plane for $\tan\beta=10$.
The  points shown in green  
satisfy the DM relic density constraint, the XENON1T $\sigma^{\rm SI}_{\chi p}$
data along with
$(g-2)_\mu$ and $(g-2)_e$ limits at $2\sigma$ level. The points also
have $y_e$ in the LTCYC zone.
 The black line indicates the current exclusion bounds in the slepton-neutralino plane
as obtained from slepton pair production data from ATLAS,
LHC\cite{ATLAS:2019lff}.
All the points pass the compressed higgsino
constraint as mentioned in the text.
Fig.\ref{tan40_allscanned_all_constraints_lsp_vs_ml_muon}:
Similar plot for $\tan\beta=40$.
}
\label{tan1040_allscanned_all_constraints_lsp_vs_ml_muon}
	\end{center}
\end{figure}

Since the LHC data is seen to affect the combined analysis of ${(g-2)}_\mu$ and
${(g-2)}_e$ with dark matter rather strongly,
we feel that it is important to go back and examine the situation of
considering only the ${(g-2)}_\mu$ constraint with dark matter
(as in Section\ref{gmuon_and_dm}) in relation to the ATLAS
constraints described above.  
The results are seen in 
Figure~\ref{tan10_allscanned_all_constraints_lsp_vs_ml_muon} and \ref{tan40_allscanned_all_constraints_lsp_vs_ml_muon}.
Figure~\ref{tan10_allscanned_all_constraints_lsp_vs_ml_muon} shows the 
ATLAS data\cite{ATLAS:2019lff} satisfied appropriate mass values of sparticles in the
$m_L - m_{\tilde{\chi}^0_1}$ plane. As before, we also impose the
constraint from the compresed higgsino scenario.  
For $\tan\beta=10$, satisfying all the constraints, valid $m_{\tilde{\chi}^0_1}$
ranges from 400 GeV to 750~GeV for any value of $m_L$. 
On the other hand when $m_L>700$~GeV, $m_{\tilde{\chi}^0_1}$ 
assumes values up to the same upper limit 
namely 750 GeV.
For $\tan\beta=40$, as seen 
in Figure~\ref{tan10_allscanned_all_constraints_lsp_vs_ml_muon} the above numbers for
$m_{\tilde{\chi}^0_1}$
are in between 400 to 775 GeV, whereas for $m_L>700$~GeV, valid $m_{\tilde{\chi}^0_1}$ zone extends to 775 GeV.
Apart from the above, the LSP mass values for valid (green) parameter points outside and near the 
black contour in its lower side are correlated with $m_L$ in an approximately
linear fashion.  One finds for both the values of $\tan\beta$, $m_{{\tilde \chi}^0_1}$ can be as large as 
400~GeV when $m_L=550$~GeV. 

We now show two representative points of our analysis in Table~\ref{bptable}
for two values of $\tanbeta$.
Referring to Figure~\ref{tan10_allscanned_all_constraints_lsp_vs_ml_both}
that corresponds to satisfying both $(g-2)_\mu$ and $(g-2)_e$ constraints,  
the two points are chosen based on the ATLAS provided contours\cite{ATLAS:2019lff} for slepton pair production shown in black.
We choose $tan\beta=10$ and 40 for specifying two benchmark points BP1 and BP2 both of which correspond to 
moderate values of $\mlspone$ and slepton mass parameter $m_L$. The points
are far away from having any effect due to the constraints from compressed
spectra. 
We remind that all the left and right slepton mass parameters
are taken to be equal. The LSP is almost a 
higgsino in nature. The relic density is satisfied via
 like $\tilde \chi_1^0-
\tilde \chi_2^0$ or $\tilde \chi_1^0-\tilde \chi_1^{\pm}$ coannihilation
mechanisms.
Apart from other constraints like ${(g-2)}_{e,\mu}$ and dark matter,
the constraints from B-physics are easily satisfied because of
large squark masses of the
third generation. With a larger stop-mixing via $A_t$,
the Higgs mass is satisfied within the
limits of Eq.\ref{higgsmassdata}. Notably, $a_e$ value of BP2
for $\tan\beta=40$ is similar to that of BP1 corresponding to $\tan\beta=10$.
This is because, as discussed earlier, in NHSSM, the SUSY contributions
to $a_e$ has a milder 
$\tan\beta$ dependence unlike a proportional relationship as seen in MSSM. 
With LSP as a higgsino it is a candidate of 
a subdominant component of DM. Hence we use a scale factor $\xi$ (Eq.\ref{scalefactorequation}) that further reduces the  
SI and SD direct detection cross sections. Certainly, the valid
parameter space would be larger for both small and large $\tan\beta$
if we had chosen only $(g-2)_\mu$ among the two magnetic moment
constraints.

\begin{center}
	\begin{table}[!htbp]
		\caption{Representative Points in NHSSM
                corresponding to SUSY scale input of soft parameters
                that satisfy all the constraints considered in the analysis
                along with $a_\mu$ and $a_e$
              (using data from ${}^{133}{\rm Cs}$-based measurement).
                All the dimensionful parameters are in GeV. Here, $T_e^\prime=y_e A_e^\prime$ and $T_t=y_t A_t$.}
		\label{bptable}  
		\centering
		
		\begin{tabular}{|c||c|c|}
			
			\hline\hline
			
			Parameters  &  BP1 & BP2  \\ [0.5ex]
			\hline
			$M_1$  & $553$ & $539$ \\
			$\mu$ & $411$ & $401$ \\
                        $B_\mu$ & $6.19\times 10^5$ & $1.56 \times 10^5$ \\                        
			$\tan\beta$ &  10 & 40  \\
$T_t$ & $-3500$ & $-3500$\\ 
 $T_e^{\prime}, T_{\mu}^{\prime}$ &  $-3.7$, $180.8$ & $-3.1$, $190.1$ \\
$m_L \equiv M_{\tilde{L}_{11,22}}=M_{\tilde{e}_{11,22}} \equiv m_R$ & 594 & 526\\
			
			\hline\hline
			$a_{\mu}^{SUSY}$ & $1.87 \times 10^{-9}$ & $3.17 \times 10^{-9}$\\
			$a_{e}^{SUSY}$  & $-1.60 \times 10^{-13}$ & $-1.63 \times 10^{-13}$\\
			\hline
			$m_h$ & 122.8 & 123.9\\
			$m_H,m_A,m_{H^{\pm}}$ & {2481,2481,2482} & {2373,2373,2377}\\
			$m_{\tilde t_{1,2}}$  & { 3947, 4111} & { 3943, 4102} \\
			$m_{\tilde b_{1,2}}$  & { 4056, 4091} & { 4043, 4067}\\
			$m_{\tilde e_{1,2}}$  & { 609, 640} & { 543, 577}\\
			$m_{\tilde \mu_{1,2}}$  & { 593, 652} & { 522,592}\\
			$m_{\tilde \tau_{1,2}}$  & { 2004, 2016} & { 1983, 2008} \\
			$m_{\tilde \chi_{1,2}^0}$  & { 410, 424} & { 401, 414} \\
			$m_{\tilde \chi_{1,2}^{\pm}}$  & {421,1534} & {411, 1534}\\
			$m_{\tilde g}$  & 3026 & 3026 \\
			\hline
			$BR(B \rightarrow X_s +\gamma)$  & $3.19\times10^{-4}$ & $3.13\times10^{-4}$ \\
			$BR(B_s \rightarrow \mu^+ \mu^-)$  & $3.22\times10^{-9}$ &  $3.18\times10^{-9}$\\
			\hline
			$\Omega_{DM} h^2$ & $0.0243$ & $0.0230$ \\
			$\sigma_{\chi p}^{SI}$  & $1.53\times10^{-9}$ & $1.29\times10^{-9}$\\
		$\xi \sigma_{\chi p}^{SI}$  & $3.15\times10^{-10}$ & $2.50\times10^{-10}$\\
			$\sigma_{\chi n}^{SD}$ & $1.50 \times10^{-5}$ & $1.72 \times10^{-5}$\\
				$\xi \sigma_{\chi n}^{SD}$  & $3.09\times10^{-6}$ & $3.35\times10^{-6}$\\
			\hline\hline 
		\end{tabular}
	\end{table}
\end{center}

\FloatBarrier  
\section{Results for $a_e$ constraint from the ${}^{87}{\rm Rb}$-based experiment}
\begin{figure}[hbt]
	\begin{center}
		\subfigure[]{%
			\includegraphics[width=0.45\textwidth]{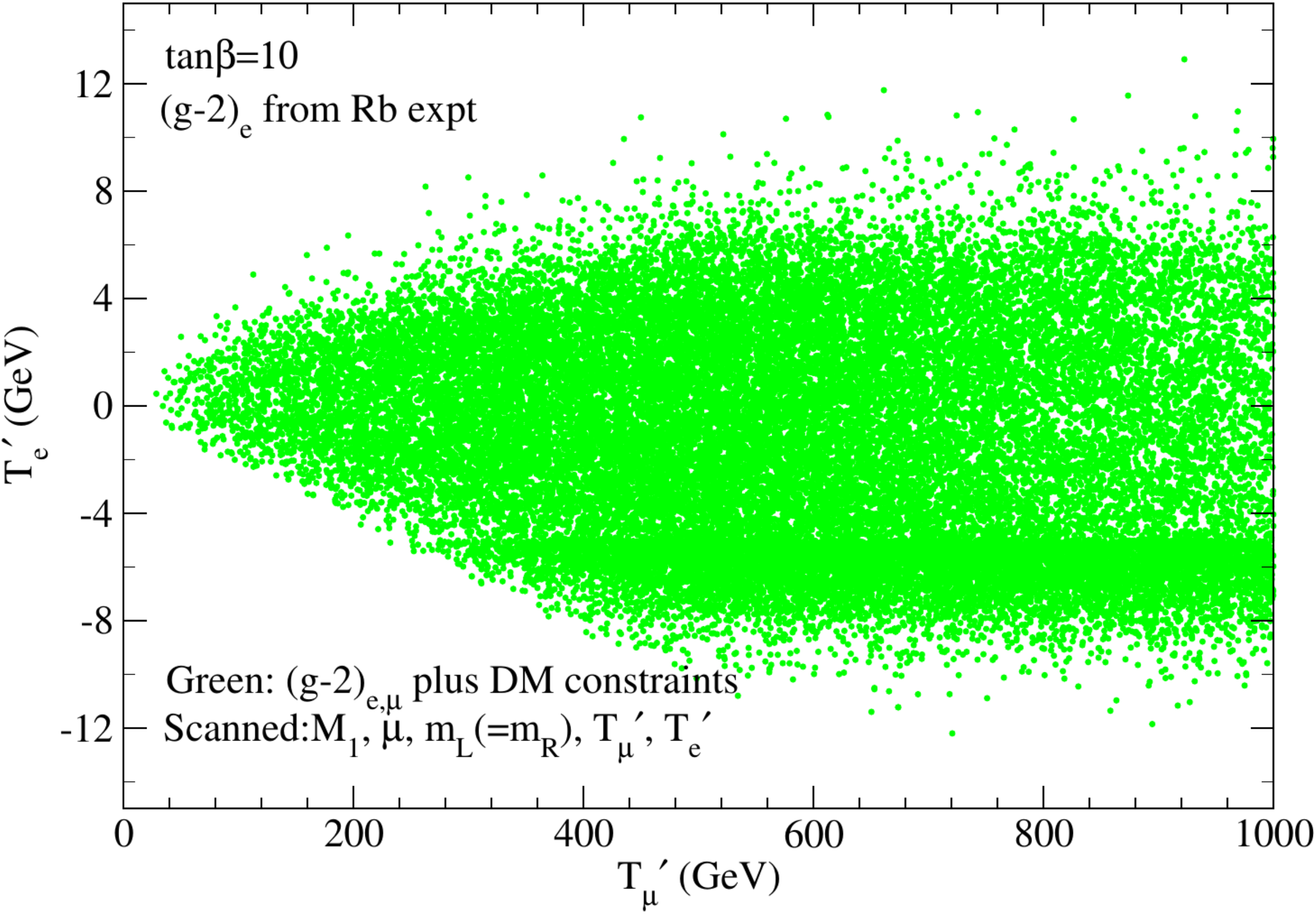}
          \label{tanb10_bothmagmom_withDM_Amuprime-Aeprime_new_Rb}              
		}%
		\hskip 30pt
		\subfigure[]{%
			\includegraphics[width=0.45\textwidth]{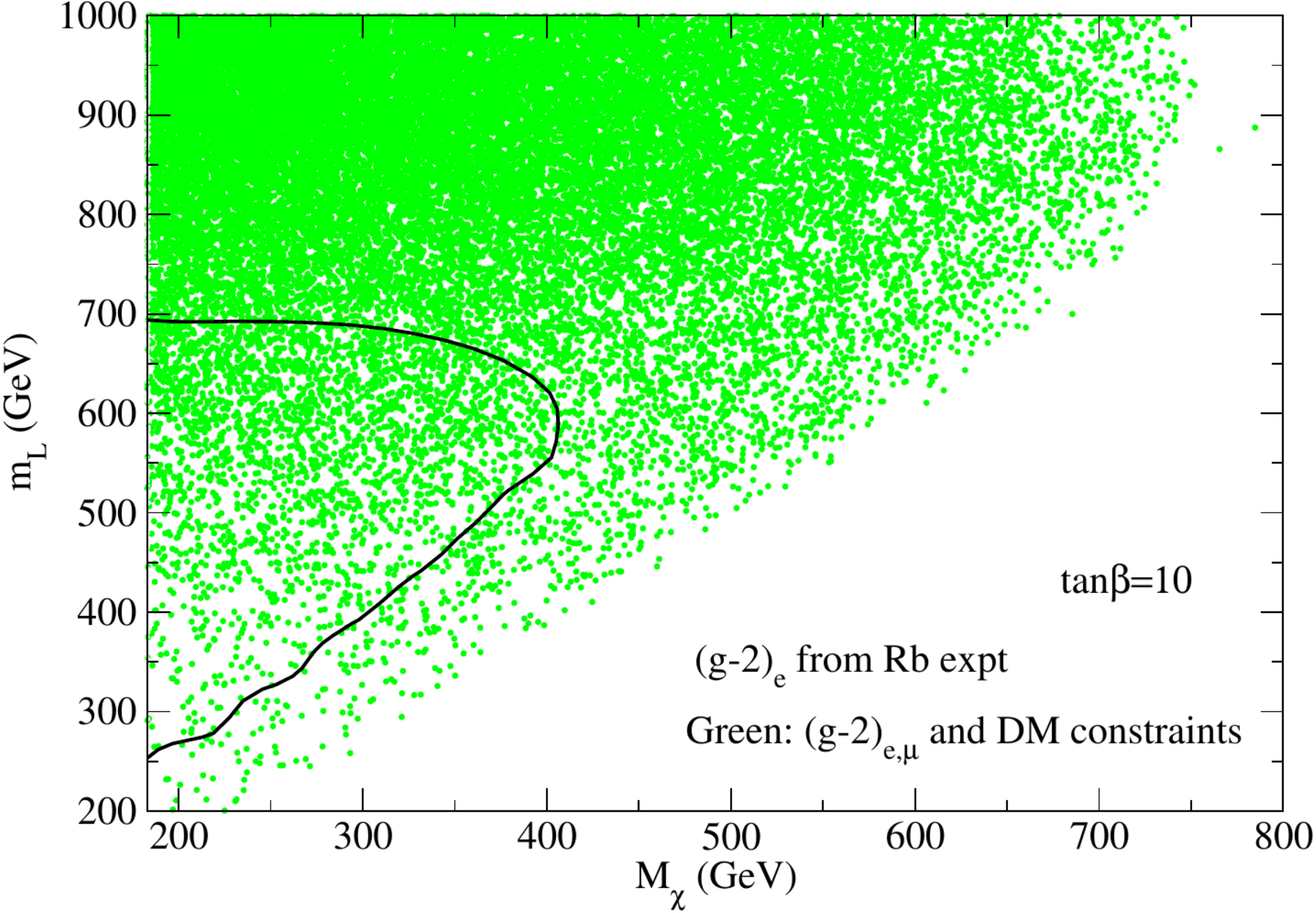}
        \label{tan10_allscanned_all_constraints_lsp_vs_ml_both_Rb}
}
\caption{Fig.\ref{tanb10_bothmagmom_withDM_Amuprime-Aeprime_new_Rb}:
Analysis with ${}^{87}{\rm Rb}$ data for $a_e$ for scatter-plots in the 
$T^{\prime}_e - T^{\prime}_{\mu}$ plane $\tan\beta=10$.
The color scheme and constraints imposed are same as that of
Figure\ref{bothmagmom_withDM_Amuprime-Aeprime_new}.
Fig.\ref{tan10_allscanned_all_constraints_lsp_vs_ml_both_Rb}:
Analysis with ${}^{87}{\rm Rb}$ data for $a_e$ for scatter-plots in the 
$m_{\tilde{\chi}^0_1}-m_L$ plane for $\tan\beta=10$.
Further details are same as those of
Figure \ref{tan1040_allscanned_all_constraints_lsp_vs_ml_both}.  
}
      \label{Analyses_with_Rb}          
	\end{center}
\end{figure}

Finally, we briefly discuss the effects of using the $a_e$
constraint from Eq.\ref{delta_aerb}, i.e. from the ${}^{87}{\rm Rb}$-based
experiment as mentioned earlier. There is a 1.6$\sigma$ deviation from
the SM result and the spread is more toward the positive side, unlike the
case of Eq.\ref{delta_ae} corresponding to the experiment using
${}^{133}{\rm Cs}$ matter-wave interferometry.  
Focusing on only $\tan\beta=10$, the 
scatter-plot of Figure \ref{tanb10_bothmagmom_withDM_Amuprime-Aeprime_new_Rb}
in the $T^{\prime}_e - T^{\prime}_{\mu}$ plane shows the requirement of
both positive and negative signs of $T^{\prime}_e$ in relation to the
results shown in Figure \ref{ye_and_ae_vsAeprime_also_muon}.
The  points shown in green  
satisfy the DM relic density constraint, the XENON1T $\sigma^{\rm SI}_{\chi p}$
data along with
$(g-2)_\mu$ and $(g-2)_e$ limits at $2\sigma$ level. The points also
have $y_e$ in the LTCYC zone.
A milder deviation corresponding to Eq.\ref{delta_aerb} 
makes the result to have 
similarity with the case of using only the $a_\mu$ constraint.
Similarly, we plot Figure \ref{tan10_allscanned_all_constraints_lsp_vs_ml_both_Rb} indicating parameter points in the $m_{\tilde{\chi}^0_1}-m_L$ plane.
Further details regarding constraint
from the ATLAS data are same as those of
Figure \ref{tan1040_allscanned_all_constraints_lsp_vs_ml_both}.
Expectedly, the result is closer to that of
Figure \ref{tan10_allscanned_all_constraints_lsp_vs_ml_muon} for the only
$\gmin2$ case.

\FloatBarrier  

\section{Conclusion}
\label{conclusionsection}
The Standard Model of particle physics has achieved a
remarkable degree of success ever since the discovery of the Higgs Boson. Apart from the 
above discovery itself, the measurements of Higgs couplings, various
results from the ATLAS and the CMS experiments of the LHC, several
results from low energy precision measurements involving
flavor physics also point towards a strong validity of SM. However,
theoretical issues such as the gauge hierarchy problem,
and observational evidences like the dark matter, baryon asymmetry,
non-zero neutrino mass etc. demands the SM to be extended beyond its current boundary.
Hence, a search for a BSM physics is
highly relevant in the present era. In this connection,
we note that the recent Fermilab experiment on
muon $g-2$ measurement has confirmed the deviation from the SM
value obtained in the previous measurement at BNL.
Assuming that the SM result that
depends on low energy $e^+ e^-$ data will not get drastically
altered in the near future, or in other words, it would take a while
for the lattice-based computations of the hadronic error to reach a definite
conclusion, it is important to 
pursue the implications of the Fermilab announcement from the BSM physics perspective.

The combined discrepancy of the Fermilab and the
BNL data is at an impressive level of 4.2$\sigma$.
The anomaly amount itself is of the order of
electroweak corrections in SM. Thus, it is very crucial to probe a BSM
physics scenario that may produce a large contribution to the above 
observable at the same energy scale. SUSY, undoubtedly one of the most
important candidates for new physics, is well-known to be capable of producing a large 
correction to ${(g-2)}_\mu$. There are regions of parameter
space in MSSM with appropriate wino, bino, higgsino and left and right
handed smuon masses that can explain the Fermilab
result, typically for a large $\tan\beta$. There are also dedicated
beyond the MSSM models that can
enhance the $(g-2)_\mu$ contribution by virtue of the model 
properties, including new symmetries.
In this analysis, staying within the MSSM sparticle spectra, we
augment MSSM with non-holomorphic
trilinear soft breaking interactions.  We also pointed out that
unlike MSSM soft terms, the non-holomorphic terms which for MSSM
are soft SUSY breaking in nature
do not have the privilege of being supported by
popular UV complete models.
The appropriate NH term generates an enhancement of the SUSY
contributions to $(g-2)_\mu$ via a coupling effect
which helps to easily accommodate the data.
The model neither demands any
flavor unfriendly choice of different right and left slepton masses
nor does it 
demand a large $\tan\beta$ or a light chargino. One uses a large  
bino-smuon loop contribution with stronger mixing of
the L-R smuons to produce the desired effect.  Using the Fermilab data to constrain the model,
we further impose constraints due to dark matter related observables. It is found that a
higgsino-dominated dark matter having an underabundant contribution to the DM relic
density can be consistent with direct detection cross-section as well as ${(g-2)}_\mu$. 
For ${(g-2)}_\mu$, a scenario with dominant bino-smuon loop contribution
can satisfy the data via a large L-R mixing caused by the wrong
Higgs couplings of the non-holomorphic terms.

We further impose the relevant direct search bounds from the LHC, e.g. the
constraints from the slepton pair production searches from ATLAS as well as the constraint coming from the
compressed higgsino searches.
The conclusion of the combined ${(g-2)}_\mu$ analysis with dark matter and
collider limits is the following
(Figure~\ref{tan1040_allscanned_all_constraints_lsp_vs_ml_muon}).
For $\tan\beta=10$ and $40$, satisfying all the constraints, valid
$m_{\tilde{\chi}^0_1}$ 
ranges from 400 GeV to 750~GeV for any value of $m_L$.
On the other hand when $m_L>700$~GeV, $m_{\tilde{\chi}^0_1}$
stays in between the lower limit of $\sim 185$ GeV to
the same upper limit namely 750 GeV.
Apart from the above, the LSP mass values for the valid parameter points outside and near the
ATLAS provided contour in its lower side are correlated with $m_L$ in an approximately
linear fashion.  One finds for both the values of $\tan\beta$, $m_{{\tilde \chi}^0_1}$ at its minimum can be around 
400~GeV when $m_L \simeq 600$~GeV.

We then extend the muon $g-2$ analysis in NHSSM
to include the electron $g-2$ data as derived from the
fine-structure constant measurement based on 
${}^{133}{\rm Cs}$ matter-wave interferometry.
The resulting ${(g-2)}_e$ shows an approximately
$2.5\sigma$ level of discrepancy, but interestingly, it comes with a negative sign.
It becomes very challenging to accommodate the muon and electron g-2 anomalies simultaneously in MSSM.
This measurement is in contrast to the ${}^{87}{\rm Rb}$-based measurement
which shows a smaller positive deviation. Here, we use the more unfriendly ${}^{133}{\rm Cs}$-based data in our work to compare
our results with previous MSSM-based analyses and for probing the potential 
of the non-standard soft terms in relation to any related phenomenology.
The present work approaches to accommodate $a_e$ via
appropriately enhancing the 
Yukawa threshold corrections for $y_e$ by using minimum possible 
$T^\prime_e$ values for the chosen mass spectra. 
The analysis satisfies constraints from vacuum stability and 
charge and color breaking minima.   
The negative sign for the SUSY contributions to
$a_e$ could be generated by negative $T^\prime_e$. An interesting feature of $a_e$ in NHSSM is that
one cannot enhance it appreciably by increasing $\tan\beta$. $a_e$ at best can be a slowly increasing
function of $\tan\beta$, but unlike MSSM, it is no longer proportional to $\tan\beta$,
an effect that arises out of a large threshold correction to $y_e$.

The conclusion of the combined ${(g-2)}_\mu$ and ${(g-2)}_e$ analysis with dark matter and collider limits is
the following (see Figure~\ref{tan1040_allscanned_all_constraints_lsp_vs_ml_both}).
For $\tan\beta=10$, satisfying all the constraints, valid $m_{\tilde{\chi}^0_1}$
ranges from 400 GeV to 500~GeV for any value of $m_L$.
On the other hand when $m_L>700$~GeV, $m_{\tilde{\chi}^0_1}$
stays in between the lower limit of $\sim$185 GeV to the same upper
limit namely 500 GeV.
For $\tan\beta=40$, the ATLAS contour engulfs
a lot of parameter space with larger mass of the LSP leaving a small
window of $\mlspone$ between 400 to 425 GeV for any $m_L$. 
For $m_L>700$~GeV, the valid $m_{\tilde{\chi}^0_1}$ zone
is from 185 GeV to 275 GeV.
Apart from the above, the LSP mass values for valid parameter points outside and near the
ATLAS provided contour in its lower side are correlated with $m_L$ in an approximately
linear fashion.  One finds that for $\tan\beta=10$, $m_{{\tilde \chi}^0_1}$ can be as large as 
410~GeV for $m_L$ below 550~GeV.  For $\tan\beta=40$, $m_{{\tilde \chi}^0_1}$ can be as large as
425~GeV when $m_L$ is below 440~GeV.
Finally, we presented two representative points
that satisfy all the constraints considered in this analysis. We also
discussed briefly the results of considering the limits of
$a_e$ as coming from the ${}^{87}{\rm Rb}$-based experiment indicating a
smaller deviation from the SM result. 

\section*{Acknowledgements}
IA would like to thank the Council of Scientific and Industrial Research, India for financial support.
UC would like to acknowledge the use of the High Performance Cluster at IACS.
The work of MC is supported by the project AstroCeNT:
Particle Astrophysics Science and Technology Centre,  carried out within
the International Research Agendas programme of
the Foundation for Polish Science financed by the
European Union under the European Regional Development Fund.

\newcommand\jnl[1]{\textit{\frenchspacing #1}}
\newcommand\vol[1]{\textbf{#1}}

\end{document}